\documentclass[fleqn,times,usenatbib]{mnras}

\usepackage{newtxtext,newtxmath}
\usepackage{mathptmx}
\usepackage{amsmath}

\usepackage{amssymb}
\usepackage{caption}
\usepackage[figuresright]{rotating}
\usepackage{multicol}

\usepackage[T1]{fontenc}
\usepackage{cuted}

\DeclareRobustCommand{\VAN}[3]{#2}
\let\VANthebibliography\thebibliography
\def\thebibliography{\DeclareRobustCommand{\VAN}[3]{##3}\VANthebibliography}

\usepackage{graphicx}	
\newcommand{\mJyb}{mJy beam$^{-1}$}

\title[VLBI Observations of PG  quasars II]{VLBI Observations of a sample of Palomar-Green quasars II: characterising the parsec-scale radio emission} 

\author[Wang et al.]{
Ailing Wang,$^{1,2}$
Tao An,$^{1,2,3}$\thanks{E-mail: antao@shao.ac.cn}
Yingkang Zhang$^{1,3}$,
Xiaopeng Cheng$^{4}$,
Luis C. Ho$^{5,6}$, \newauthor
Kenneth I. Kellermann$^{7}$,
Willem A. Baan$^{8,9}$ \\
$^{1}$ Shanghai Astronomical Observatory, CAS, 80 Nandan Road, Shanghai 200030, China \\
$^{2}$ School of Astronomy and Space Sciences, University of Chinese Academy of Sciences, No. 19A Yuquan Road, Beijing 100049, China  \\
$^{3}$ Key Laboratory of Radio Astronomy and Technology, Chinese Academy of Sciences, A20 Datun Road, Beijing, 100101, P. R. China \\
$^{4}$ Korea Astronomy and Space Science Institute, 776 Daedeok-daero, Yuseong-gu, Daejeon 34055, Korea \\
$^{5}$ Kavli Institute for Astronomy and Astrophysics, Peking University, Beijing 100871, China \\
$^{6}$ Department of Astronomy, School of Physics, Peking University, Beijing 100871, China \\
$^{7}$ National Radio Astronomy Observatory, 520 Edgemont Rd., Charlottesville, VA 22903, USA \\
$^{8}$ Xinjiang Astronomical Observatory, Chinese Academy of Sciences, 150 Science 1-Street, 830011 Urumqi, P.R. China \\
$^{9}$ Netherlands Institute for Radio Astronomy ASTRON, NL-7991 PD Dwingeloo, the Netherlands  \\
}

\date{Accepted XXX. Received YYY; in original form ZZZ}

\pubyear{2015}

\begin{document}
\label{firstpage}
\pagerange{\pageref{firstpage}--\pageref{lastpage}}
\maketitle

\begin{abstract}

This study uses multi-frequency Very Long Baseline Interferometry (VLBI) to study the radio emission from 10 radio-quiet quasars (RQQs) and four radio-loud quasars (RLQs). The diverse morphologies, radio spectra, and brightness temperatures observed in the VLBI images of these RQQs, together with the variability in their GHz spectra and VLBI flux densities, shed light on the origins of their nuclear radio emission. The total radio emission of RQQs appears to originate from non-thermal synchrotron radiation due to a combination of active galactic nuclei and star formation activities. 
However, our data suggest that the VLBI-detected radio emission from these RQQs is primarily associated with compact jets or corona, with extended emissions such as star formation and large-scale jets being resolved by the high resolution of the VLBI images. Wind emission models are not in complete agreement the VLBI observations. Unlike RLQs, where the parsec-scale radio emission is dominated by a relativistically boosted core, the radio cores of RQQs are either not dominant or are mixed with significant jet emission. RQQs with compact cores or core-jet structures typically have more pronounced variability, with flat or inverted spectra, whereas jet-dominated RQQs have steep spectra and unremarkable variability. Future high-resolution observations of more RQQs could help to determine the fraction of different emission sources and their associated physical mechanisms.

\end{abstract}




\section{Introduction}

Active galactic nuclei (AGN) are divided into two classes based on the ratio of radio to optical flux densities: radio-loud (RL) and radio-quiet (RQ) \citep{1989AJ.....98.1195K,1994AJ....108.1163K,1993MNRAS.263..425M}. The radio emission from RLAGN originates mainly from relativistic jets \citep{2015MNRAS.452.1263P,2019ARA&A..57..467B} that can extend up to megaparsec (Mpc) scales \citep[e.g.][]{2022NatAs...6..109M}, far beyond the host galaxy. These jets have a profound impact on galactic evolution and the interstellar environment. However, the origins of the radio emission from RQAGN are more complex and include galactic-scale star formation activity, accretion disk winds, shocks generated by collisions between the winds and the interstellar medium (ISM) in galactic nucleus regions, corona emission, and low-power jets \citep[reviewed by][]{2019NatAs...3..387P}. These components operate at different physical scales and radio frequencies, and together they constitute the overall radio emission from RQAGN. Given that RQAGN make up the majority of the AGN population, \textit{a detailed study of the radio emission mechanisms of RQAGN is critical to understanding the AGN emission process}.

The origins of radio emission in RQQs are multifaceted, with different origins exhibiting distinct observational characteristics at different frequencies and resolutions. This diversity allows us to employ diagnostic tools to separate these origins and determine their respective contributions. For instance, star-forming activity, which is generally distributed throughout the entire galaxy \citep[e.g. ][]{1983ApJS...53..459C}, is associated with relatively diffuse radio emission. Some extreme starbursts have been observed in the central regions of ultra-luminous infrared galaxies (with sizes from 1 to a few kpc), but the radio emission from nuclear starbursts remains unresolved in 1.4 GHz arcsec-resolution images obtained with the Very Large Array (VLA) \citep{1991ApJ...378...65C}. However, most of the radio emission associated with star formation is resolved in the higher resolution VLA images and milliarcsecond (mas) resolution Very Long Baseline Interferometry (VLBI) images, leaving only a small fraction of the flux density in young supernovae and supernova remnants \citep[e.g.][]{1987ApJ...323..505B,2006ApJ...647..185L,2019A&A...623A.173V}. \textit{Therefore, the compactness of the radio emission structure is a useful tool for distinguishing whether it originates from star formation or is driven by AGN.} 

In addition, the brightness temperature of a radio source is an indicator that distinguishes between star formation and AGN. While the brightness temperatures of star-forming galaxies do not exceed $10^5$~K \citep{1992ARA&A..30..575C,2013ApJ...768...37C},  the brightness temperatures of AGN-related radio emission typically exceeds $10^8$~K, especially in VLBI observations \citep[e.g.][]{1998MNRAS.299..165B,2005ApJ...621..123U,2022ApJ...936...73A,2023MNRAS.518...39W}. 

Extended accretion disk winds and jets are the main sources of radio emission in RQAGN on parsec to sub-kpc scales. \textit{These two processes can be distinguished from the radio morphology in high-resolution images}: radio jets are generally continuous, compact and narrow (some are collimated, while others are highly bent), whereas winds have large opening angles and are usually clumpy and irregular in shape.

On parsec scales, compact jets, winds, and corona can all contribute to radio emission. Relativistic jets and mildly relativistic winds may coexist in AGN on sub-parsec scales \citep{2014ApJ...780..120F}; however, observational evidence for relativistic winds has only been found in RLAGN \citep[e.g.][]{2012MNRAS.424..754T,2014MNRAS.443.2154T,2017A&A...600A..87G} and radio-intermediate AGN \citep[e.g.][]{2021MNRAS.504.3823W,2023ApJ...944..187W}. Alternatively, the jet may be transversely stratified, with the inner region being a highly collimated, fast, and low-density flow and the outer region exhibiting wind-like characteristics and  interacting with the external ISM. Both the wind and the jet carry significant momentum fluxes to exert important feedback on the surrounding environment. However, the coexistence of winds and jets is rarely observed in RQAGN, so whether this model is applicable to RQAGN still needs to be investigated. \textit{The radio spectra of jet, corona and disk wind are different and can be used as a diagnostic tool}: optically thin jet and disc wind have steep spectra, while optically thick jets and corona have flat or inverted spectra.

Other observational characteristics, such as variability, can also help to distinguish the radio emission mechanisms of RQAGN. For example, the radio emission from star formation in an AGN, primarily driven by massive stars, has lifetimes of millions of years. In contrast, the AGN-driven radio emission can exhibit variability on much shorter timescales, ranging from days to years. This discrepancy in variability timescales raises intriguing questions about the nature of these emissions. Some RQAGN have demonstrated radio variability on timescales of a few months \citep{2005ApJ...618..108B}, suggesting that their radio emission may originate from a compact structure in the nuclear region. Furthermore, radio variability observations, particularly in the radio and X-ray bands, can help to determine the size of the emission zone and its distance from the nucleus \citep{2022MNRAS.510..718P}. 

VLBI observations can reveal the parsec-scale radio emission from the galactic nucleus and help to distinguish the different origins of radio emission associated with AGN. In our Paper \textsc{I} \citep{2023MNRAS.518...39W}, we chose a nearby optically selected AGN sample, i.e., a subset of the Palomar-Green (PG) quasar sample \citep{1983ApJ...269..352S,1992ApJS...80..109B}. These quasars have rich multi-band observational data and are suitable for studying the correlation between the radio emission and the black hole properties. The selection criteria are: (1) proximity, $z<0.5$. This criterion allows us to obtain the radio structures on scales of a few parsecs and with a high enough resolution to decouple the optical and X-ray emission between the AGN and its host galaxy; and (2) bright in radio bands. The flux densities of their radio cores in the 5-GHz VLA A-array images are greater than 1 mJy. 

Based on these two selection criteria, we constructed a sample including 16 RQQs and 4 RLQs. These 20 PG quasars were observed at 5 GHz with the US Very Long Baseline Array (VLBA). Their radio structures are obtained with parsec resolutions, successfully detecting ten of the 16 RQQs (a detection rate of 62.5 per cent) and all four RLQs \citep{2023MNRAS.518...39W}. Among the ten VLBA-detected RQQs, six sources show only a single unresolved core; one source (PG~0003+199) shows a core and two-sided jet structure; one source (PG~1351+640) shows two compact components, being either a core-jet or a compact symmetric object (CSO); the remaining two sources (PG~0050+124 and PG~0157+001) are detected only with some faint clumps at distances of about 11.41 mas and 24.08 mas away from the optical nuclei, respectively. It is clear that the nature of the compact core cannot be determined from the images in one frequency alone and that additional information on the radio spectral index is needed to distinguish the weak jets from the corona. On the other hand, due to the steep spectra of extended jets and the disk winds, the 1.5 GHz (L band) is better suited than 5 GHz (C band) to study whether RQQs have extended structures on size scales of a few parsecs and whether these extended structures are associated with jets or winds.  

To characterise the radio emission of the VLBA-detected RQQs in Paper \textsc{I},  we observed nine of the ten RQQs with the VLBA at L and C bands. PG~0050+124 was removed from the list due to its low brightness in the VLBA image, but we observed it separately with the more sensitive European VLBI Network (EVN) plus the enhanced Multi-Element Remote-Linked Interferometer Network (e-MERLIN). The new VLBA observations were carried out at three frequencies, 1.5, 4.7 and 7.6 GHz, to obtain the radio spectra of the selected RQQs, and to compare them with the 2015 observations in Paper \textsc{I} to study the variability of the parsec-scale radio structures. The following text is organised as follows: Section \ref{sec:obs} describes the VLBA observations and data processing; Section \ref{sec:result} presents the observational results; the discussion of the radiation mechanism of the radio emission in RQQ is shown in Section \ref{sec:discussion}; Section \ref{sec:summary} summarises the main results of the paper.

\section{Observations and data processing} \label{sec:obs}

\begin{table}
    \centering \setlength{\tabcolsep}{4pt}
    \caption{VLBA observation information.}
    \begin{tabular}{cccccl} \hline\hline
Session  & Date          & $\nu_{\rm obs}$  & Bandwidth & Telescopes & \\ 
         & yyyy-mm-dd    & (GHz)  & (MHz) &   \\  \hline
EY037   &2020-11-17&4.9  &256& EVN + e-MERLIN  \\
BW138a1 &2021-12-20&1.6  &512&VLBA (no HN, NL, SC) \\ 
BW138a2 &2022-01-22&1.6  &512&VLBA (no HN, NL) \\ 
BW138a3 &2021-12-05&1.6  &512&VLBA (no MK)\\ 
BW138a4 &2021-12-30&1.6  &512&VLBA (no NL)\\ 
BW138b1 &2021-12-31&4.7,7.6&256,256&VLBA (no NL, LA)\\
BW138b2 &2022-01-23&4.7,7.6&256,256&VLBA (no NL)\\ 
BW138b3 &2021-12-04&4.7,7.6&256,256&VLBA (no BR)\\ 
          \hline
    \end{tabular}
    \label{tab:obs}
\end{table}

\subsection{Observations} \label{sec:vlbaobs}

\begin{table*}
\centering
\caption{Image parameters and radio properties. 
Columns (1): source name; 
Columns (2): observation date (yyyy-mm-dd); 
Columns (3): observation session; 
Columns (4): observation frequency (GHz); 
Columns (5): major axis and minor axis of the restoring beam (in unit of mas), and the position angle (in unit of degree) of the major axis, measured from north to east; 
Columns (6): peak flux density (mJy); 
Columns (7): root-mean-square noise (mJy beam$^{-1}$) of the image, measured in off-source regions;
Columns (8): offset between the optical Gaia nucleus position and the radio core, in unit of mas.
Columns (9): total flux density (mJy); 
Columns (10): the flux density of the core component (mJy); 
Columns (11): component size (full width at half-maximum of the fitted Gaussian component, mas); 
Columns (12): brightness temperature (in unit of log(K) of the core component, expressed in the logarithm form. Note: PG 0157+001 is marked with an asterisk, denoting that it lacks an identifiable core, so some parameters related to core component are not shown here.
Columns (13): contour levels, where $\sigma$ refers to the rms noise listed in Column 7;
} 
\setlength{\tabcolsep}{2pt}
\label{tab:imagepar} 
\begin{tabular}{lcccccccccccl} 
\hline\hline
Name           &Obs data       &Session     &Freq.    &Beam                     &S$_{\rm peak}$  &\textit{rms} &$\Delta_{\rm OR}$ &$\rm S_{total}$ &$\rm S_{core}$	&$\theta_{\rm FWHM}$   &log($T_{B}$)     & Contours   \\           
(1)            &(2)            &(3)         &(4)      &(5)                      &(6)             &(7)          &(8)       &(9)         &(10)              &(11)               &(12)             &(13) \\    \hline
\multicolumn{13}{c}{radio-quiet quasars} \\
PG~0003+199    &2021-12-05     &BW138A3     &1.6      &13.21$\times$6.49, 79    &1.83            &0.037        &...        &6.99 	   &1.81 $\pm$ 0.09   &5.73 $\pm$ 0.30 	   &7.43 $\pm$ 0.12  &$3\sigma \times (1, \sqrt{2}, ..., 16)$            \\
               &2021-12-04     &BW138B3     &4.7      &4.00$\times$1.73, 90     &0.25            &0.022        &0.62       &1.49       &0.31 $\pm$ 0.02   &$<$0.85 			&$>$7.38 	 	  &$3\sigma \times (1, \sqrt{2}, 2, 2\sqrt{2})$       \\
               &2021-12-04     &BW138B3     &6.2      &3.04$\times$1.25, 88     &0.21            &0.015        &0.40       &0.87 	    &0.28 $\pm$ 0.03   &0.93 $\pm$ 0.18     &7.02 $\pm$ 0.40  &$3\sigma \times (1, \sqrt{2}, ..., 4)$	          \\
PG~0050+124    &2020-11-17     &EY037       &4.9      &5.83$\times$1.39, 8.9    &0.12            &0.009        &0.38       &1.82       &0.25 $\pm$ 0.02   &3.78 $\pm$ 0.22     &5.98 $\pm$ 0.12  &$3\sigma \times (1, \sqrt{2}, ..., 4)$             \\
PG~0157+001$^{*}$ &2021-12-05  &BW138A3     &1.6      &14.80$\times$6.34, 78    &2.79            &0.028        &...        &8.90       &...               &...      	       &...              &$3\sigma \times (1, \sqrt{2}, ..., 32)$       	 \\
               &2021-12-04     &BW138B3     &4.7      &4.08$\times$1.72, 91     &0.23            &0.021        &...        &1.56       &...               &...      	        &...              &$3\sigma \times (1, \sqrt{2}, 2, 2\sqrt{2})$       \\
               &2021-12-04     &BW138B3     &6.2      &4.45$\times$2.46, 91     &0.16            &0.026        &...        &0.86 	    &...               &...      	        &...              &$\sigma \times (3, 4, 5, 6)   $                    \\
PG~0921+525    &2021-12-20     &BW138A1     &1.6      &14.01$\times$4.97, 65    &1.92            &0.034        &...        &2.57       &2.36 $\pm$ 0.12   &3.19 $\pm$ 0.21 	   &8.06 $\pm$ 0.14  &$3\sigma \times (1, \sqrt{2}, ..., 16)$ 	         \\
               &2021-12-31     &BW138B1     &4.7      &4.23$\times$1.43, 68     &0.95            &0.035        &0.97       &1.21       &1.12 $\pm$ 0.06   &0.75 $\pm$ 0.14     &8.06 $\pm$ 0.37  &$3\sigma \times (1, \sqrt{2}, ..., 8)$   	      \\
               &2021-12-31     &BW138B1     &6.2      &3.32$\times$1.13, 68     &0.81            &0.028        &0.80       &1.07       &0.94 $\pm$ 0.05   &0.61 $\pm$ 0.02     &7.93 $\pm$ 0.08  &$3\sigma \times (1, \sqrt{2}, ..., 8)$             \\
               &2021-12-31     &BW138B1     &7.6      &2.59$\times$0.86, 67     &0.69            &0.036        &1.03       &0.90 	    &0.88 $\pm$ 0.05   &0.58 $\pm$ 0.11 	&7.76 $\pm$ 0.39  &$3\sigma \times (1, \sqrt{2}, ..., 4\sqrt{2})$	  \\
PG~1149-110    &2021-12-20     &BW138A1     &1.6      &18.18$\times$5.06, 108   &0.77            &0.032        &...        &1.28       &1.28 $\pm$ 0.07   &5.53 $\pm$ 0.47     &7.32 $\pm$ 0.18  &$3\sigma \times (1, \sqrt{2}, ..., 8)$	         \\
               &2021-12-31     &BW138B1     &4.7      &3.95$\times$1.51, 84     &0.51            &0.030        &3.26       &0.83       &0.57 $\pm$ 0.03   &$<$0.78   		    &$>$7.74  	      &$3\sigma \times (1, \sqrt{2}, ..., 4\sqrt{2})$     \\
               &2021-12-31     &BW138B1     &6.2      &2.96$\times$1.18, 82     &0.42            &0.028        &3.14       &0.47       &0.46 $\pm$ 0.02   &0.26 $\pm$ 0.02     &8.35 $\pm$ 0.13  &$3\sigma \times (1, \sqrt{2}, ..., 4)$             \\
               &2021-12-31     &BW138B1     &7.6      &2.35$\times$0.91, 81     &0.40            &0.037        &3.11       &0.42 	    &0.42 $\pm$ 0.04   &0.28 $\pm$ 0.22     &8.08 $\pm$ 1.60  &$3\sigma \times (1, \sqrt{2}, 2, 2\sqrt{2})$       \\
PG~1216+069    &2021-12-20     &BW138A1     &1.6      &11.33$\times$4.92, 93    &1.49            &0.056        &...        &1.90       &1.78 $\pm$ 0.09   &2.93 $\pm$ 0.39     &8.12 $\pm$ 0.27  &$3\sigma \times (1, \sqrt{2}, ..., 8)$	         \\
               &2021-12-31     &BW138B1     &4.7      &3.59$\times$1.45, 87     &8.06            &0.030        &1.49       &8.40       &8.17 $\pm$ 0.41   &0.19 $\pm$ 0.01     &10.24 $\pm$ 0.16 &$3\sigma \times (1, 2, ..., 64)$	                  \\
               &2021-12-31     &BW138B1     &6.2      &2.80$\times$1.15, 84     &10.27           &0.028        &1.42       &10.50      &10.04$\pm$ 0.50   &$<$0.56             &$>$9.14          &$3\sigma \times (1, 2, ..., 64)$	                  \\
               &2021-12-31     &BW138B1     &7.6      &2.15$\times$0.87, 83     &7.52            &0.031        &1.35       &7.83       &7.62 $\pm$ 0.38   &0.13 $\pm$ 0.01     &10.12 $\pm$ 0.16 &$3\sigma \times (1, 2, ..., 64)$	 	              \\
PG~1351+640    &2022-01-22     &BW138A2     &1.6      &11.16$\times$4.89, 107   &13.97           &0.029        &...        &17.30      &2.57 $\pm$ 0.13   &2.17 $\pm$ 0.14     &8.45 $\pm$ 0.14  &$3\sigma \times (1, 2, ..., 128)$	                 \\
               &2022-01-23     &BW138B2     &4.7      &4.22$\times$1.68, 110    &2.65            &0.028        &1.48       &6.07       &2.43 $\pm$ 0.12   &0.25 $\pm$ 0.04     &9.38 $\pm$ 0.43  &$3\sigma \times (1, \sqrt{2}, ..., 8\sqrt{2})$     \\
               &2022-01-23     &BW138B2     &6.2      &3.25$\times$1.31, 111    &2.45            &0.030        &1.35       &4.85       &2.24 $\pm$ 0.11   &$<$0.64             &8.31 $\pm$ 0.06  &$3\sigma \times (1, \sqrt{2}, ..., 8\sqrt{2})$     \\
               &2022-01-23     &BW138B2     &7.6      &2.62$\times$1.04, 110    &2.56            &0.032        &1.28       &3.92 	    &2.85 $\pm$ 0.15   &0.35 $\pm$ 0.05     &7.53 $\pm$ 0.19  &$3\sigma \times (1, \sqrt{2}, ..., 8\sqrt{2})$ 	  \\
PG~1612+261    &2022-01-22     &BW138A2     &1.6      &10.11$\times$4.45, 89    &0.32            &0.029        &...        &0.92       &0.26 $\pm$ 0.01   &2.77 $\pm$ 0.63     &6.90 $\pm$ 0.16  &$3\sigma \times (1, \sqrt{2}, 2, 2\sqrt{2})$ 	     \\
PG~1700+518    &2022-01-22     &BW138A2     &1.6      &9.30$\times$4.60, 97     &1.00            &0.049        &...        &2.02       &1.44 $\pm$ 0.07   &3.55 $\pm$ 0.35     &7.85 $\pm$ 0.21  &$3\sigma \times (1, \sqrt{2}, ..., 4\sqrt{2})$ 	 \\
               &2022-01-23     &BW138B2     &4.7      &3.74$\times$1.64, 83     &1.09            &0.027        &3.78       &1.33       &1.15 $\pm$ 0.06   &0.31 $\pm$ 0.10     &8.93 $\pm$0.62   &$3\sigma \times (1, \sqrt{2}, ..., 8\sqrt{2})$	  \\
               &2022-01-23     &BW138B2     &6.2      &2.89$\times$1.28, 84     &0.71            &0.023        &3.76       &0.96       &0.90 $\pm$ 0.05   &0.80 $\pm$ 0.02     &7.77 $\pm$ 0.07  &$3\sigma \times (1, \sqrt{2}, ..., 8\sqrt{2})$     \\
               &2022-01-23     &BW138B2     &7.6      &2.32$\times$1.02, 84     &0.38            &0.028        &3.79       &0.51       &0.51 $\pm$ 0.03   &0.71 $\pm$ 0.14 	&7.44 $\pm$ 0.40  &$3\sigma \times (1, \sqrt{2}, ..., 4)$             \\
PG~2304+042    &2021-12-05     &BW138A3     &1.6      &14.73$\times$6.57, 77    &0.64            &0.038        &...        &1.03       &0.78 $\pm$ 0.04   &3.66 $\pm$ 0.79     &7.46 $\pm$ 0.43  &$3\sigma \times (1, \sqrt{2}, ..., 4)$	         \\
               &2021-12-04     &BW138B3     &4.7      &4.08$\times$1.63, 92     &0.48            &0.029        &0.70       &0.71       &0.59 $\pm$ 0.03   &1.03 $\pm$ 0.22     &7.51 $\pm$ 0.42  &$3\sigma \times (1, \sqrt{2}, ..., 4)$	          \\
               &2021-12-04     &BW138B3     &6.2      &3.05$\times$1.32, 88     &0.35            &0.021        &0.52       &0.62       &0.45 $\pm$ 0.02   &0.79 $\pm$ 0.04     &7.37 $\pm$ 0.11  &$3\sigma \times (1, \sqrt{2}, ..., 4)$             \\
               &2021-12-04     &BW138B3     &7.6      &2.66$\times$1.01, 96     &0.24            &0.027        &0.67       &0.30       &0.30 $\pm$ 0.03   &0.36 $\pm$ 0.26     &7.71 $\pm$ 1.43  &$3\sigma \times (1, \sqrt{2}, 2, 2\sqrt{2})$     \\
\hline
\multicolumn{13}{c}{radio-loud quasars} \\
PG~1004+130    &2021-12-20     &BW138A1     &1.6      &11.60$\times$4.66, 82    &16.43           &0.034        &...        &21.68      &17.65 $\pm$0.89   &1.71 $\pm$ 0.02      &9.55 $\pm$ 0.06 &$3\sigma \times (1, 2, ..., 128)$		             \\
               &2021-12-31     &BW138B1     &4.7      &3.62$\times$1.45, 82     &18.08           &0.028        &...        &22.49      &14.18 $\pm$0.71   &0.71 $\pm$ 0.01      &9.29 $\pm$ 0.05 &$3\sigma \times (1, 2, ..., 128)$	              \\
               &2021-12-31     &BW138B1     &6.2      &2.80$\times$1.11, 82     &14.90           &0.023        &...        &19.65      &13.26 $\pm$0.66   &0.28 $\pm$ 0.01      &9.83 $\pm$ 0.05 &$3\sigma \times (1, 2, ..., 128)$	              \\
               &2021-12-31     &BW138B1     &7.6      &2.19$\times$0.88, 83     &11.49           &0.028        &...        &22.51      &10.73 $\pm$0.54   &0.18 $\pm$ 0.02      &9.94 $\pm$ 0.08 &$3\sigma \times (1, 2, ..., 128)$			      \\
PG~1048-090    &2021-12-20     &BW138A1     &1.6      &11.91$\times$4.79, 99    &35.77           &0.034        &...        &45.78      &34.53$\pm$ 1.73   &1.71 $\pm$ 0.01      &9.88 $\pm$ 0.05 &$3\sigma \times (1, 2, ..., 256)$	                 \\
               &2021-12-31     &BW138B1     &4.7      &3.83$\times$1.47, 89     &30.92           &0.031        &...        &42.54      &24.27$\pm$ 1.22   &0.73 $\pm$ 0.01      &9.52 $\pm$0.05  &$3\sigma \times (1, 2, ..., 128)$	              \\
               &2021-12-31     &BW138B1     &6.2      &2.71$\times$1.06, 87     &27.17           &0.029        &...        &37.25      &24.54$\pm$ 1.23   &$<$0.52              &$>$9.58         &$3\sigma \times (1, 2, ..., 128)$                  \\
               &2021-12-31     &BW138B1     &7.6      &2.17$\times$0.84, 87     &29.21           &0.028        &...        &38.50 	    &25.68 $\pm$1.28   &0.42 $\pm$ 0.01      &9.62 $\pm$ 0.05 &$3\sigma \times (1, 2, ..., 128)$	   	          \\
PG~1425+267    &2021-12-20     &BW138A1     &1.6      &11.45$\times$4.18, 89    &17.34           &0.028        &...        &32.54      &28.76$\pm$ 1.44   &4.98 $\pm$ 0.03      &8.88 $\pm$ 0.05 &$3\sigma \times (1, 2, ..., 128)$	  	             \\
               &2021-12-31     &BW138B1     &4.7      &3.70$\times$1.42, 85     &18.52           &0.028        &...        &24.97      &21.18$\pm$ 1.06   &0.78 $\pm$ 0.01      &9.42 $\pm$ 0.05 &$3\sigma \times (1, 2, ..., 128)$	  	          \\
               &2021-12-31     &BW138B1     &6.2      &2.87$\times$1.08, 83     &13.32           &0.028        &...        &21.28       &18.26$\pm$ 0.91   &0.87 $\pm$ 0.01      &9.05 $\pm$ 0.05 &$3\sigma \times (1, 2, ..., 128)$                  \\
               &2021-12-31     &BW138B1     &7.6      &2.18$\times$0.85, 84     &10.50           &0.035        &...        &16.87      &14.89$\pm$ 0.75   &0.72 $\pm$ 0.01      &8.92 $\pm$ 0.05 &$3\sigma \times (1, 2, ..., 128)$  	              \\
PG~1704+608    &2022-01-22     &BW138A2     &1.6      &10.00$\times$4.75, 89    &9.55            &0.039        &...        &10.74 	   &10.47$\pm$ 0.53   &1.59 $\pm$ 0.04      &9.43 $\pm$ 0.07 &$3\sigma \times (1, 2, ..., 64)$	                 \\
               &2022-01-23     &BW138B2     &4.7      &3.85$\times$1.66, 78     &7.32            &0.028        &...        &7.90 	    &7.90 $\pm$ 0.40   &0.61 $\pm$ 0.01      &9.20 $\pm$ 0.07 &$3\sigma \times (1, 2, ..., 64)$	                  \\
               &2022-01-23     &BW138B2     &6.2      &2.94$\times$1.23, 79     &5.96            &0.024        &...        &7.24       &7.01 $\pm$ 0.35   &0.65 $\pm$ 0.01      &8.86 $\pm$ 0.05 &$3\sigma \times (1, 2, ..., 64)$                   \\ 
               &2022-01-23     &BW138B2     &7.6      &2.37$\times$1.02, 78     &3.24            &0.030        &...        &4.34 	    &4.34 $\pm$ 0.22   &0.79 $\pm$ 0.02      &8.31 $\pm$ 0.07 &$3\sigma \times (1, 2, ..., 32)$	                  \\
\hline 
\end{tabular}
\end{table*}

We used the VLBA of the National Radio Astronomy Observatory (NRAO) to observe nine RQQs and four RLQs between December 2021 and January 2022 (code: BW138). The ten VLBA telescopes are Brewster (BR), Fort Davis (FD), Hancock (HN), Kitt Peak (KP), Los Alamos (LA), Mauna Kea (MK), North Liberty (NL), Owens Valley (OV), Pie Town (PT), and Saint Croix (SC). The data were correlated using the DiFX software correlator \citep{2011PASP..123..275D} at Socorro, USA, with an average time of 2s and 128 frequency channels per IF. Details of the observations are given in Table~\ref{tab:obs}. Observation dates are given in the format yyyy-mm-dd. The observations were conducted at L-band (the central frequency of 1.6 GHz) and C-band (except for PG~1612+261). The wide C-band was split into two sub-bands with central frequencies of 4.7 GHz and 7.6 GHz, respectively, to facilitate the measurement of the in-band spectral index. Due to the wide distribution of the right ascension (RA) of the sample sources, the sample was divided into 3--4 groups to optimise the (u,v) coverage for each source. In seven observing sessions, 7--9 telescopes participated, with individual telescopes absent from some observations due to maintenance or other technical problems. The codes of these missing telescopes are given in brackets in column 5 of Table~\ref{tab:obs}. The recording bandwidth of the observations was 512 MHz, with a data recording rates of 4 Gbps at 1.6 GHz and 2 Gbps at each 4.7 and 7.6 GHz sub-bands. Since most of the target sources were very weak in our previous 5 GHz VLBA observations (Paper \textsc{I}), we used phase-referencing in all observations. At 1.6 GHz, we used a phase-referencing cycle of ``calibrator (1 min) -- target (3.5 min)''. At 4.7 and 7.6 GHz, we used a shorter cycle of ``calibrator (1 min) -- target (3 min)''.  The integration time for correlation processing is 30 s, resulting in a typical VLBA baseline sensitivity of $\sim$1.7 mJy. Most of the calibrators had a signal-to-noise ratio of $>50$. The four radio-loud PG quasars have flux densities between 10 and 40 mJy, and their structure is compact enough to allow fringe search without the need for phase-referencing. The average on-source time for each RQQ source is about 80 min. The mean root mean square (\textit{rms}) noise of each RQQ image was $\sim$0.025 \mJyb\ for C band and 0.021 \mJyb\ for L band (Table~\ref{tab:imagepar}). The actual noise on the images is at or close to the expected values. The angular resolutions of the radio maps are also presented in Table \ref{tab:imagepar}.

Due to its complex radio structure, PG~0050+124 (also known as \textsc{I} Zwicky 1, Mrk~1502) was not included in the sample of the VLBA snapshot observations described above. Instead, it was observed separately for a longer duration using the EVN+e-MERLIN at a central frequency of 4.9 GHz on 17 November 2020  (code: EY037). A total of 19 telescopes participated in the observations: Jodrell Bank MK II (Jb), Westerbork single antenna (Wb), Effelsberg (Ef), Medicina (Mc), Onsala (O8), Tianma (T6), Torun (Tr), Yebes (Ys), Hartebeesthoek (Hh), Svetloe (Sv), Zelenchukskaya (Zc), Badary (Bd), Kunming (Km), Irebene 32 m (Ir), Cambridge (Cm), Darnhall (Da), Defford (De), and Knockin (Kn), Pickmere (Pi). These observations were made in electronic EVN (e-EVN) mode, with a data rate of 2 Gbps, and the data were transferred in quasi-real time via the internet to the EVN software correlator \citep{2015ExA....39..259K} at the Joint Institute for VLBI ERIC in the Netherlands. The integration time for the data correlation is 1 s, and the frequency resolution is 0.5 MHz. The phase-referencing calibrator is J0056+1341, which was the same as in our first round of VLBA observations \citep{2023MNRAS.518...39W}. The total observing time was 8 h, and the time used for PG~0050+124 was $\sim$5.9 h.

\subsection{Data Reduction}

The correlated data were downloaded to the China Square Kilometre Array Regional Centre \citep{2019NatAs...3.1030A} and were processed through a pipeline\footnote{The VLBI pipeline in github: \url{https://github.com/SHAO-SKA/vlbi-pipeline}}  written in Python and Parseltongue \citep{2022SCPMA..6529501A}. The pipeline was executed in a scripted manner by calling programs integrated into the Astronomical Image Processing System \citep[AIP:][]{2003ASSL..285..109G}. We first manually checked the data  and corrected for ionosphere-induced dispersion delays in the visibility data using a model of the electron content obtained by the Global Positioning System. The visibility magnitudes were then calibrated using information extracted from the gain curve, system temperature and weather tables for all antennas. The subsequent calibration process was automated in a scripted manner, and the method is described in detail in \citet{2023MNRAS.518...39W} and will not be repeated here.

The gain solutions obtained from the phase calibrators were interpolated to the corresponding target sources. The data of each source were then exported from AIPS. The data in channels of the same intermediate frequency (IF) were combined to reduce the amount of data. Since the precise coordinates of the target sources have been obtained from our previous observations \citep{2023MNRAS.518...39W} and the emission of the target sources is concentrated within a few tens of mas of the image centre, the channel-averaged data do not introduce bandwidth smearing and time smearing effects. It is worth noting that the C-band VLBA data were recorded in two sub-bands, 4.7 and 7.6 GHz. Their data calibration was carried out separately, allowing us to obtain images for both 4.7 and 7.6 GHz.

\section{Results}
 \label{sec:result}

\renewcommand{\arraystretch}{1.1}
\begin{table}
\centering
\caption{Spectral Index of PG quasars. Column (1): source name; Column (2): the spectral index of the entire VLBI structure; Column (3): the spectral index of the core component; Note: Most of the spectral index calculations are given in Table \ref{tab:imagepar}, except for PG 0050+124 and PG 1612+261. The spectral index of PG 0050+124 was obtained from \citep{2022ApJ...936...73A}. The spectral index of PG 1612+261 was calculated by 1.6 GHz (Table \ref{tab:imagepar}) and 4.9 GHz \citep{2023MNRAS.518...39W} flux densities. PG 1216+069 exhibits a peaked spectrum, which can be represented by the SSA model. Although its spectral index information is not provided here, the optimal fitting parameters can be found in Appendix \ref{context:PG1216}. } 
\label{table:Spectral_Index}
\begin{tabular}{|c|l|l|}
\hline
\hline
Name           & $\alpha_{\rm total}$                     & $\alpha_{\rm core}$          \\
(1)            & (2)                                  & (3)                      \\
\hline
\multicolumn{3}{c}{radio-quiet quasars} \\
PG 0003+199    &$\alpha^{6.2}_{1.6}=-1.50\pm0.09 $    & $\alpha^{6.2}_{4.7}=-0.37\pm0.03 $   \\
PG 0050+124    &$\alpha^{4.8}_{1.4}=-0.90\pm0.03 $    & $\alpha^{4.8}_{1.4}=-0.00\pm0.29 $   \\
PG 0157+001    &$\alpha^{6.2}_{1.6}=-1.70\pm0.09 $    & ...                                  \\
PG 0921+525    &$\alpha^{7.6}_{1.6}=-0.66\pm0.03 $    & $\alpha^{7.6}_{1.6}=-0.66\pm0.04 $   \\
PG 1149--110    &$\alpha^{7.6}_{1.6}=-0.75\pm0.18 $    & $\alpha^{7.6}_{1.6}=-0.74\pm0.03 $   \\
PG 1216+069    &$\alpha^{6.2}_{1.6}=1.26\pm0.10  $    & $\alpha^{6.2}_{1.6}=1.28\pm0.11  $   \\
               &$\alpha^{7.6}_{6.2}=-1.44\pm0.01 $    & $\alpha^{7.6}_{6.2}=-1.35\pm0.01 $   \\
PG 1351+640    &$\alpha^{7.6}_{1.6}=-0.94\pm0.02 $    & $\alpha^{7.6}_{1.6}=-0.04\pm0.12 $   \\
PG 1612+261    &...                                   & $\alpha^{4.9}_{1.6}=-0.28\pm0.03 $   \\
PG 1700+518    &$\alpha^{4.7}_{1.6}=-0.39\pm0.06 $    & $\alpha^{4.7}_{1.6}=-0.21\pm0.03 $   \\
               &$\alpha^{7.6}_{4.7}=-1.94\pm0.77 $    & $\alpha^{7.6}_{4.7}=-1.64\pm0.76 $   \\
PG 2304+042    &$\alpha^{4.7}_{1.6}=-0.35\pm0.06 $    & $\alpha^{4.7}_{1.6}=-0.26\pm0.03 $   \\
               &$\alpha^{7.6}_{4.7}=-1.66\pm1.22 $    & $\alpha^{7.6}_{4.7}=-1.36\pm0.40 $   \\
\hline
\multicolumn{3}{c}{radio-loud quasars} \\
PG 1004+130    &$\alpha^{7.6}_{1.6}=-0.04\pm0.09 $    & $\alpha^{7.6}_{1.6}=-0.25\pm0.08 $   \\
PG 1048--090    &$\alpha^{7.6}_{1.6}=-0.14\pm0.05 $    & $\alpha^{7.6}_{1.6}=-0.22\pm0.07 $   \\
PG 1425+267    &$\alpha^{7.6}_{1.6}=-0.36\pm0.09 $    & $\alpha^{7.6}_{1.6}=-0.37\pm0.07 $   \\
PG 1704+608    &$\alpha^{6.2}_{1.6}=-0.30\pm0.06 $    & $\alpha^{6.2}_{1.6}=-0.30\pm0.03 $   \\
               &$\alpha^{7.6}_{4.7}=-0.81\pm1.22 $    & $\alpha^{7.6}_{4.7}=-1.15\pm0.76 $   \\
\hline
\end{tabular}
\end{table}
\renewcommand{\arraystretch}{1}

\begin{figure*}
\centering
  \begin{tabular}{cccc}
  \includegraphics[height=4.5cm]{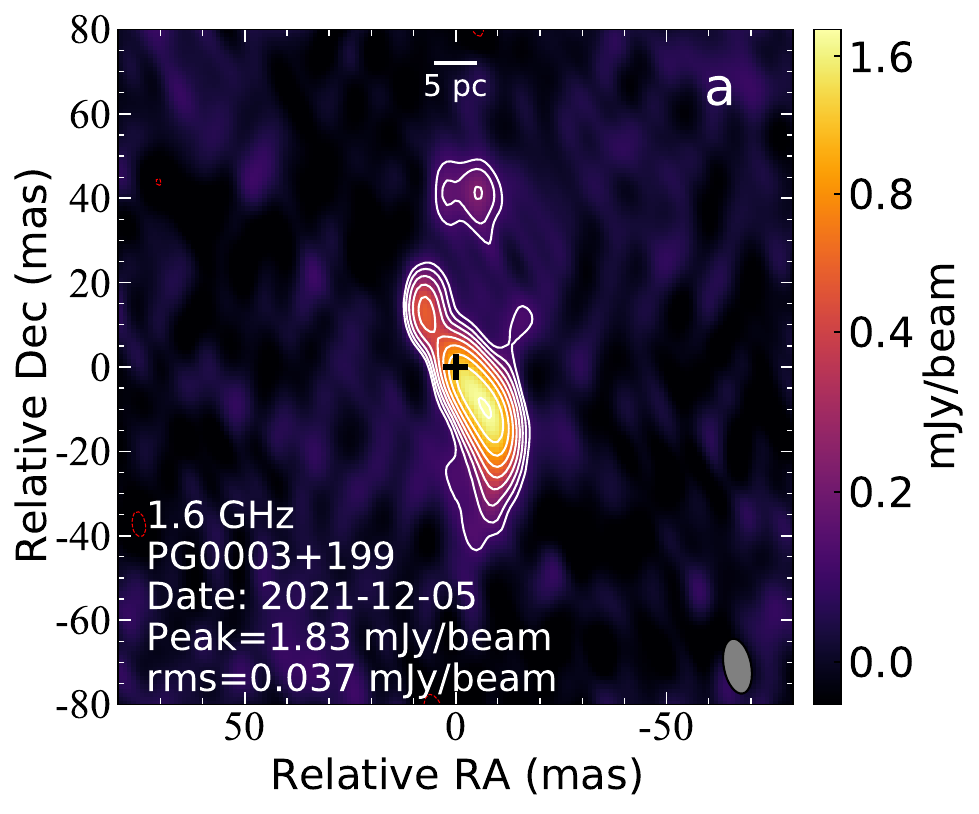}
  \includegraphics[height=4.5cm]{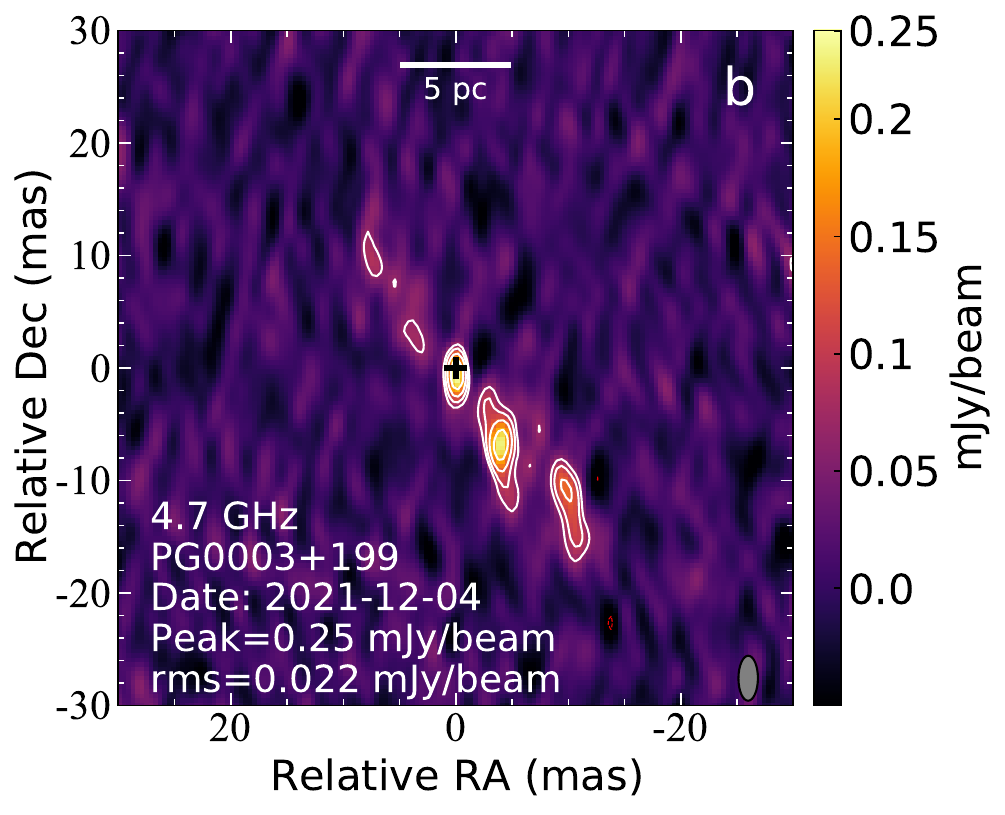}
  \includegraphics[height=4.5cm]{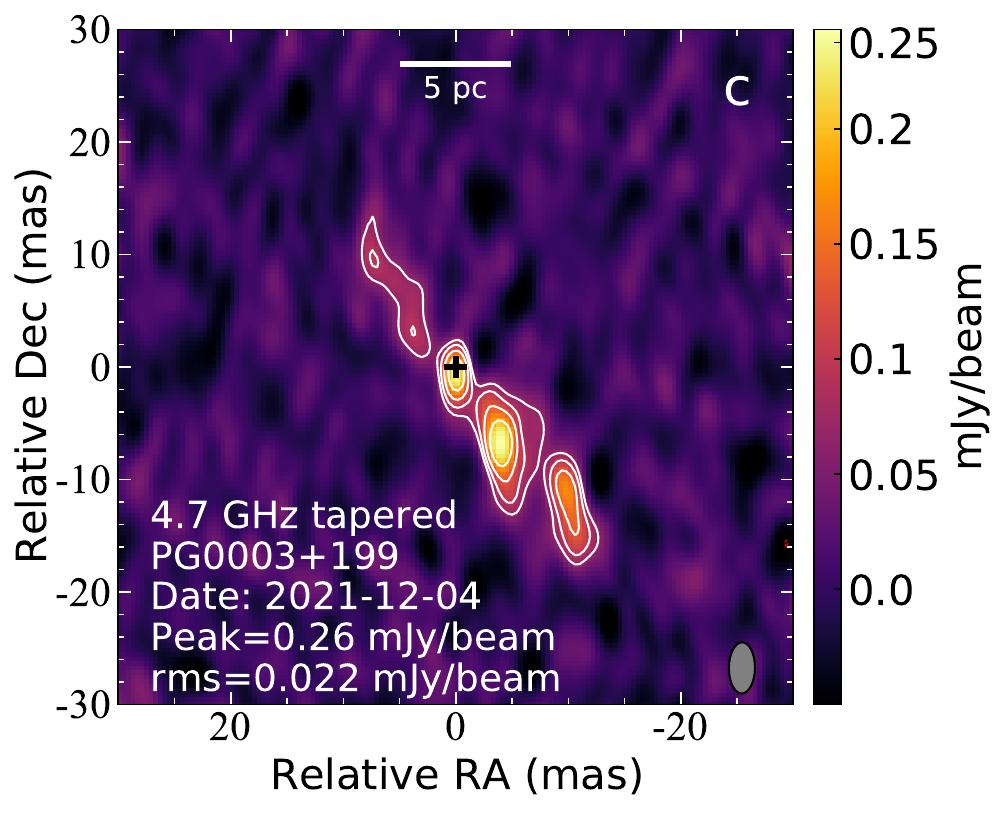} \\ 
  \includegraphics[height=4.5cm]{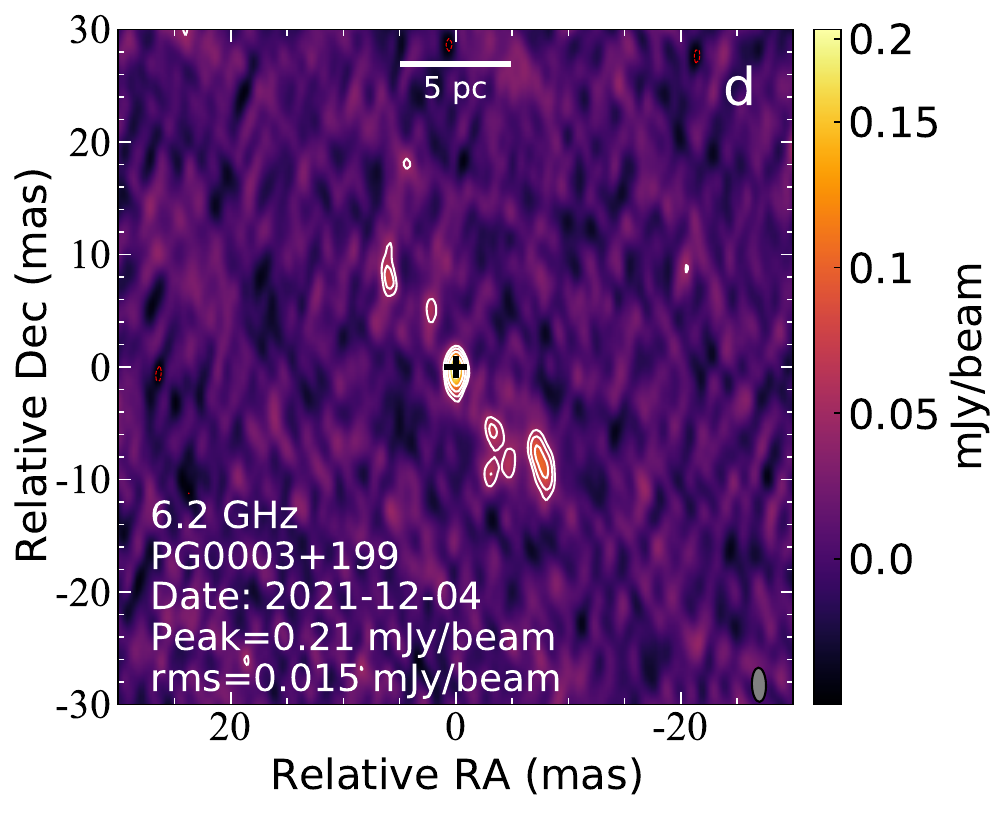}
  \includegraphics[height=4.5cm]{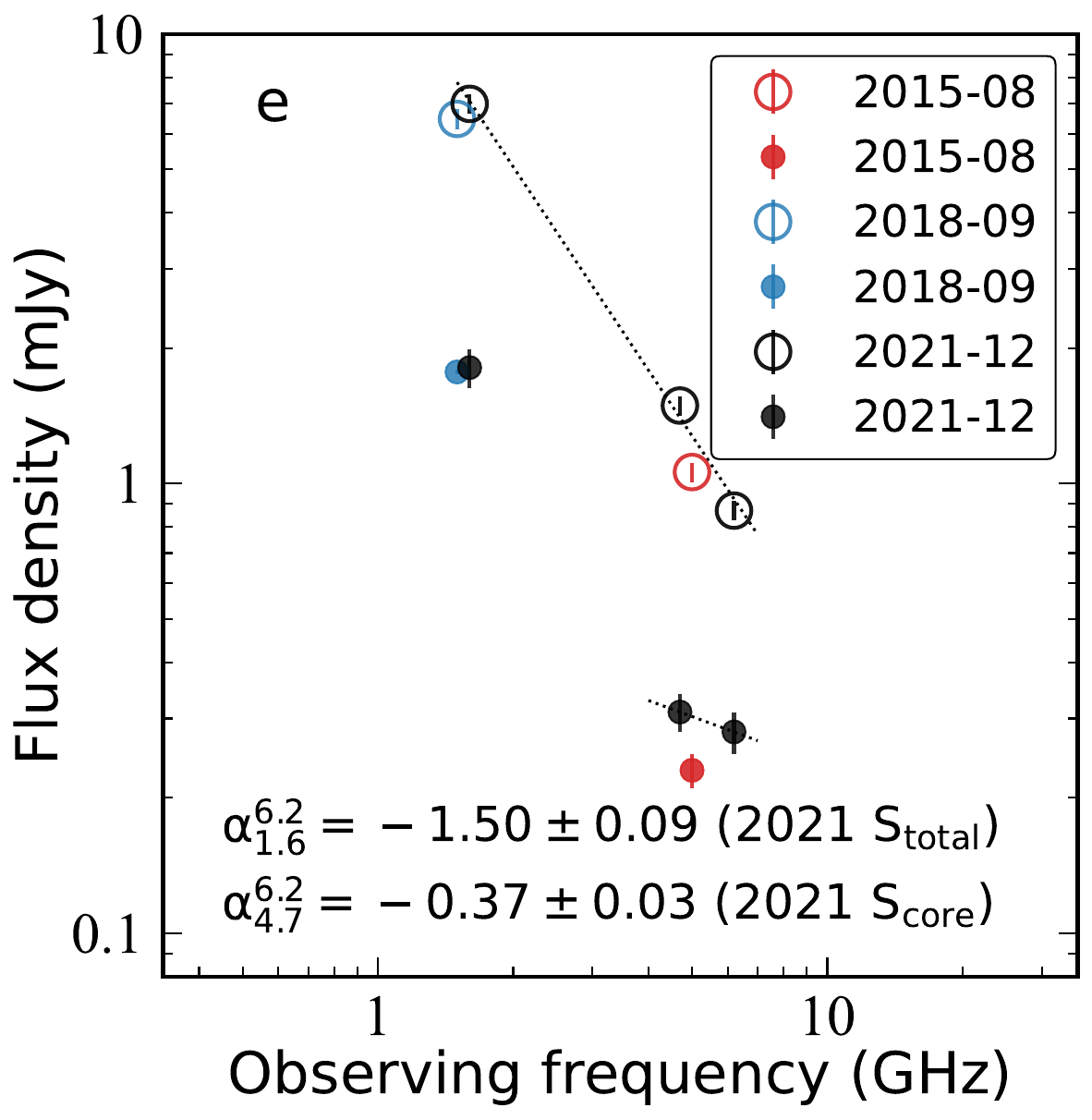}
\end{tabular}
\caption{VLBA images and radio spectrum of radio-quiet quasar PG 0003+199 (Mrk 335). VLBA images are displayed sequentially in panels a to d: 1.6 GHz image, 4.7 GHz full resolution image, 4.7 GHz tapered image and 6.2 GHz image. The colour scale represents the brightness in logarithmic form, and the contours increase in steps of $\sqrt{2}$. The cross in the centre of the image marks the \textit{Gaia} position of the optical nucleus. The observation frequency, source name, observation date, peak flux density and root-mean-square (rms) noise of the image are given in the annotations in the lower left corner. The restoring beam is shown as an ellipse in the bottom right-hand corner. Panel e shows the flux density of PG 0003+199 with respect to the observing frequency on parsec scale, including epoch 2015 \citep{2023MNRAS.518...39W}, epoch 2018 \citep{2021MNRAS.508.1305Y} and epoch 2021 (Table \ref{tab:imagepar}). The solid circular data point in Panel \textit{e} marks the flux density from the core component, while the open point represents the flux density of the entire VLBI source. The annotations illustrate the spectral indices between the different frequencies.}
\label{fig:0003}
\end{figure*}

\begin{figure*}
\centering
\includegraphics[height=5cm]{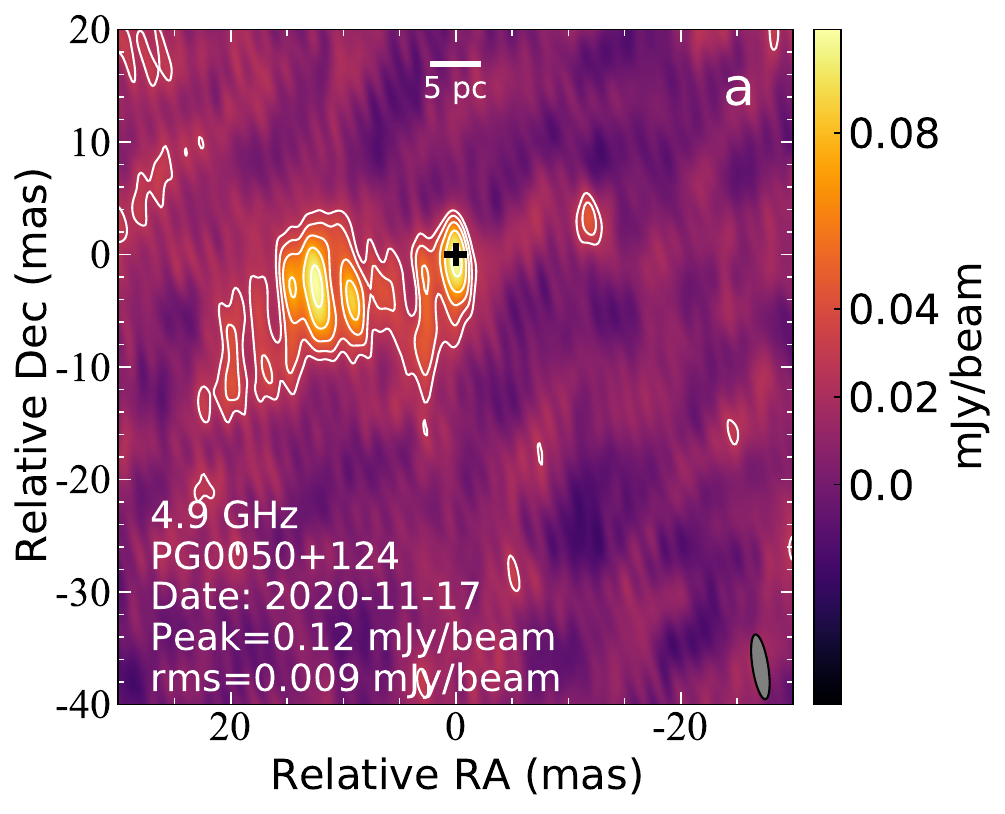}
\includegraphics[height=5cm]{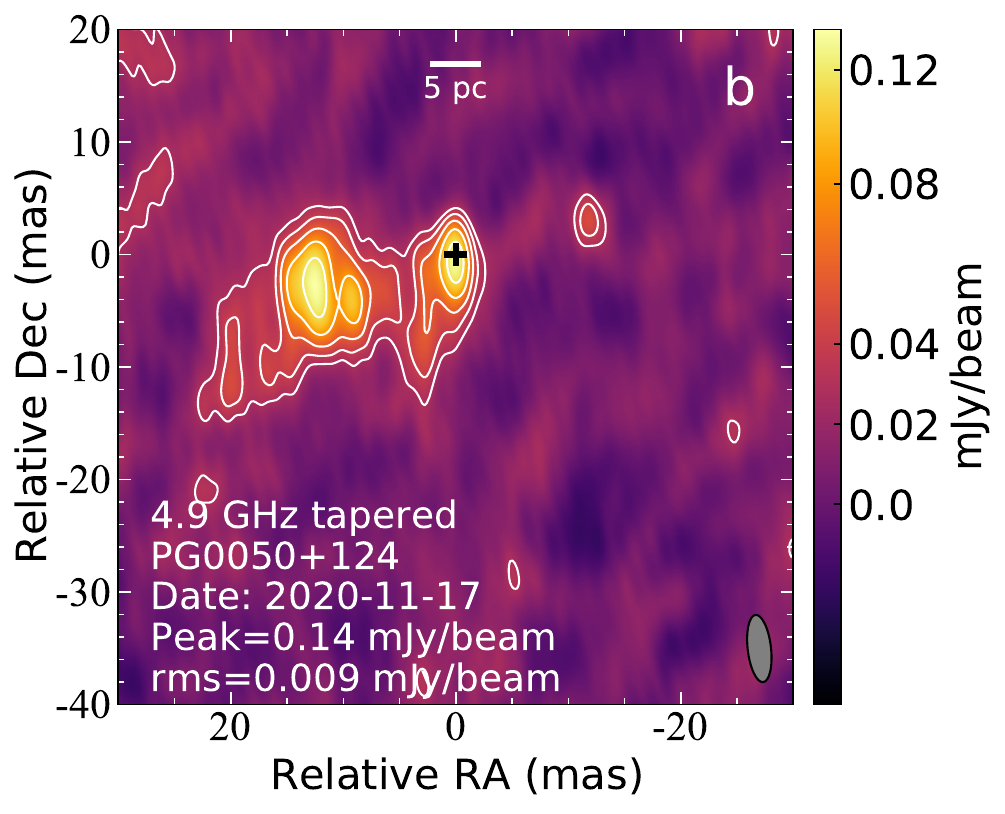}
\includegraphics[height=5cm]{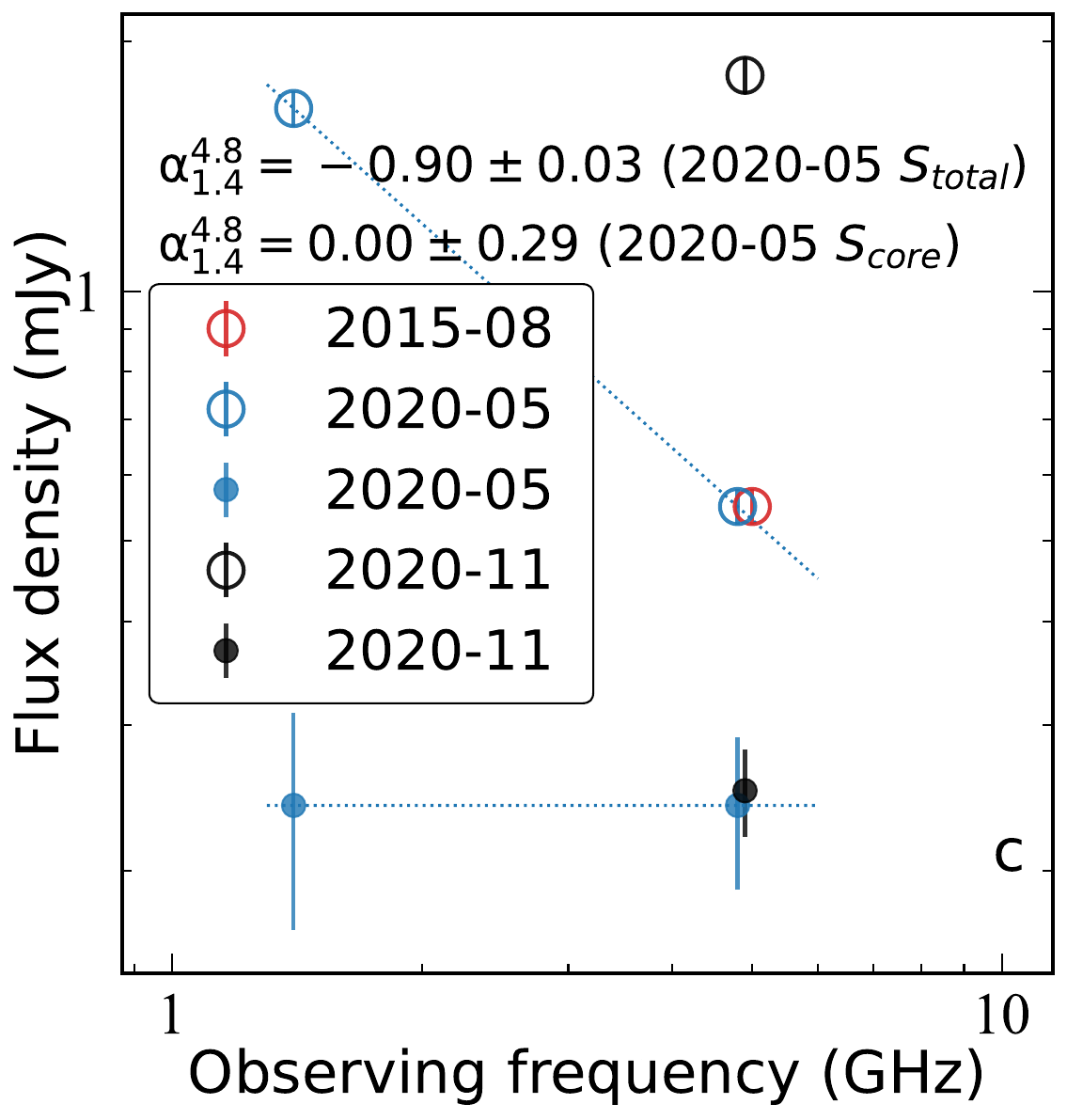}
  \caption{EVN images of radio-quiet quasar PG 0050+124 (Mrk 1502, I Zw 1) at 4.9 GHz and the radio spectrum. 
  The cross in the image centre marks the \textit{Gaia} position: 00h53m34.933s, +12d41m35.931s (J2000). 
  The flux densities used in Panel \textit{c} are from \citet[][epoch 2015-08]{2023MNRAS.518...39W},  \citet[][epoch 2020-05]{2022ApJ...936...73A}  and this paper (epoch 2020-11). 
  }
  \label{fig:0050}
\end{figure*}

\begin{figure*}
\centering
\begin{tabular}{cccc}
  \includegraphics[height=4.5cm]{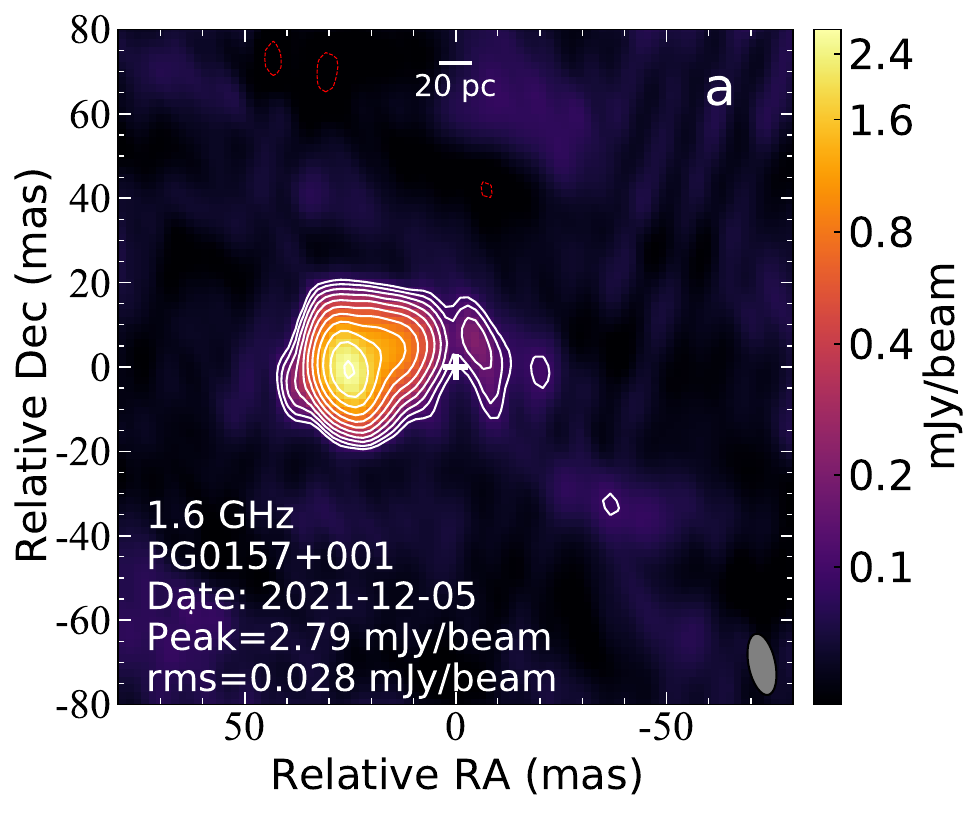}
  \includegraphics[height=4.5cm]{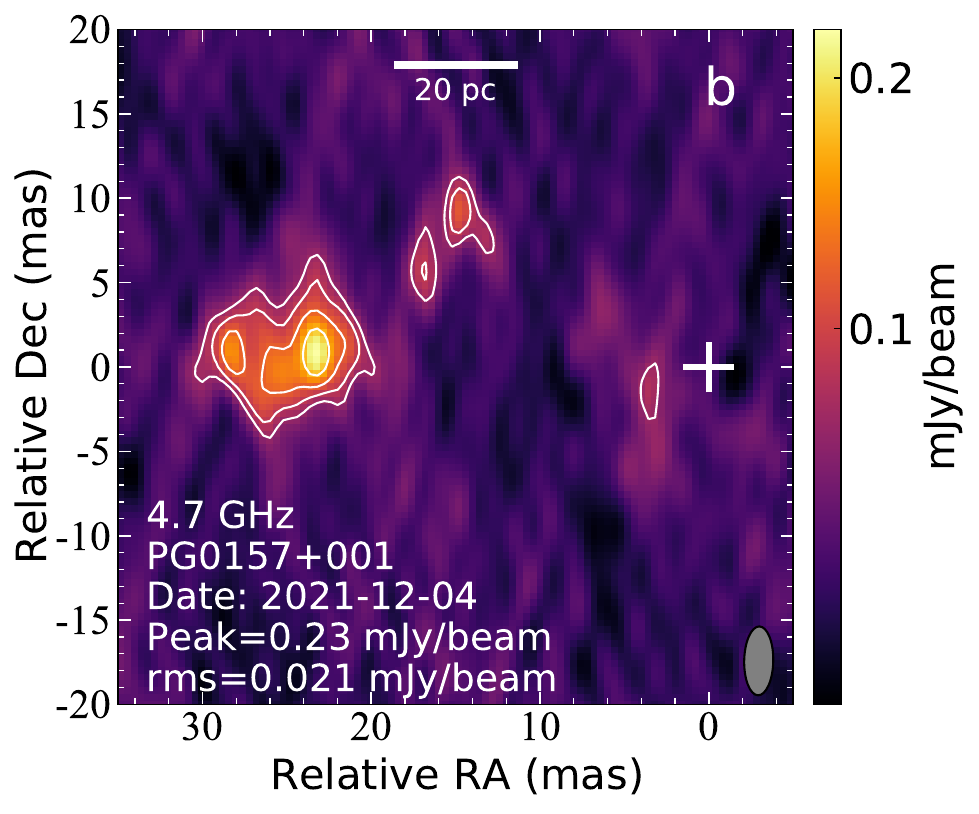}
  \includegraphics[height=4.5cm]{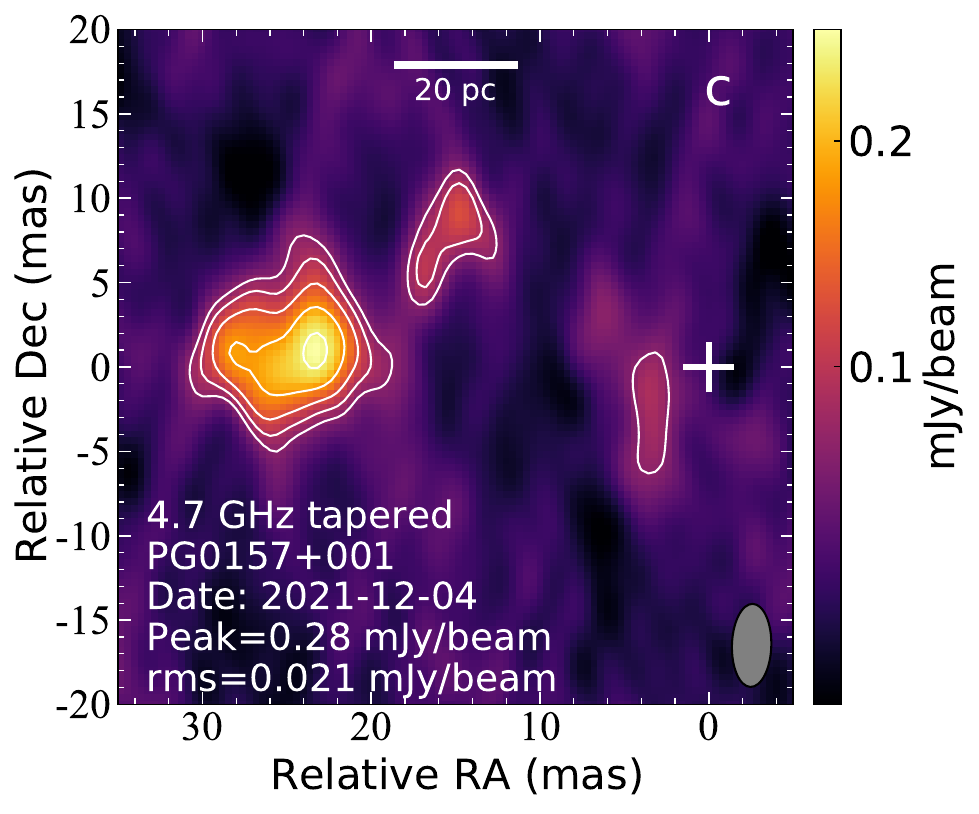} \\
  \includegraphics[height=4.5cm]{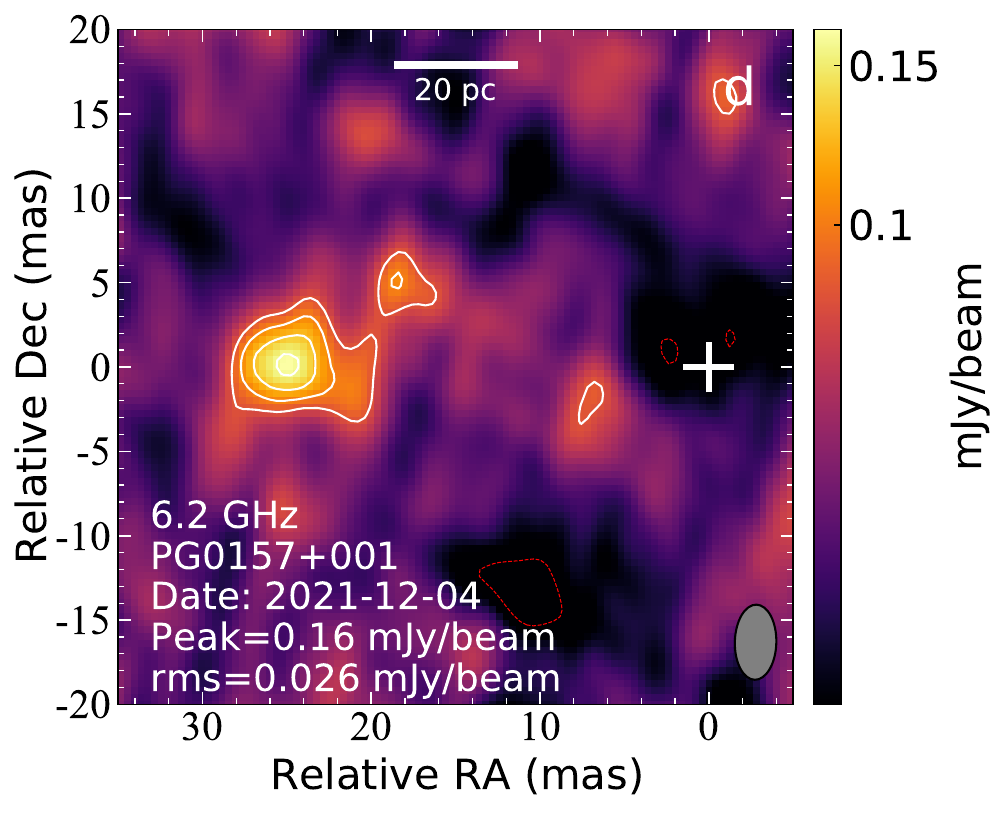}
  \includegraphics[height=4.5cm]{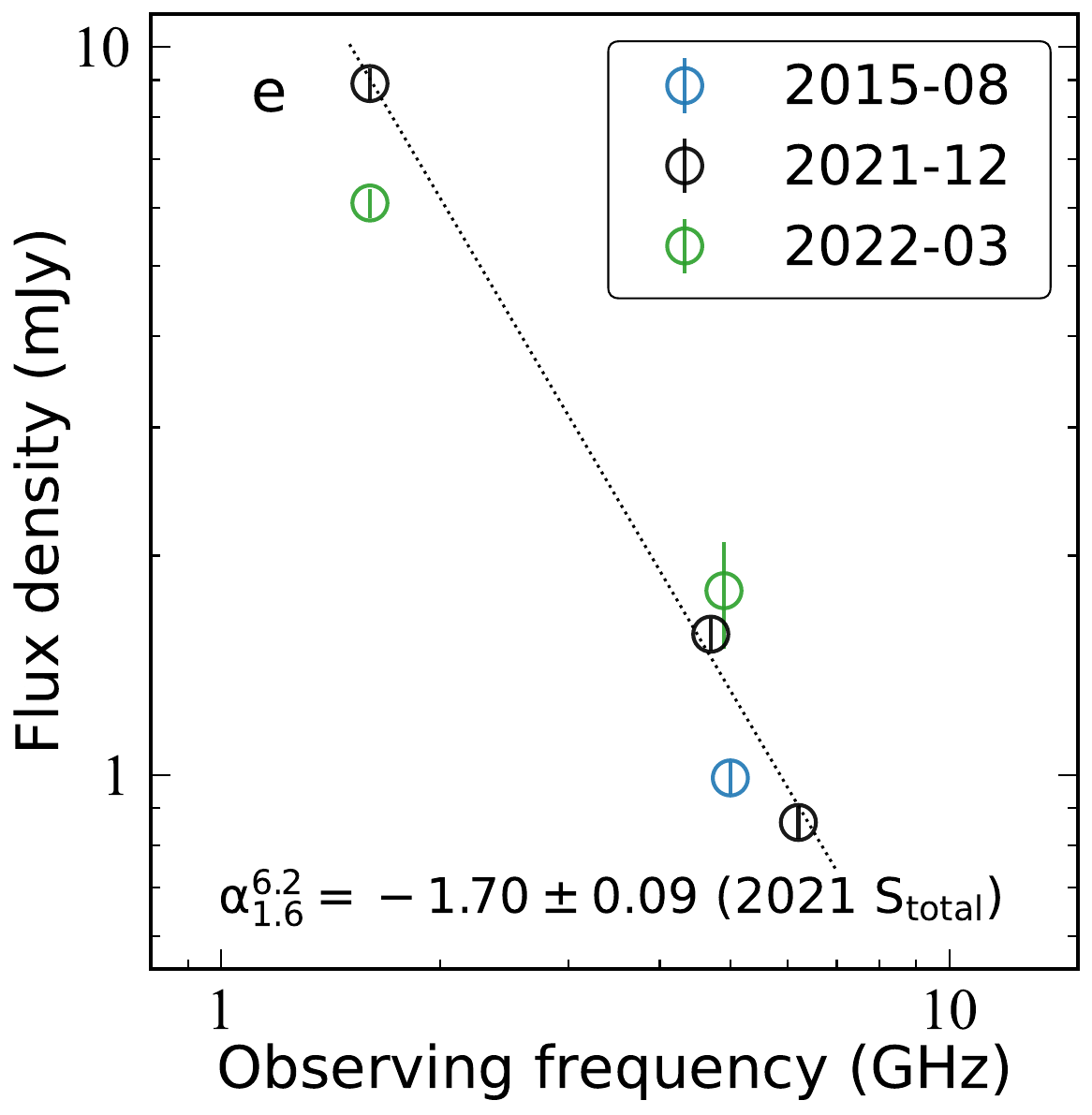}
\\
\end{tabular}
  \caption{VLBA images and radio spectrum of radio-quiet quasar PG 0157+001. 
  Flux densities used in Panel \textit{e} are from \citet[][epoch 2015]{2023MNRAS.518...39W}, this paper (epoch 2021) and \citet[][epoch 2022]{2023arXiv230713599C}. 
  }
  \label{fig:0157}
\end{figure*}

\begin{figure*}
\centering
\begin{tabular}{cccc}
  \includegraphics[height=4.5cm]{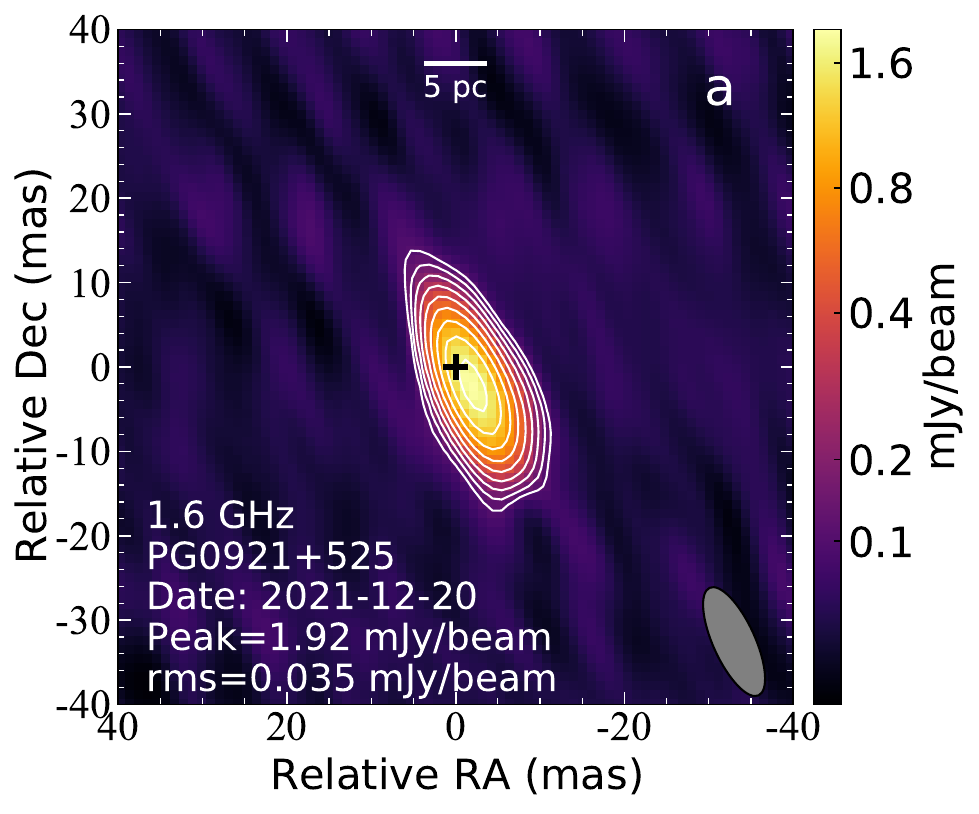}
  \includegraphics[height=4.5cm]{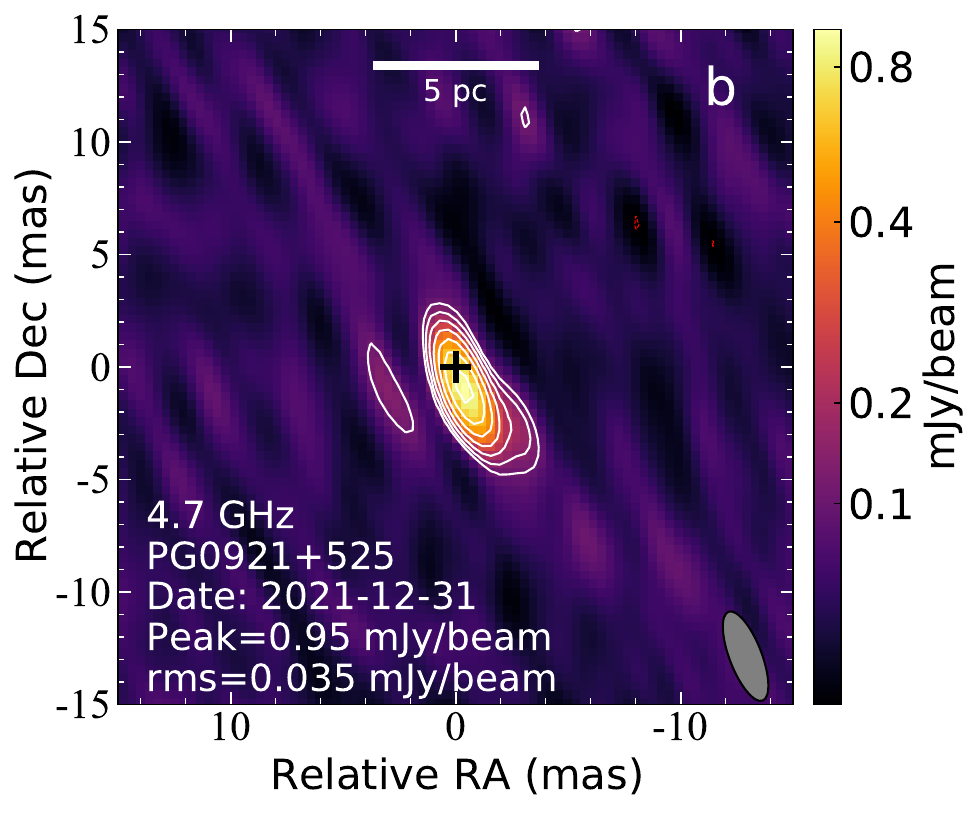}
  \includegraphics[height=4.5cm]{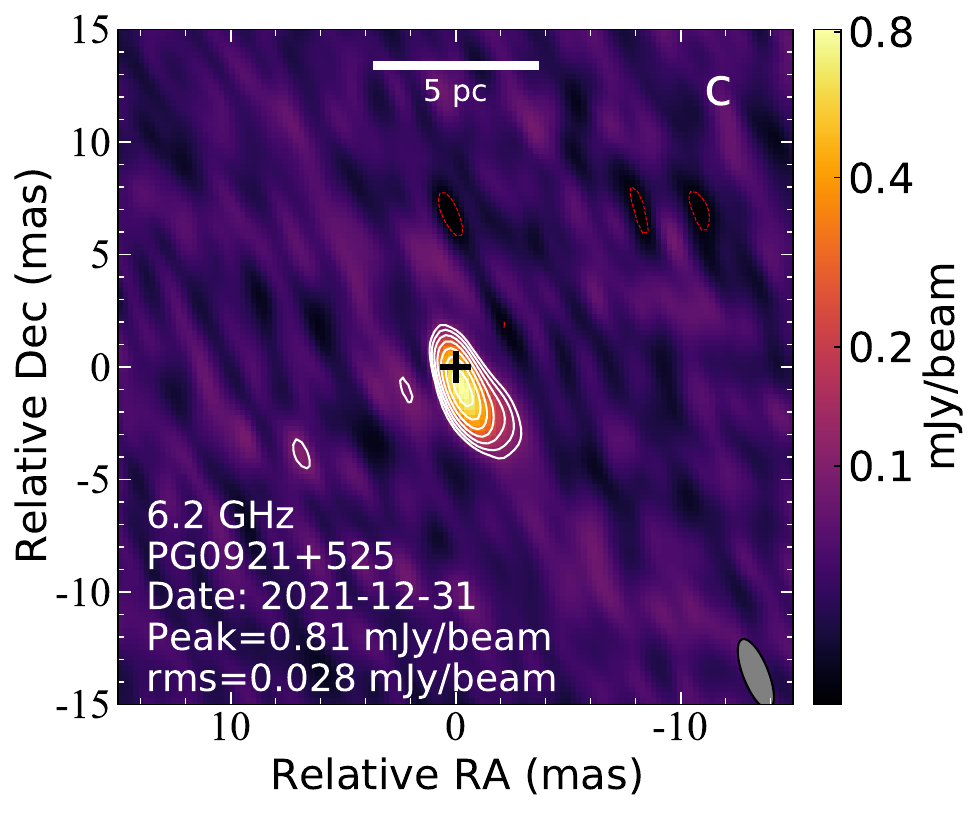} \\
  \includegraphics[height=4.5cm]{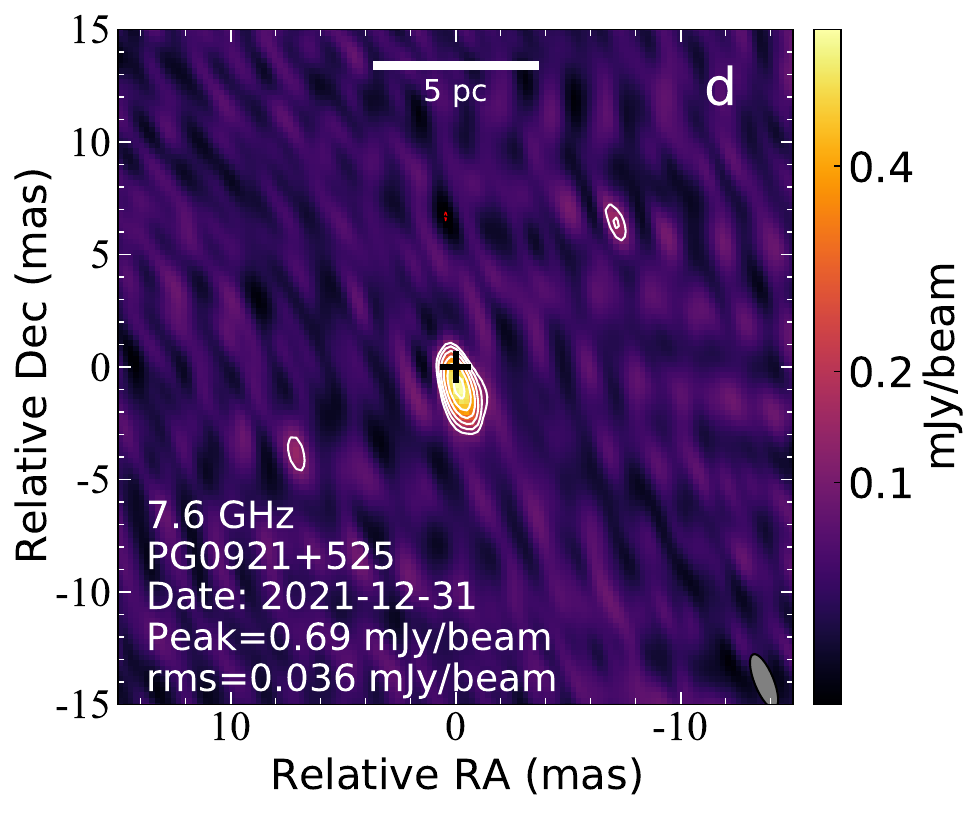}
    \includegraphics[height=4.5cm]{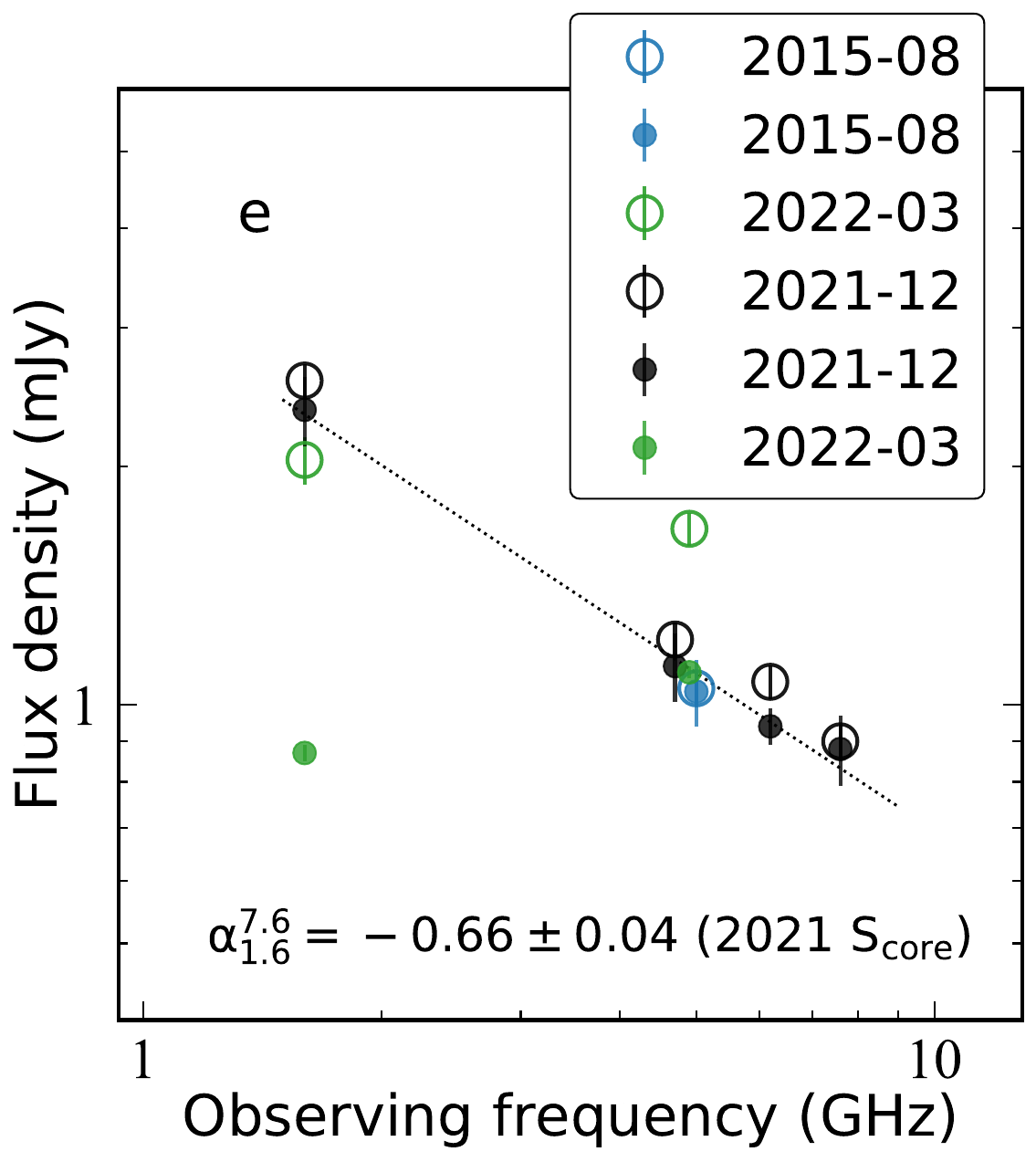}
\\
\end{tabular}
  \caption{VLBA images and radio spectrum of radio-quiet quasar PG 0921+525.  }
  \label{fig:0921}
\end{figure*}

\begin{figure*}
\centering
\begin{tabular}{cccc}
  \includegraphics[height=4.5cm]{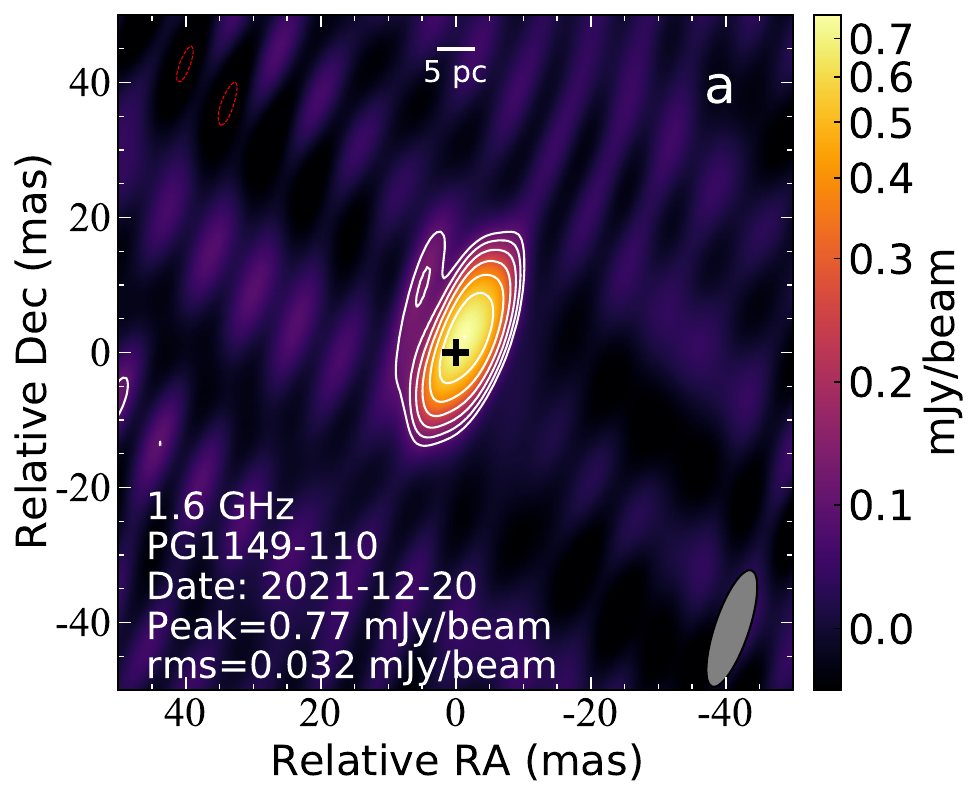}
  \includegraphics[height=4.5cm]{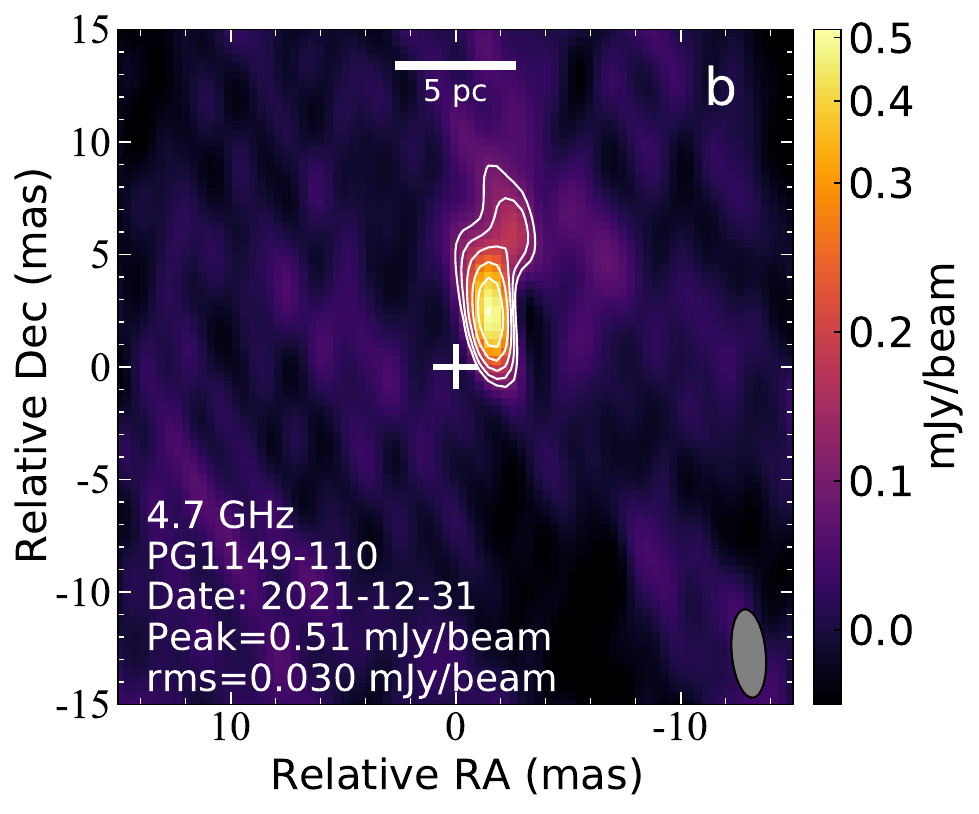}
  \includegraphics[height=4.5cm]{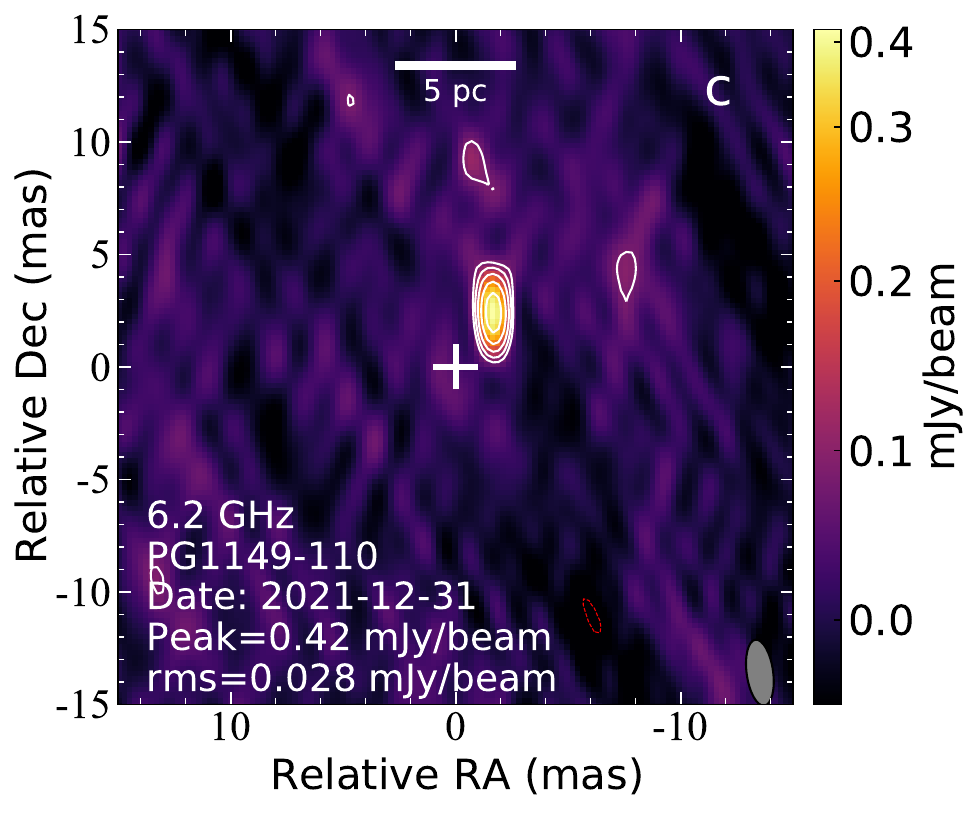} \\
  \includegraphics[height=4.5cm]{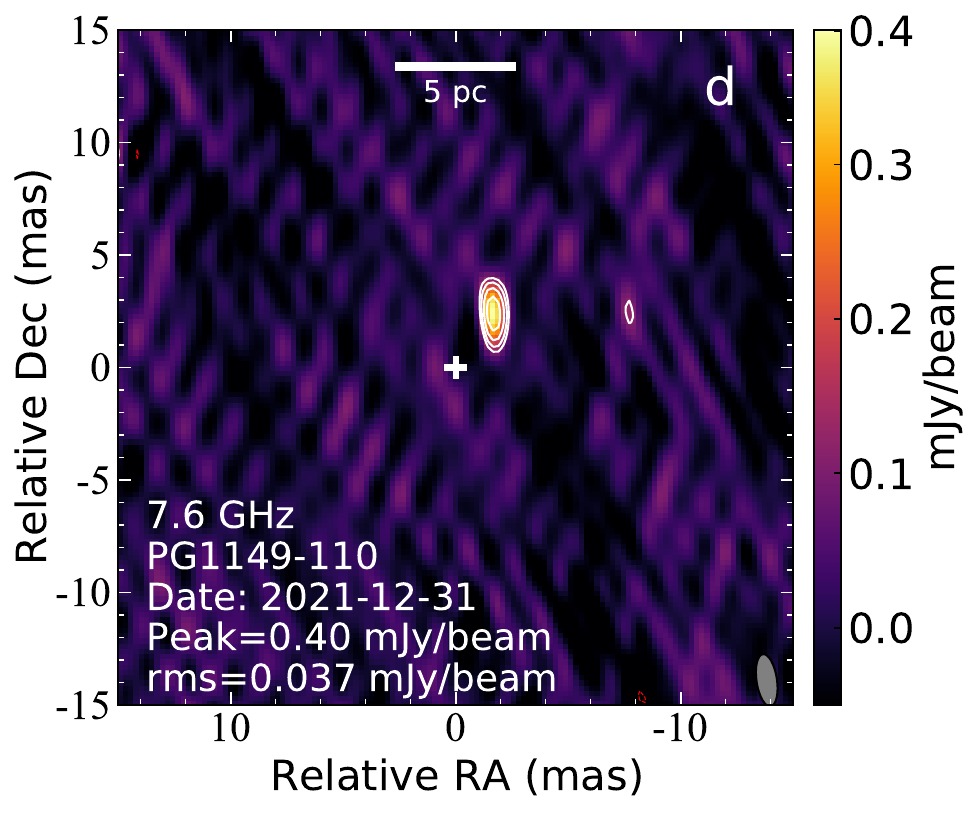}
  \includegraphics[height=4.5cm]{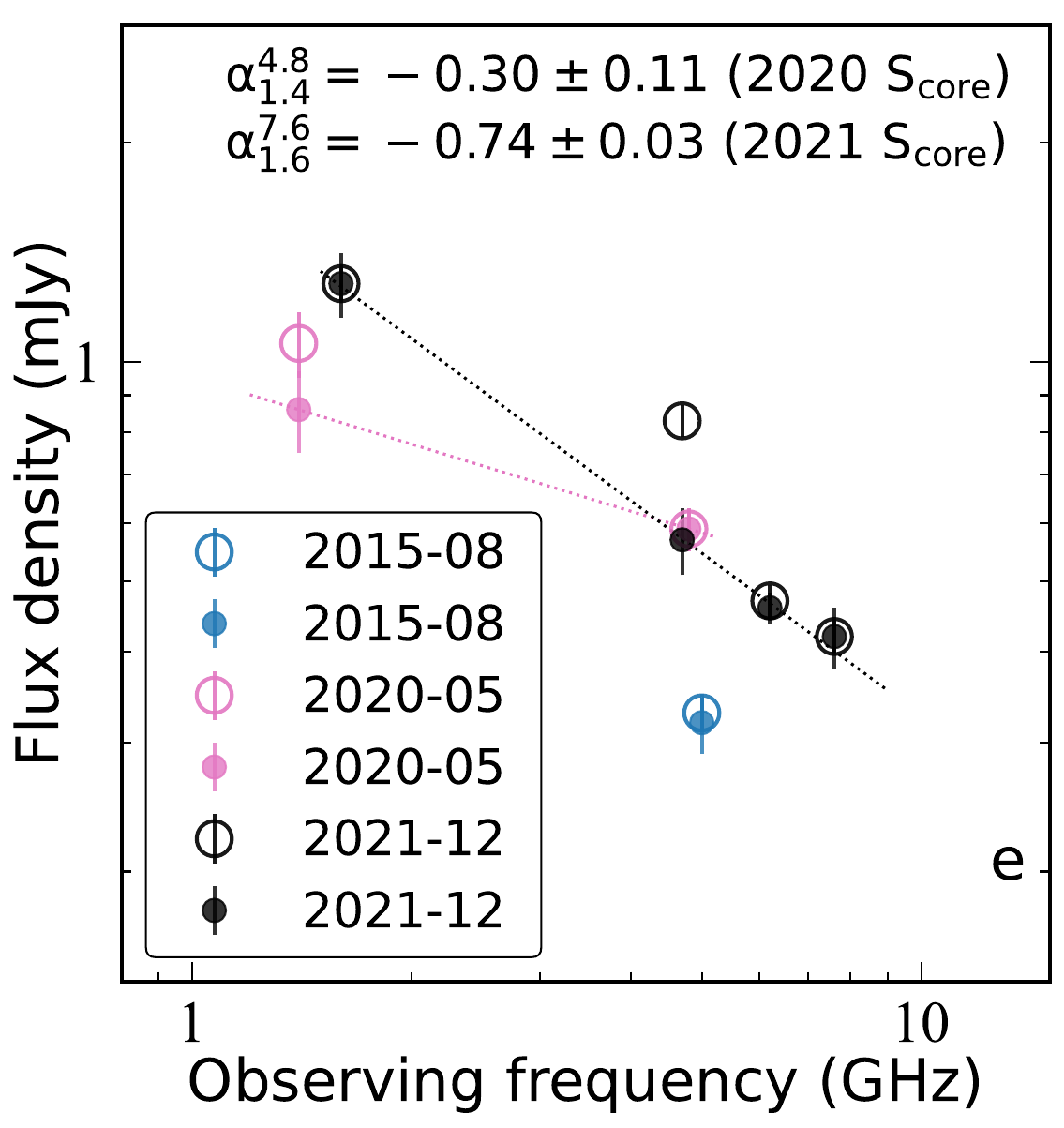}

\end{tabular}
    \caption{VLBA images and radio spectrum of radio-quiet quasar PG 1149$-$110. 
    Flux densities used in Panel \textit{e} are from \citet[][epoch 2015]{2023MNRAS.518...39W},  \citet[][epoch 2020]{2022ApJ...936...73A} and this paper (epoch 2021). }
    \label{fig:1149}
\end{figure*}

\begin{figure*}
\centering
\begin{tabular}{cccc}
  \includegraphics[height=4.5cm]{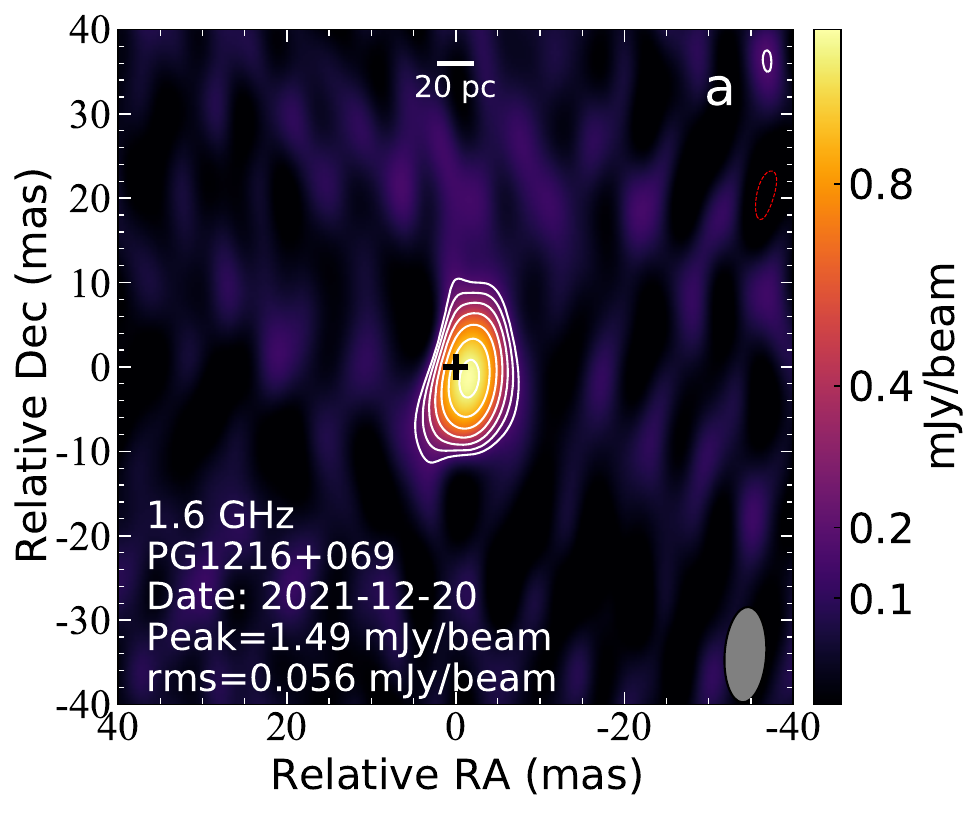}
  \includegraphics[height=4.5cm]{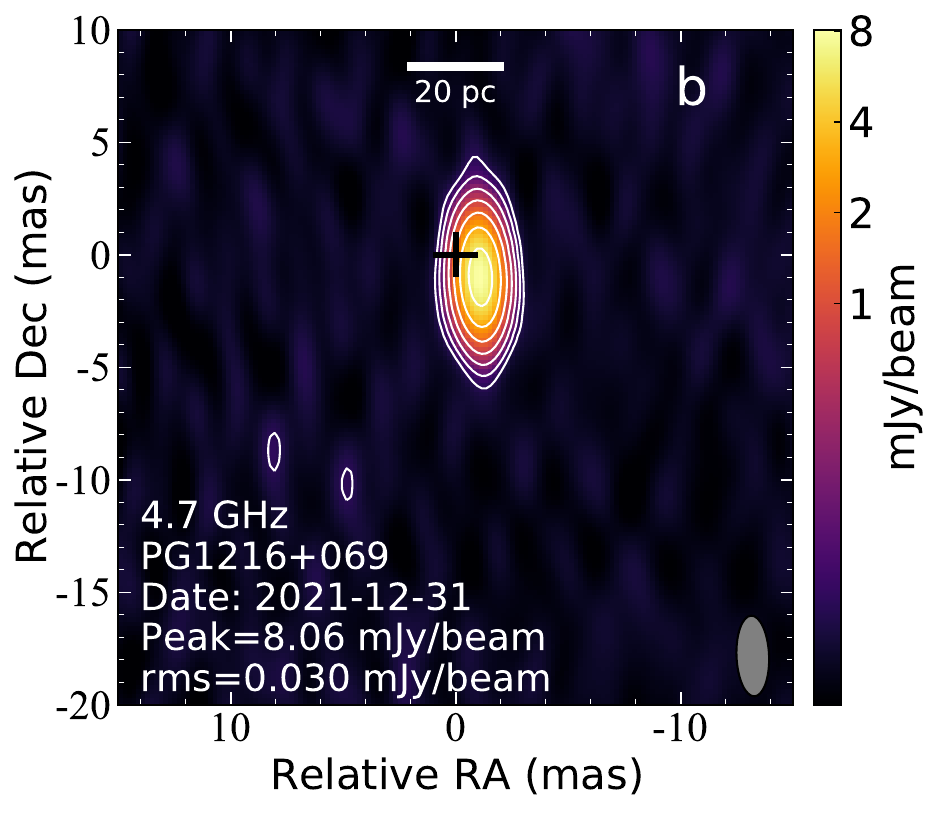}
  \includegraphics[height=4.5cm]{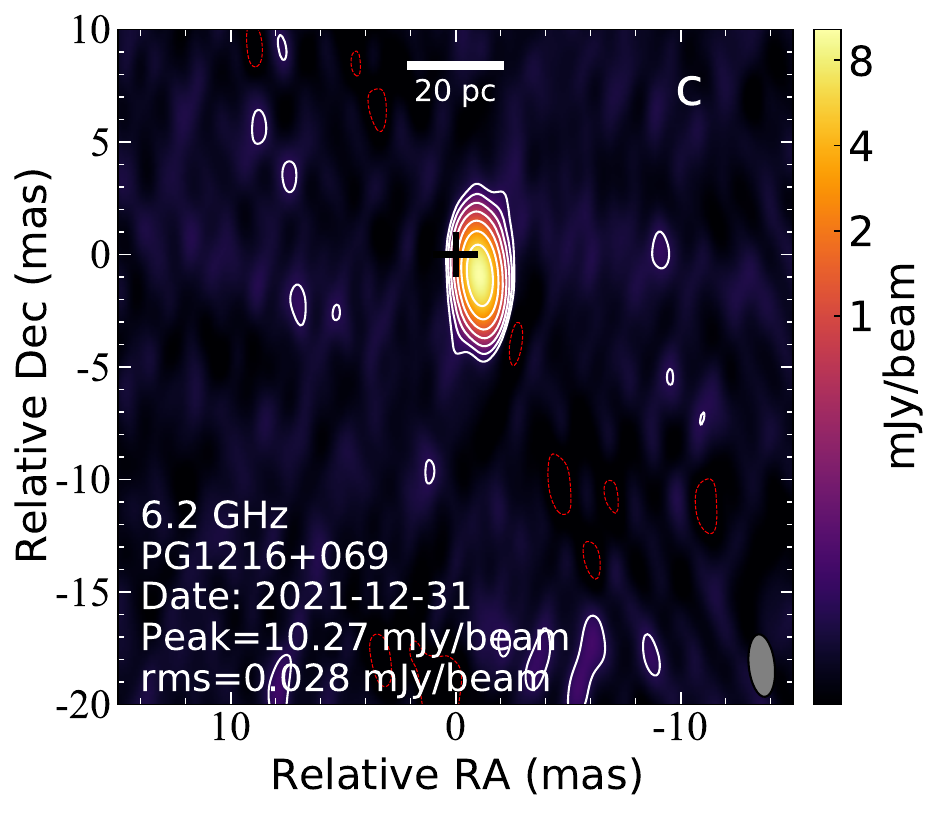} \\
  \includegraphics[height=4.5cm]{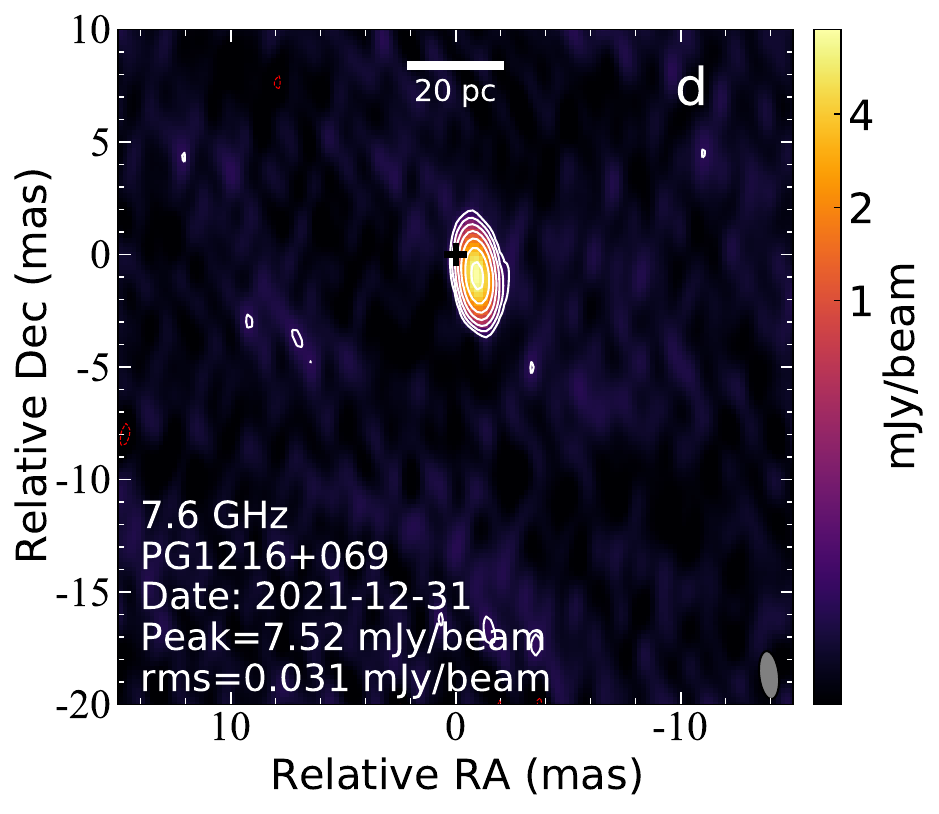}
  \includegraphics[height=4.5cm]{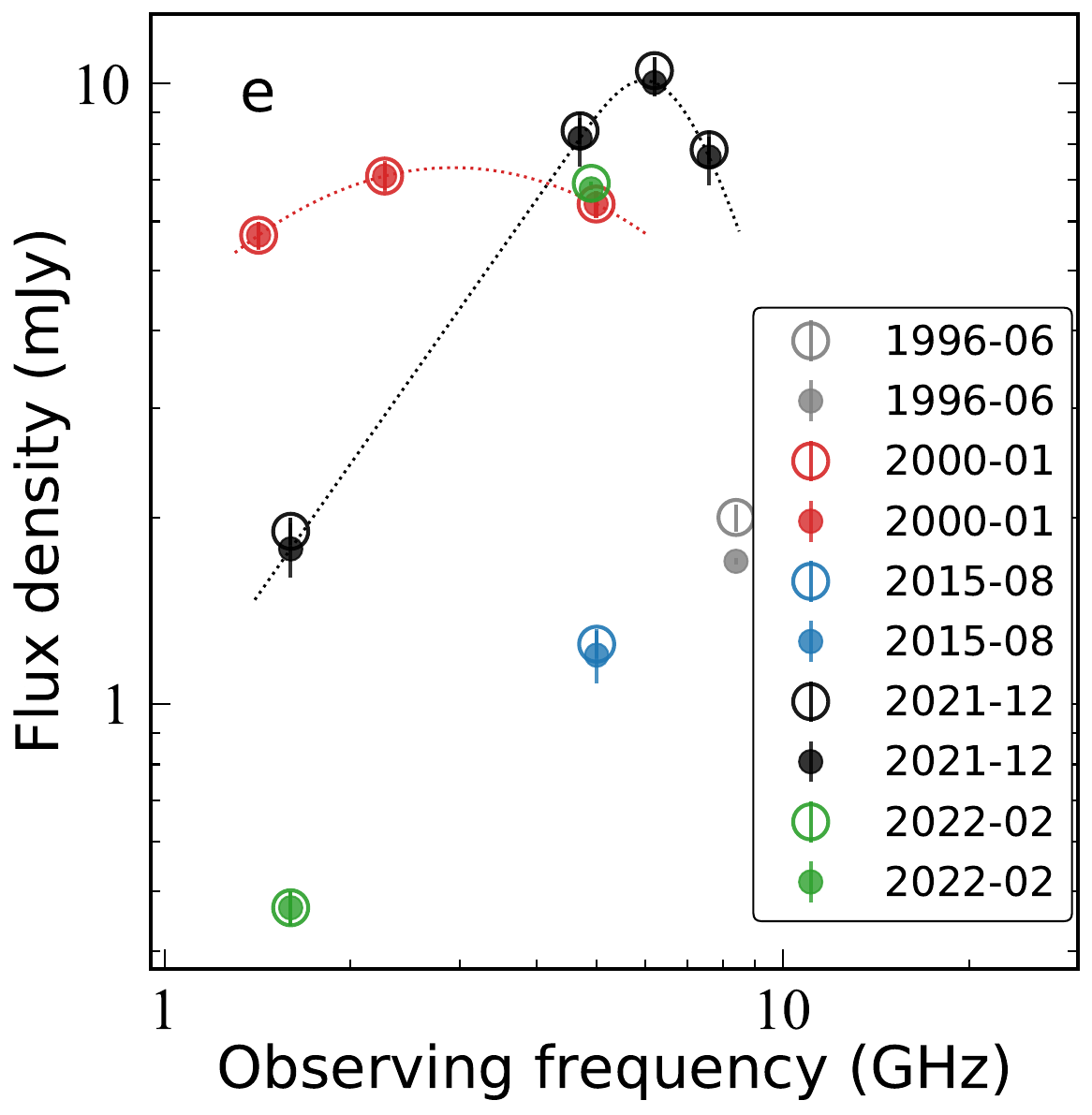}
\end{tabular}
  \caption{VLBA images and radio spectrum of radio-quiet quasar PG~1216+069. 
   Flux densities used in Panel \textit{e} are from \citet[][epoch 1996]{1998MNRAS.299..165B},  \citet[][epoch 2000]{2005ApJ...621..123U}, \citet[][epoch 2015]{2023MNRAS.518...39W}, this paper (epoch 2021) and \citet[][epoch 2022]{2023arXiv230713599C}. }
  \label{fig:1216}
\end{figure*}

\begin{figure*}
\centering
\begin{tabular}{cccc}
  \includegraphics[height=4.5cm]{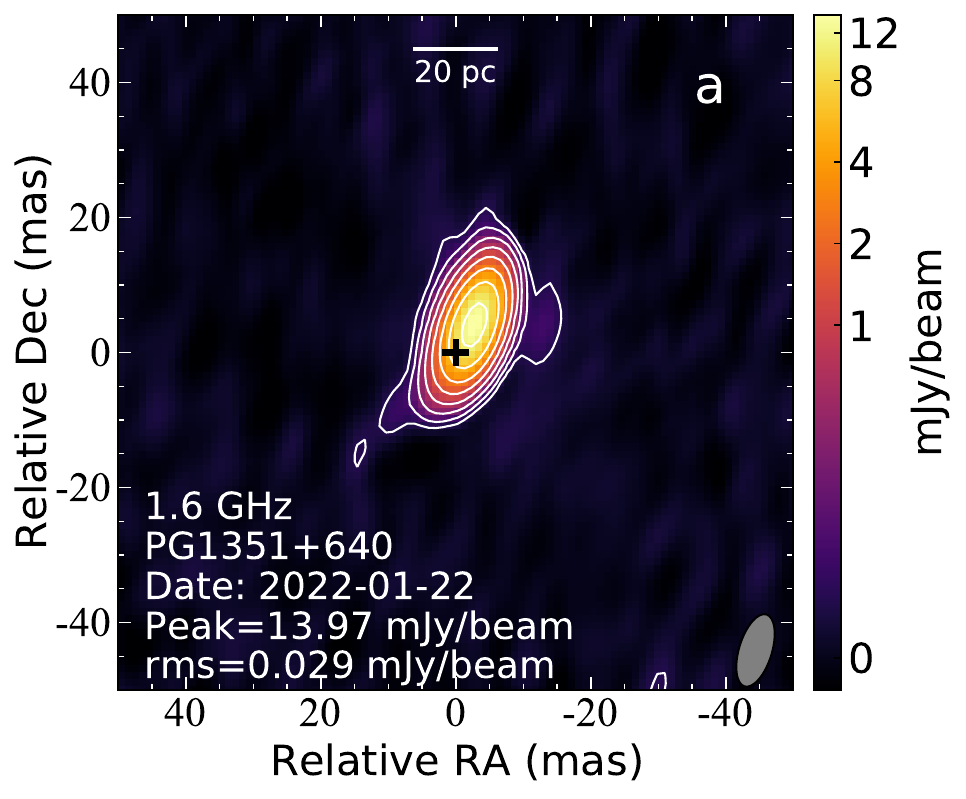}
  \includegraphics[height=4.5cm]{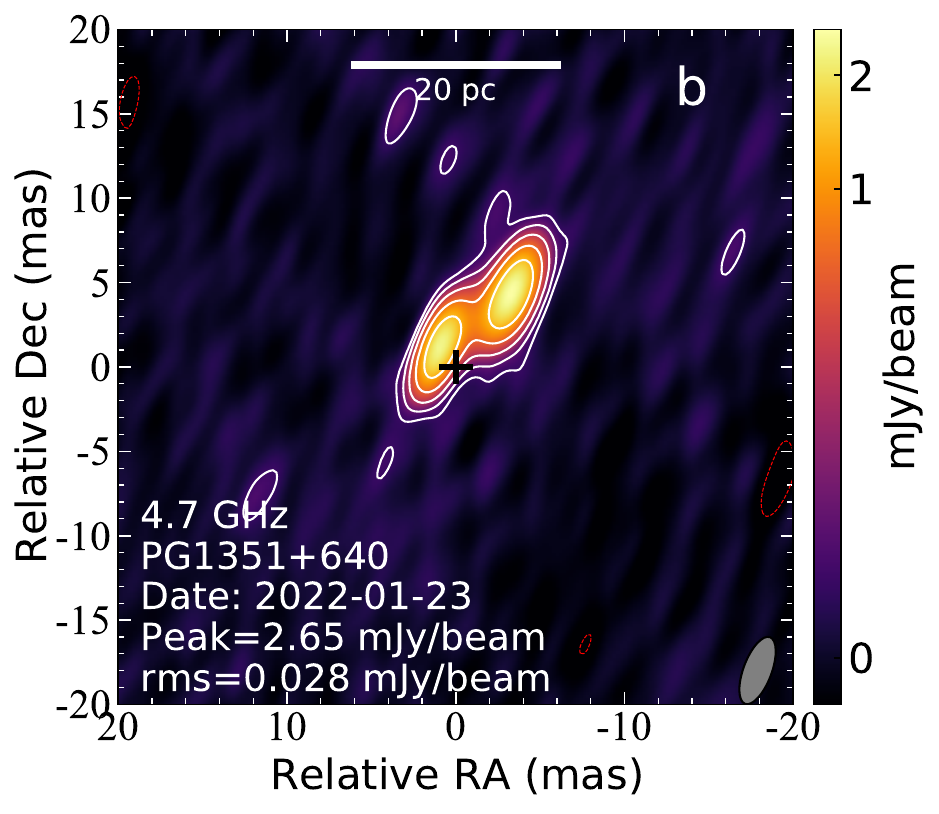}
  \includegraphics[height=4.5cm]{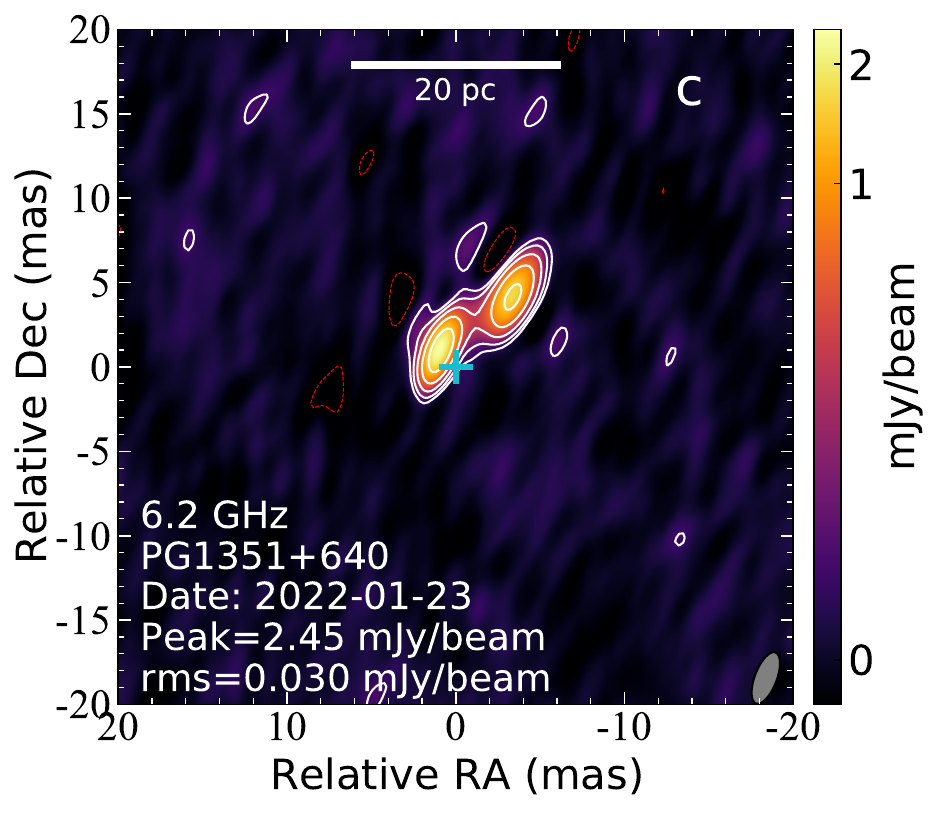} \\
  \includegraphics[height=4.5cm]{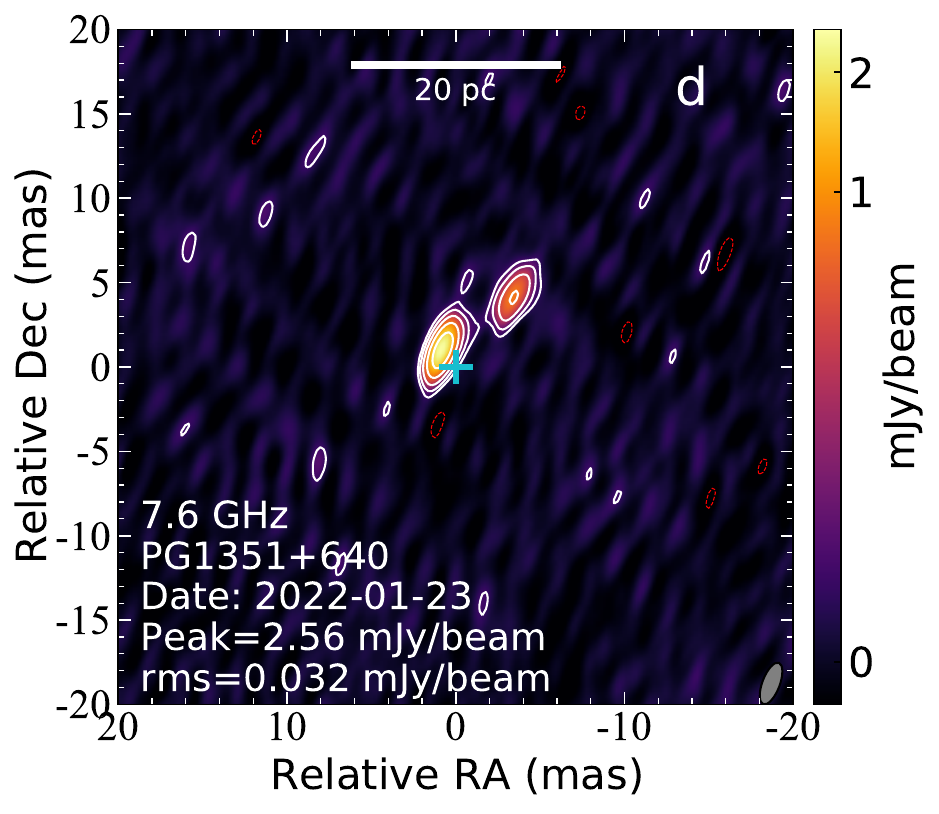}
  \includegraphics[height=4.5cm]{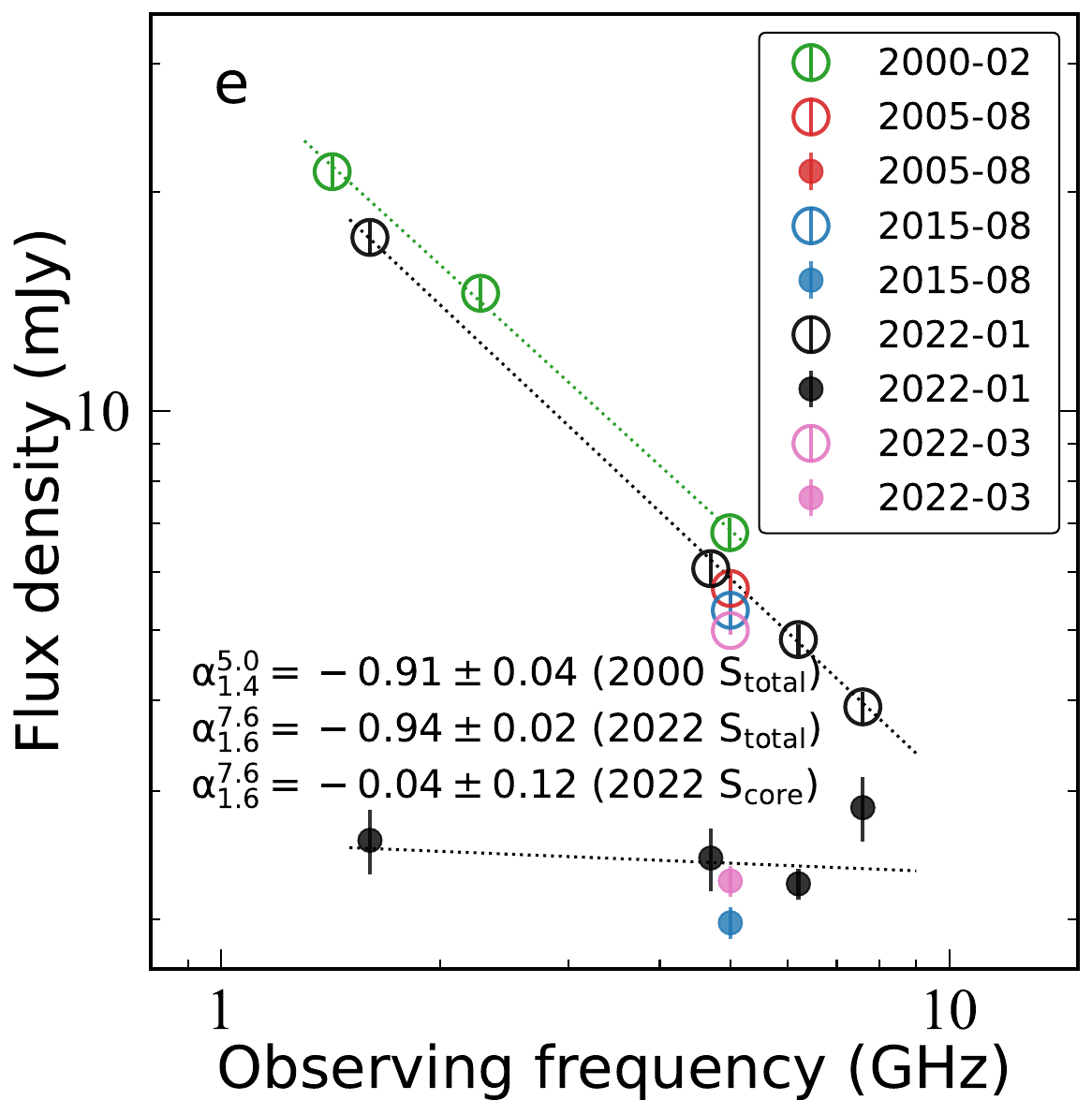}\end{tabular}
  \caption{VLBA images and radio spectrum of radio-quiet quasar PG~1351+640. 
  Flux densities used in Panel \textit{e} are from \citet[][epoch 2000]{2005ApJ...621..123U}, \citet[][epoch 2015]{2023MNRAS.518...39W}, this paper (epoch 2022-01) and \citet[][epoch 2022-03]{2023arXiv230713599C}. }
  \label{fig:1351}
\end{figure*}

\begin{figure}

\centering
  \includegraphics[width=0.45\textwidth]{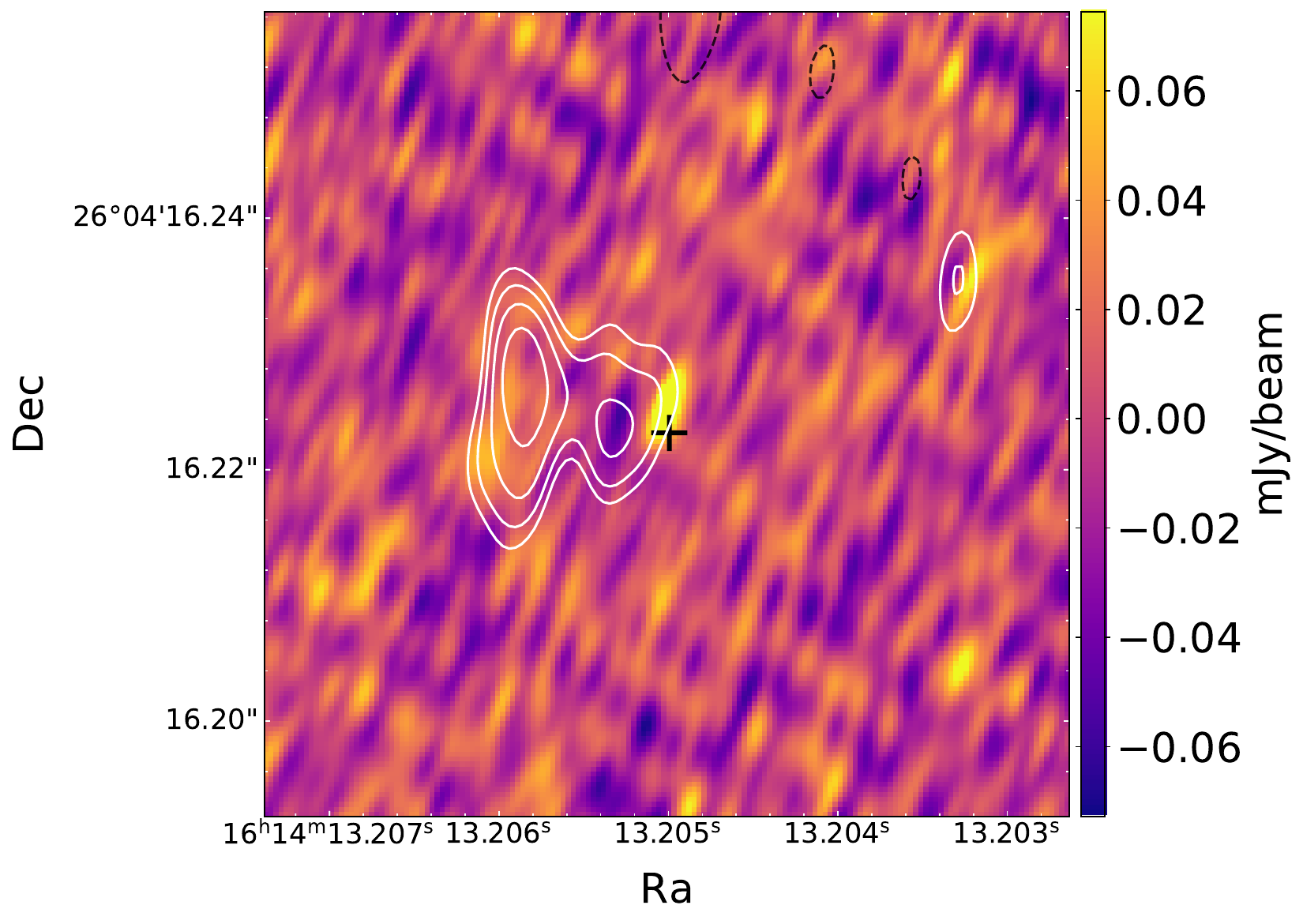}
  \includegraphics[height=5cm]{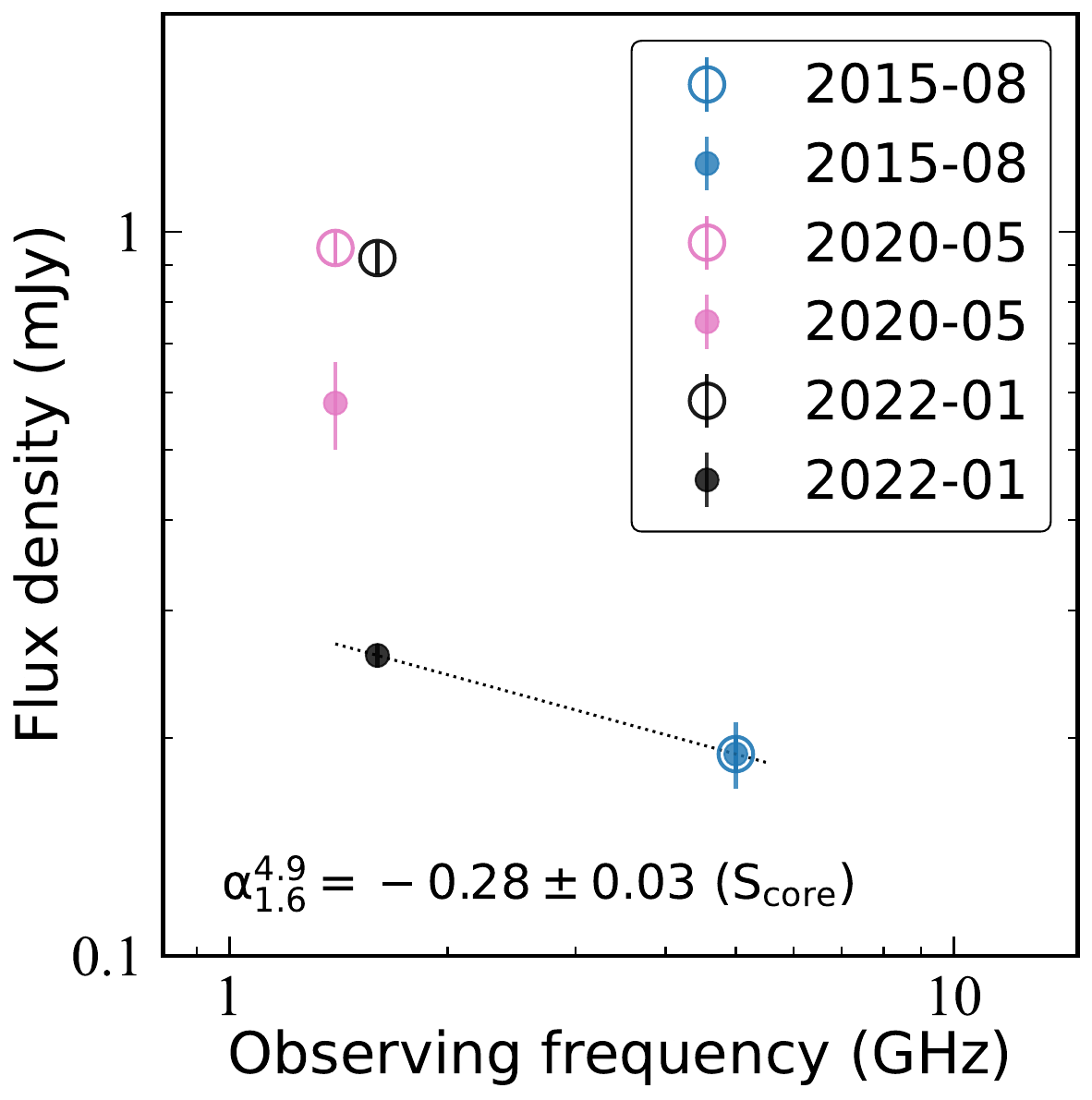}
  \caption{VLBA images and radio spectrum of radio-quiet quasar PG 1612+261.  
   Flux densities used in Panel \textit{b} are from \citet[][epoch 2015]{2023MNRAS.518...39W}, \citet[][epoch 2020]{2022ApJ...936...73A} and this paper (epoch 2022).}
  \label{fig:1612}

\end{figure}

\begin{figure*}
\centering
  \begin{tabular}{cccc}
  \includegraphics[height=4.5cm]{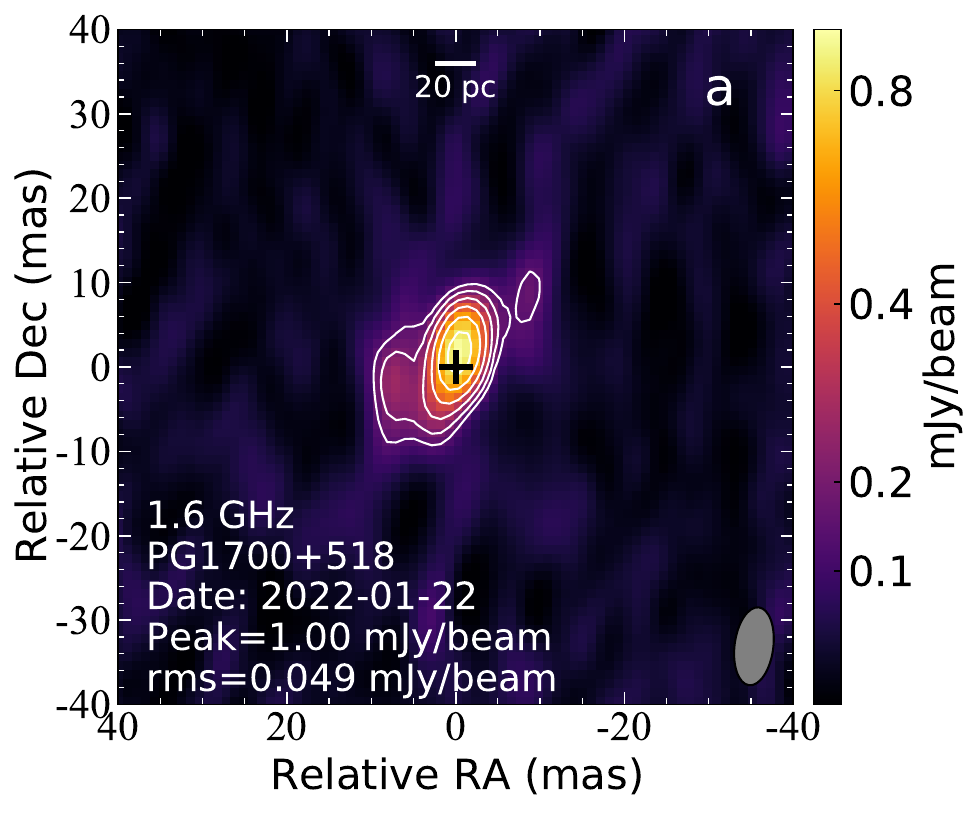}
  \includegraphics[height=4.5cm]{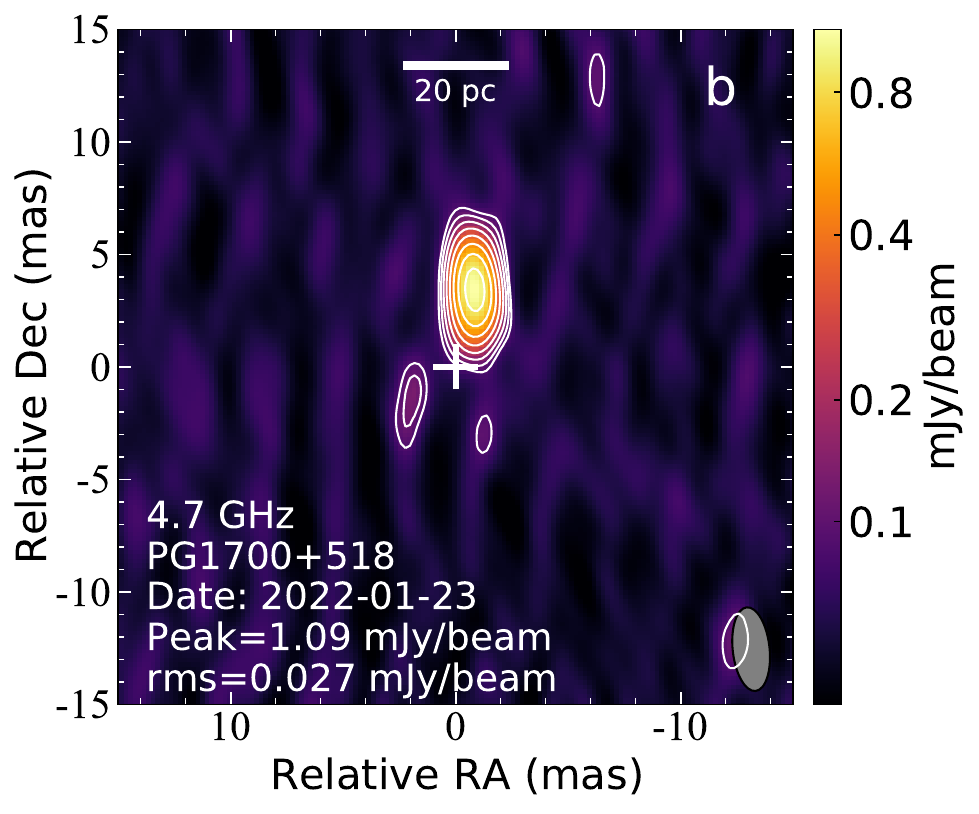}
  \includegraphics[height=4.5cm]{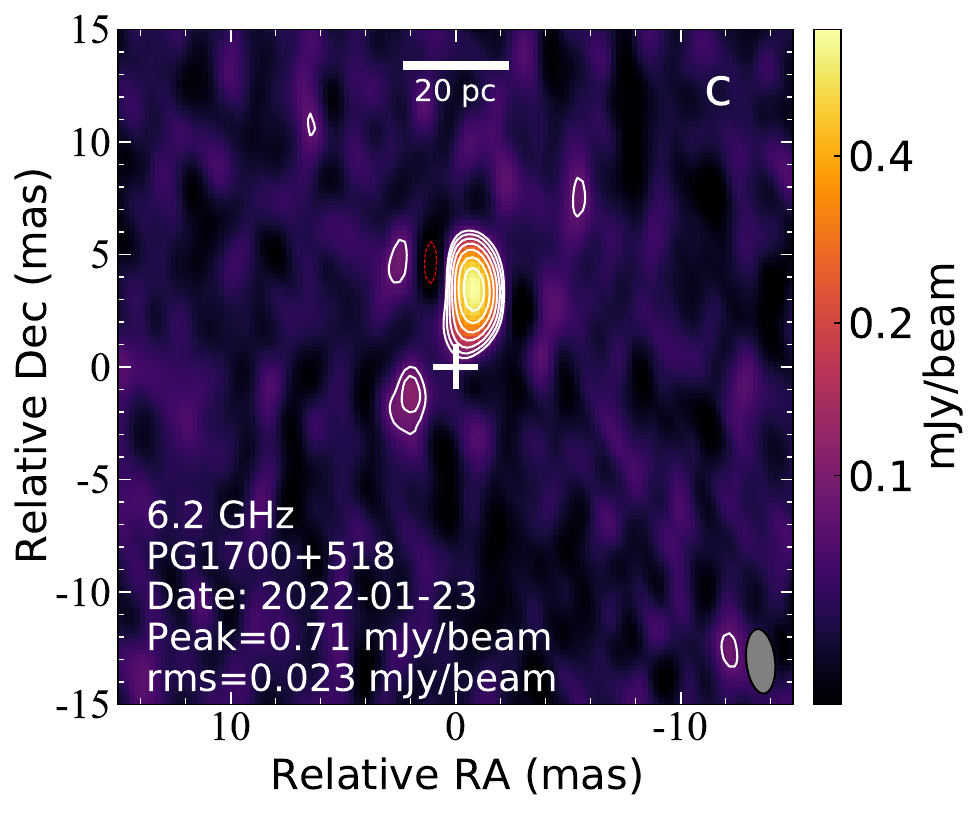} \\
  \includegraphics[height=4.5cm]{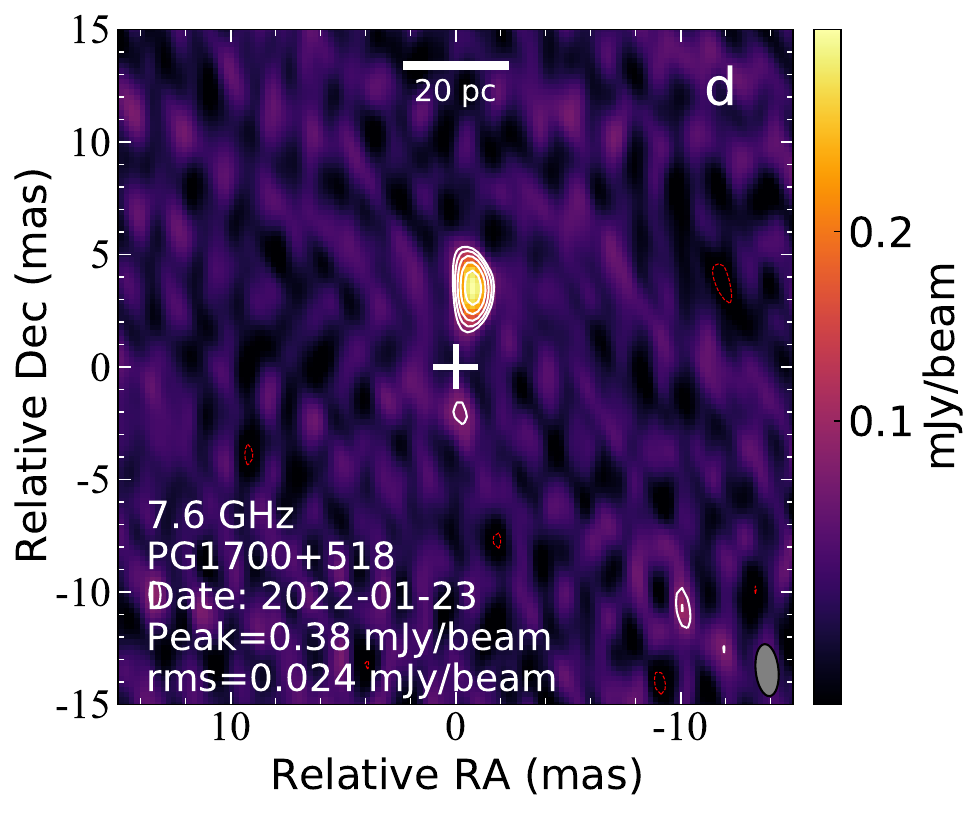}
    \includegraphics[height=4.5cm]{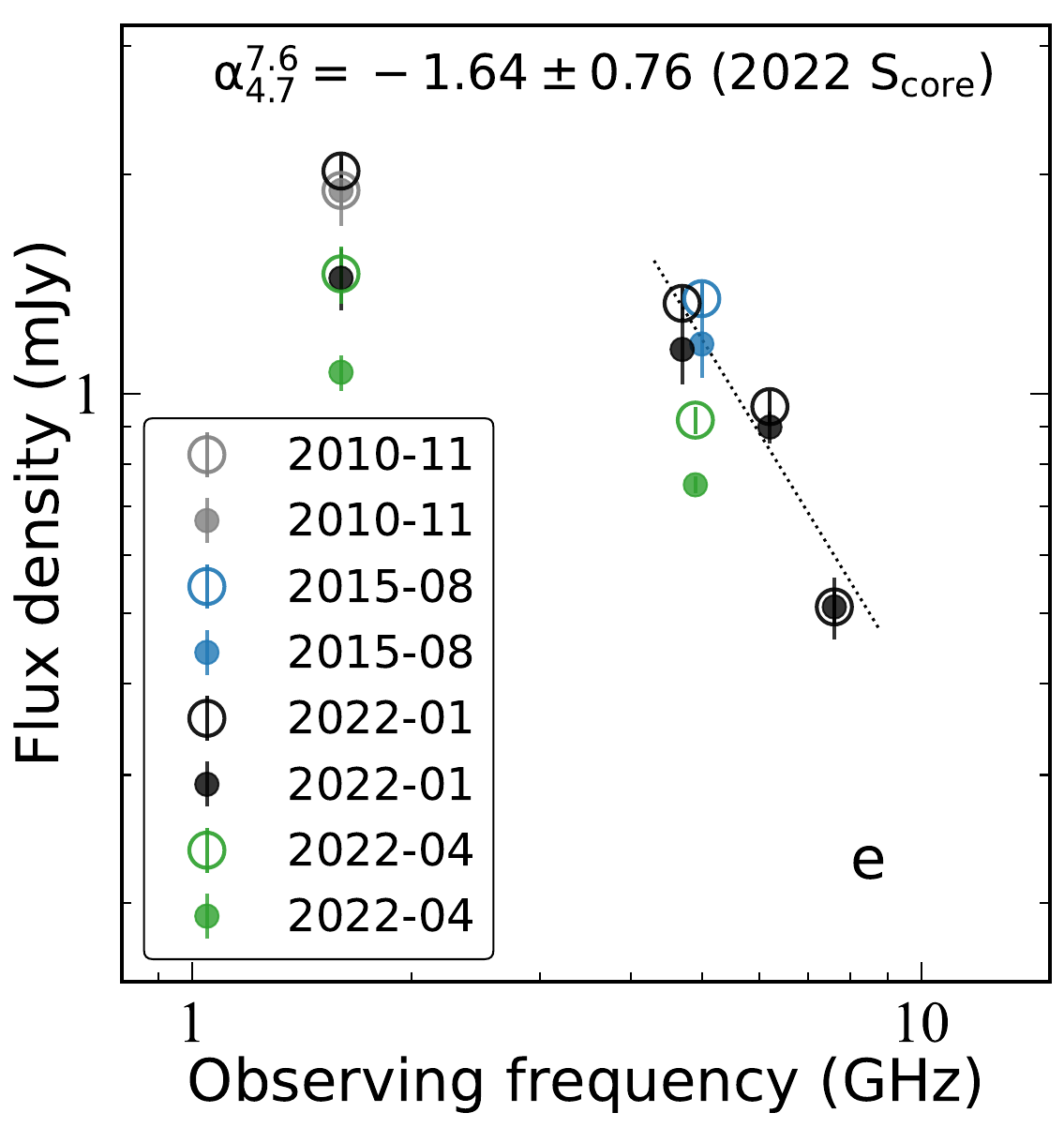}  
  \\
 \end{tabular}
  \caption{VLBA images and radio spectrum of radio-quiet quasar PG~1700+518. 
   Flux densities used in Panel \textit{e} are from \citet[][epoch 2010]{2012MNRAS.419L..74Y},  \citet[][epoch 2015]{2023MNRAS.518...39W}, this paper (epoch 2022-01) and and \citet[][epoch 2022-03]{2023arXiv230713599C}.  }
  \label{fig:1700}
\end{figure*}

\begin{figure*}
\centering
  \begin{tabular}{cccc}
  \includegraphics[height=4.5cm]{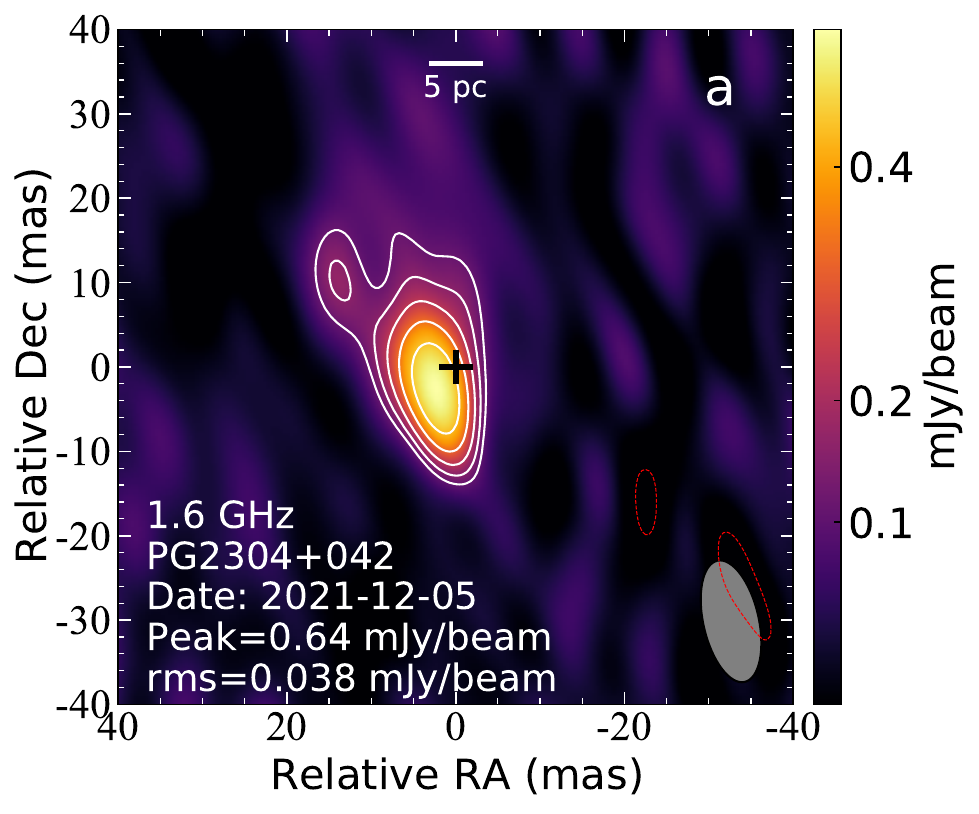}
  \includegraphics[height=4.5cm]{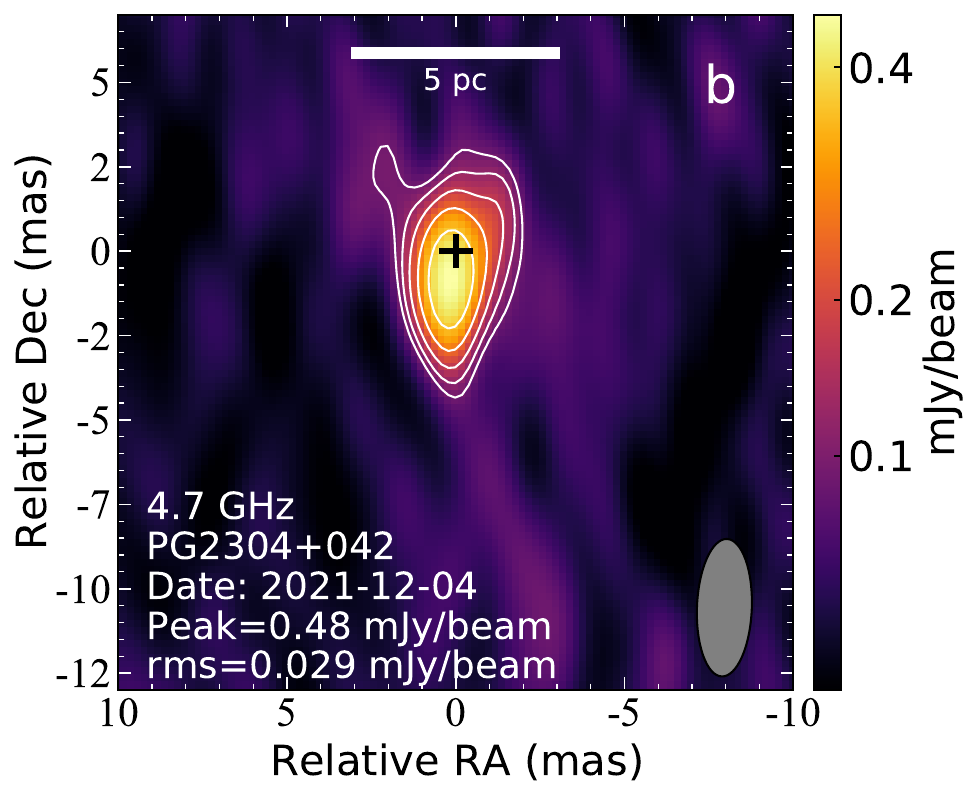}
  \includegraphics[height=4.5cm]{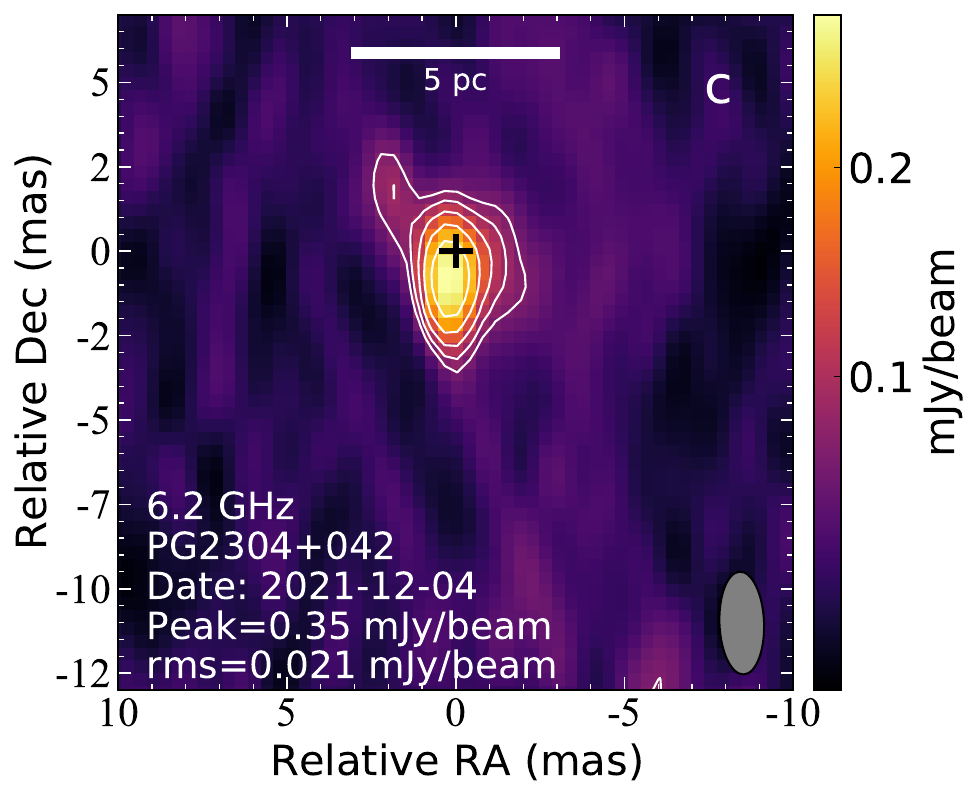} \\
  \includegraphics[height=4.5cm]{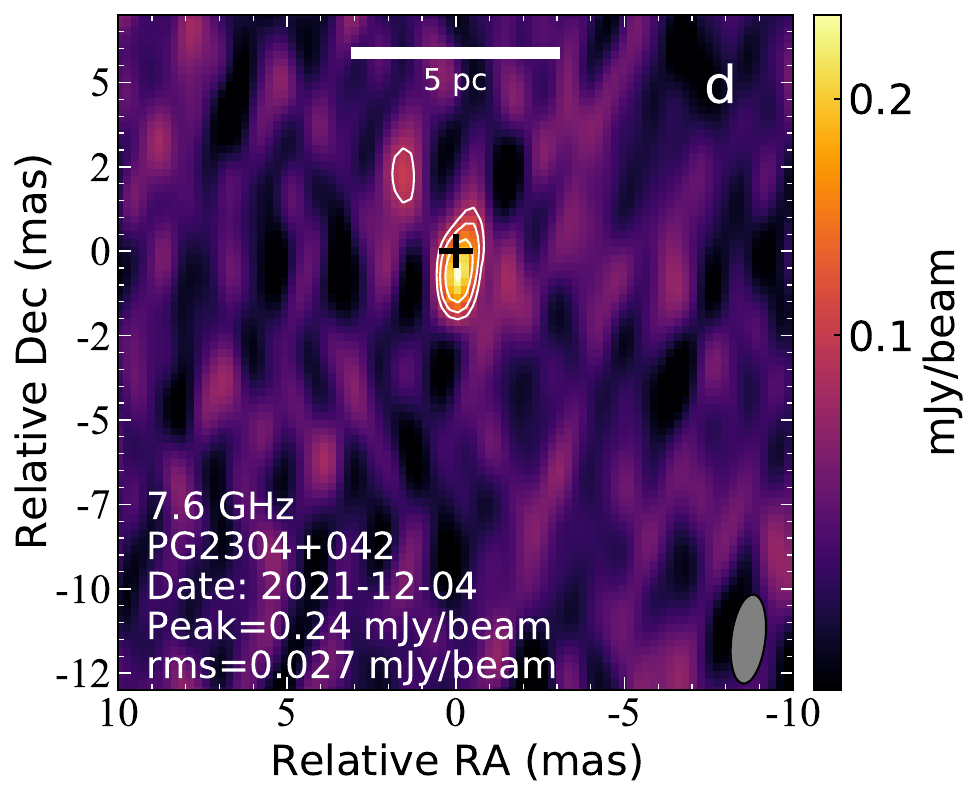}
  \includegraphics[height=4.5cm]{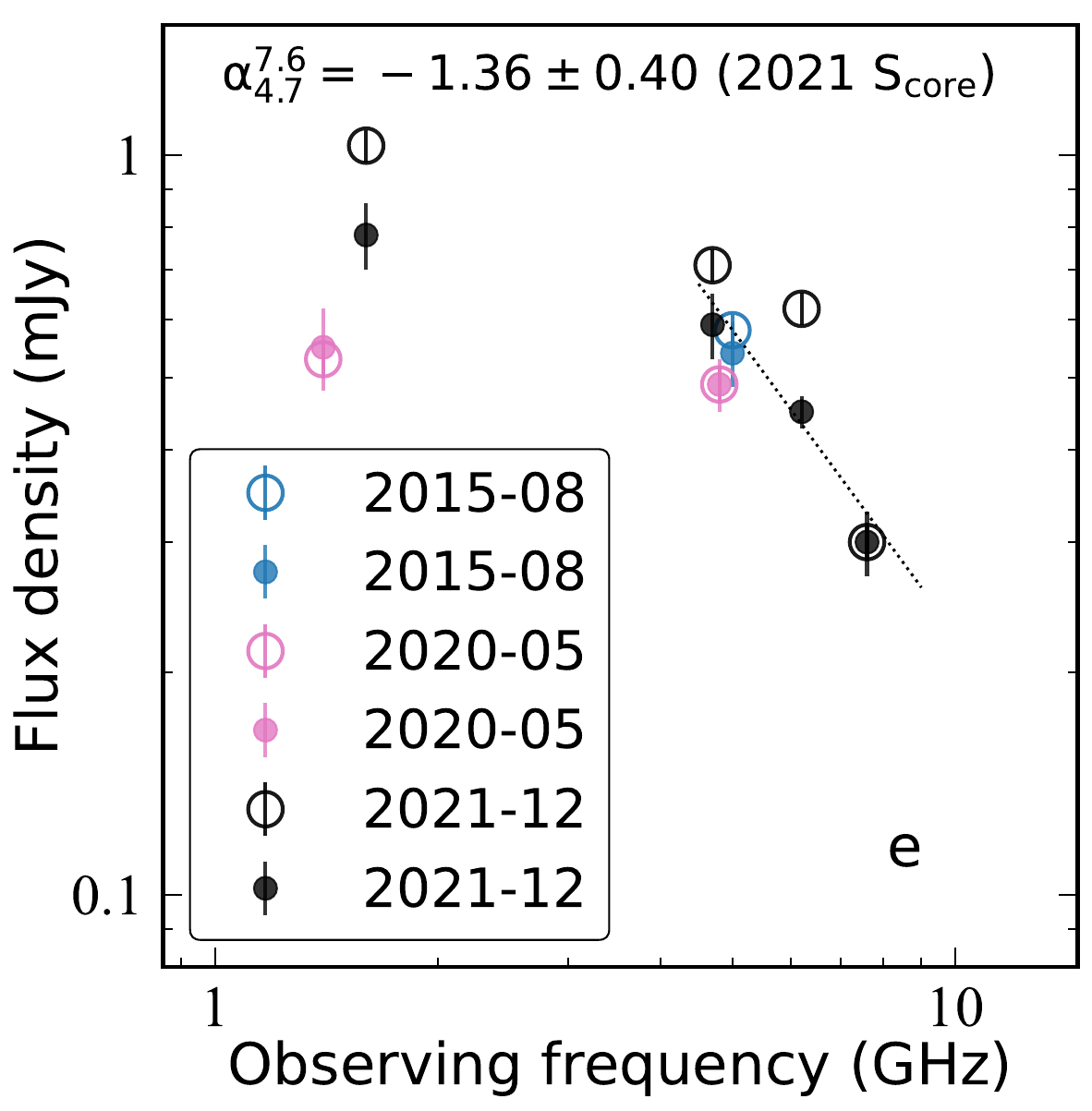}
  \end{tabular}
  \caption{VLBA images and radio spectrum of radio-quiet quasar PG~2304+042. 
  Flux densities used in Panel \textit{e} are from \citet[][epoch 2015]{2023MNRAS.518...39W},  \citet[][epoch 2020]{2022ApJ...936...73A} and this paper (epoch 2021).}
  \label{fig:2304}
\end{figure*}

Figures~\ref{fig:0003} to \ref{fig:2304} exhibit the VLBI images of these RQQs. To facilitate the identification of the radio core, the optical nucleus position, as determined by the \textit{Gaia} mission \citep{2023A&A...674A...1G}, is labelled in each image. Figure~\ref{fig:RLQ} displays the VLBI images of four RLQs.
To simplify the visualisation of the relative position of the radio and optical nuclei in RQQ, we have adjusted the (0, 0) point of each figure to coincide with the \textit{Gaia} optical position and the optical nucleus is marked with a cross. The primary source of uncertainty in the radio position is attributable to the position error of the phase reference calibrator, typically less than 1 mas, while the \textit{Gaia} position uncertainty is approximately 0.02 mas. 
Interestingly, the peak position of the brightest radio component often coincides with the AGN core. 
However, in cases such as PG~1149$-$110, PG~1216+069, PG~1351+640, and PG~1700+518, the peak radio emission shows a large offset from the image centre. 
The discrepancy between the optical position as determined by \textit{Gaia} and the radio core ($\Delta_{\rm OR}$) is presented in column 8 of Table \ref{tab:imagepar}. The significant offset ($\Delta_{\rm OR}>1$ mas) could potentially be due to, among other factors, the radio observations pointing towards a bright hotspot in the jet, rather than the central engine. Table \ref{table:Spectral_Index} presents the spectral index of PG quasars in our research sample.


\subsection{Radio quiet quasars}

Our study uses multi-frequency VLBI images of 10 RQQs and 4 RLQs to extend our previous study \citep{2023MNRAS.518...39W} and the existing literature, enabling a comprehensive analysis of their radio emission properties which include morphology,  radio spectrum, variability and ratio of radio to X-ray luminosity ($L_R$/$L_X$). The data suggest that the parsec-scale radio emission detected with VLBI in these RQQs is mainly associated with compact jets or possibly corona. The exceptional resolution provided by VLBI allows us to identify and characterise these compact features, while resolving other extended emission structures, such as star formation or large-scale diffuse jets. This finding emphasises the important role of compact jets or coronal structures as the main contributors to the observed radio emission in these RQQs on parsec scales. 
The observational results are summarised below and detailed in Table \ref{tab:radioprop}.

\begin{figure}
\centering
  \includegraphics[height=6cm]{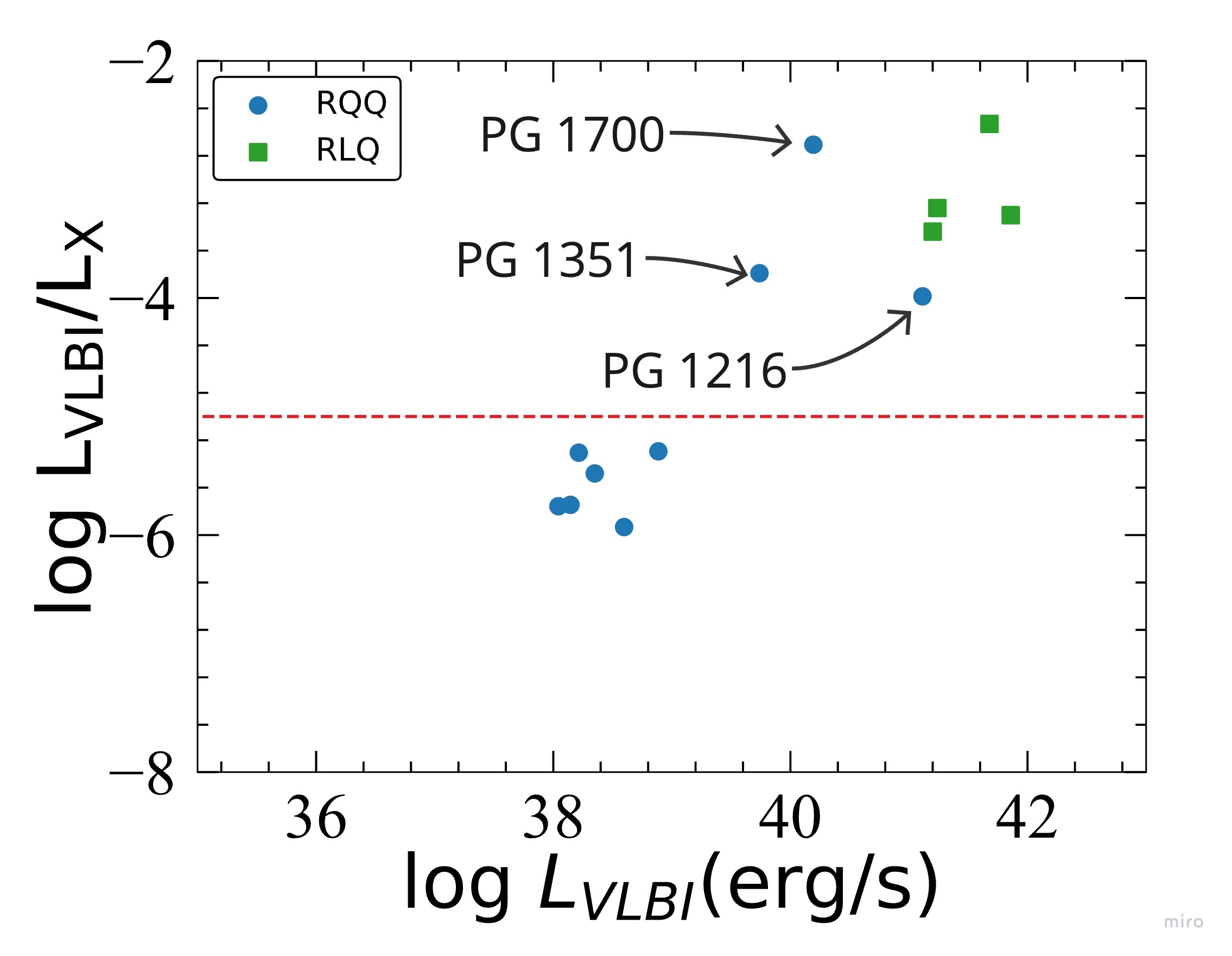}
  \caption{The correlation of radio-to-X-ray emission ratio and the radio luminosity. The radio luminosity is derived from the VLBI observations from this study, representing the parsec-scale emission structure.  The X-ray luminosity obtained from \citet{2008MNRAS.390..847L}. 
  There are significant $C_{IV}$ absorption in PG 1351+640 and PG 1700+518, which affect the X-ray leading higher luminosity than intrinsic.
  The red-coloured dotted line denotes the division line of $L_R$/$L_X$ = $10^{-5}$ between RQQs and RLQs.} 
  \label{fig:LrLx}
\end{figure}

 \begin{itemize}
 
     \item Radio structure. The VLBI images of the studied RQQs show diverse morphologies, including compact core, core with one-sided jets, and core with two-sided jets. The diversity in source compactness is also reflected in the varying brightness temperatures of the detected VLBI components. RQQs with lower brightness temperatures ($<10^{7.5}$~K), such as PG~0003+199, PG~0050+124, PG~0157+001, and PG~1612+261, show a weak core or a core accompanied by two-sided extended jets. In contrast, RQQs with higher brightness temperatures, including PG~0921+525, PG~1149-110, PG~1216+069, PG~1351+640, PG~1700+518, and PG~2304+042, exhibit either a core or a core with a compact one-sided jet. Notably, PG~0003+199 and PG~0050+124, which host black holes with high Eddington accretion rates, also exhibit active radio jets on the order of tens of parsecs, challenging traditional AGN emission models and warranting further exploration. PG~1351+640, which has a compact core-jet structure, has been discussed in detail in a separate paper \citep{2023MNRAS.523L..30W}.

    \item Radio spectrum. The GHz-frequency radio spectra, derived from the VLBI flux densities of our sample (Table \ref{table:Spectral_Index}), show significant variations and correlate with their corresponding radio structures. Sources such as  PG~0003+199, PG~0050+124 and PG~0157+001 with steep radio spectra ($\alpha_{\rm total}<0.9$) display extended jet-dominated structures, while sources with flat or inverted spectra (PG~1216+069, PG~1700+518, PG~2304+042) are characterised by naked cores or compact core-jet structures. 
    The core spectral indices of PG~0921+525 and PG~1149+110 are $\alpha^{7.6}_{1.6} = -0.66\pm0.03$ and $\alpha^{7.6}_{1.6} = -0.74\pm0.18$, respectively, which are identical to their $\alpha_{\rm total}$. This is consistent with their compact and unresolved core structure.  These spectral index values are slightly steeper than those conventionally defined for an AGN core, suggesting that the radio emission from these cores is a mixture of optically thick and thin jet emission. 
    Our VLBA observations reveal a peaked spectrum in PG~1216+069, with a turnover occurring between 6--7 GHz. This feature is suggestive of a self-absorbed jet base or corona. The physical process causing this spectrum turnover, as well as its temporal variation, deserves further study. The spectra of PG~1700+518 and PG~2304+042 show differences below and above 4.7 GHz: a steep power law above 4.7 GHz; the flux density at 1.6 GHz is substantially lower than the extrapolation from the spectrum above 4.7 GHz. Their spectra are similar to the peaked spectrum of PG~1216+069, but the turnover frequencies are not well defined due to insufficient data points. The total VLBI emission of PG~1351+640 has a steep spectrum, while its radio core exhibits a typical flat spectrum.

\item Variability. The variability of the parsec-scale radio emission, including its timescale and fractional amplitude, sheds light on the origin of the radio emission and the dynamics in the nuclear region. 
The VLBA cores of PG 0003+199 and PG 0050+124 exhibit negligible variability, consistent with the fact that their cores do not dominate the VLBA structure.
The total VLBA flux density of PG~0157+001 underwent a significant change of $\sim$50 per cent between epochs 2015 and 2021, mainly from the eastern hotspot.
The radio morphology alone doesn't provide enough information to ascertain whether the parsec-scale emission of PG 1612+261 originates from the jet or the wind. However, the variability of the VLBA flux density of PG 1612+261 over a period of several years suggests that the structure observed by the VLBA is more likely related to the jet rather than to the disk wind. The core of PG~1351+640 shows remarkable variability between two 5-GHz VLBA observations.
 PG~1216+069 exhibits extreme variability, with its flux density increasing by a factor of $\sim$7 from 2015 to 2021. Such extreme variability may indicate a blazar-like core \citep{1996ApJ...471..106F}, similar to observations in RIQs such as \textsc{III}~Zw~2 \citep{1999ApJ...514L..17F,2023ApJ...944..187W}. 

    \item Radio to X-ray luminosity ratio ($L_R$/$L_X$).  $L_R$/$L_X$ in RQQs can serve as a valuable indicator of the origin of the radio emission. A high radio-to-X-ray luminosity ratio may indicate that the radio emission is powered by jet-related processes, similar to those observed in radio-loud quasars. On the other hand, a low  $L_R$/$L_X$ ($\leq 10^{-5}$) may suggest that the radio emission is powered by other mechanisms, such as magnetically heated corona \citep{2008MNRAS.390..847L}. Figure \ref{fig:LrLx} shows the radio to X-ray emission ratio of the observed RQQs and RLQs. PG 1700+518, PG1351+640 and PG 1216+069 lie above the division line. The intrinsic X-ray luminosity of PG 1700+518 and PG~1351+640 is likely influenced by $C_{IV}$ absorption, making their actual $L_R$/$L_X$ ratios smaller than the observed values, but not by 1--2 orders of magnitude. Therefore, we are inclined to believe that the radio emission from PG~1700+519 and PG~1351+640 is dominated by jets. The high value of $L_R$/$L_X$ in PG 1216+069 supports the view that its parsec-scale radio emission is dominated by the jet.  PG 0921+525 and PG 1149$-$110 fall below the division line, but their steep spectra are inconsistent with the corona.   Even though PG 0003+199 and PG 0050+124 lie below the division line, their other observed properties, such as their steep radio spectra and extended radio structures, support their jet-radiation origin. PG~2304+042 has a $L_R$/$L_X$ of $10^{-5.74}$, and none of its other observational features (e.g., morphology, spectral shape) can aid in distinguishing the corona from the jet.
 \end{itemize}

\subsection{Radio loud quasars}
\begin{figure*}
\centering
  \begin{tabular}{cccc}
  \includegraphics[height=3.8cm]{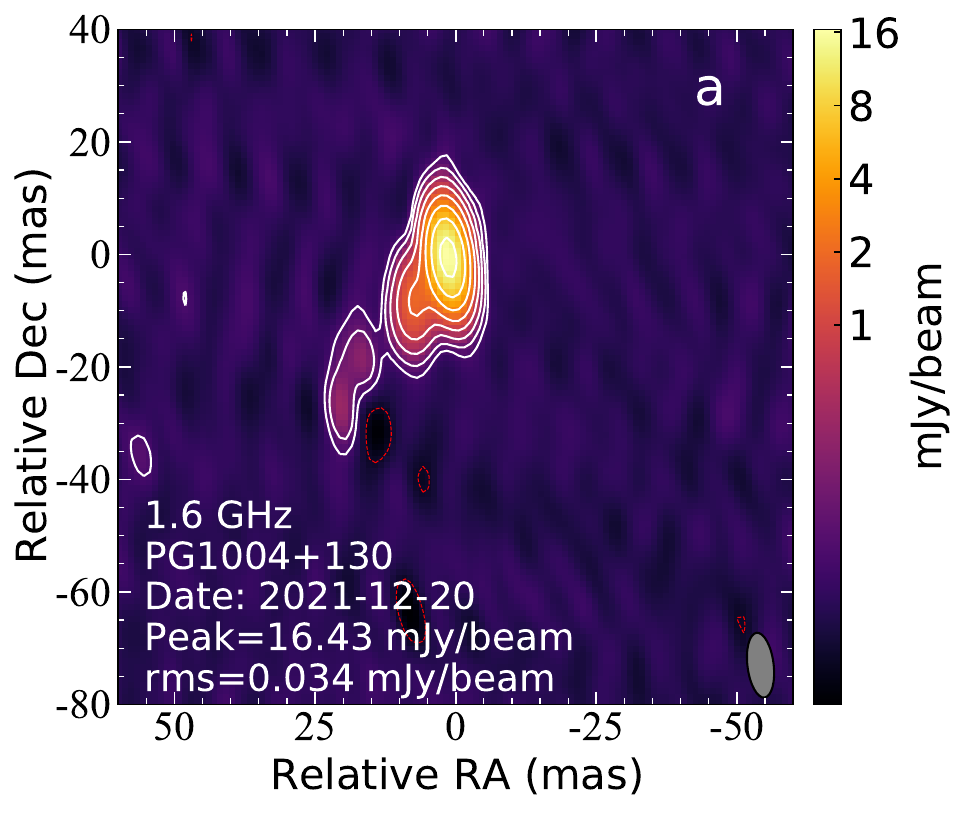}
  \includegraphics[height=3.8cm]{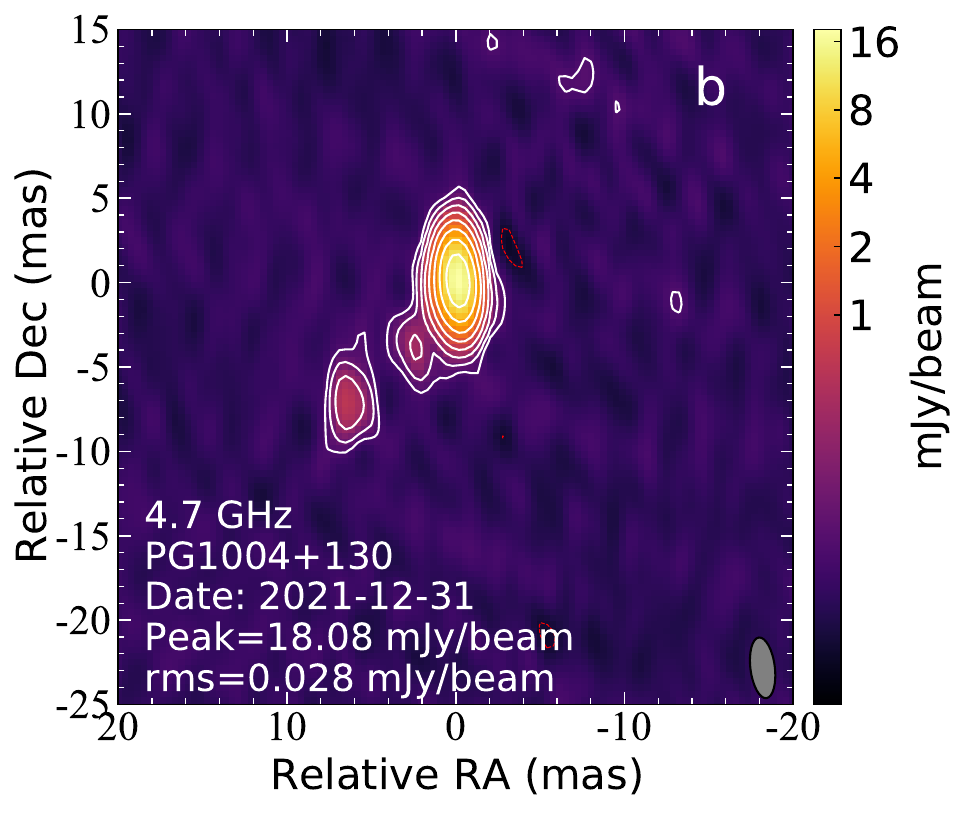}
  \includegraphics[height=3.8cm]{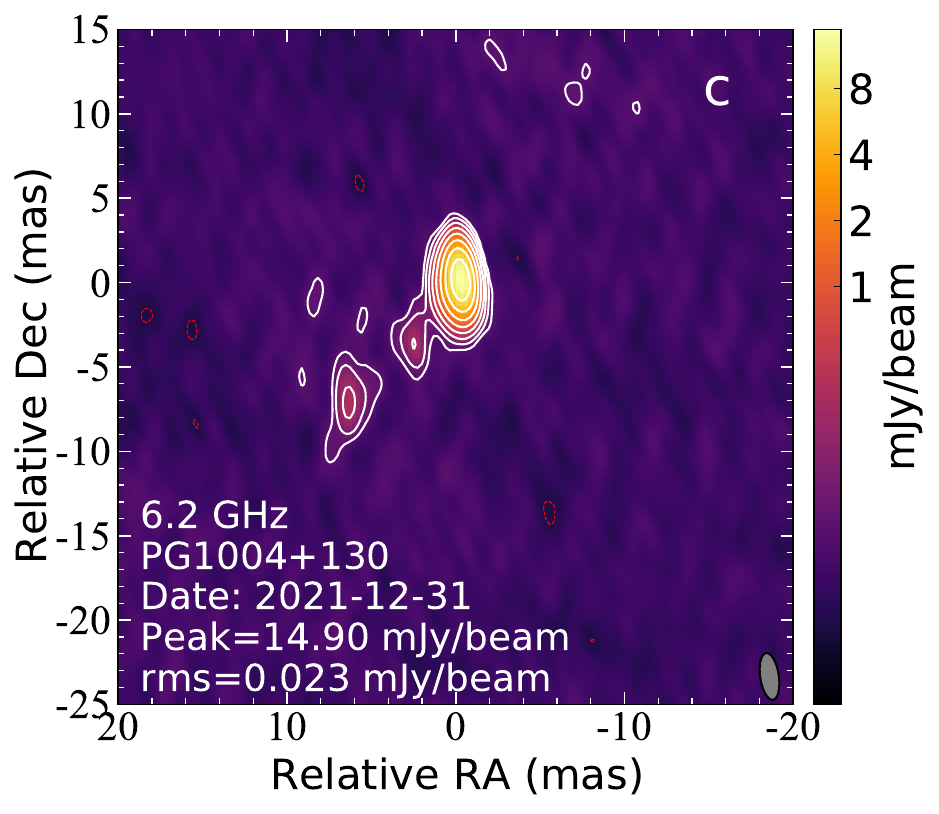}
  \includegraphics[height=3.8cm]{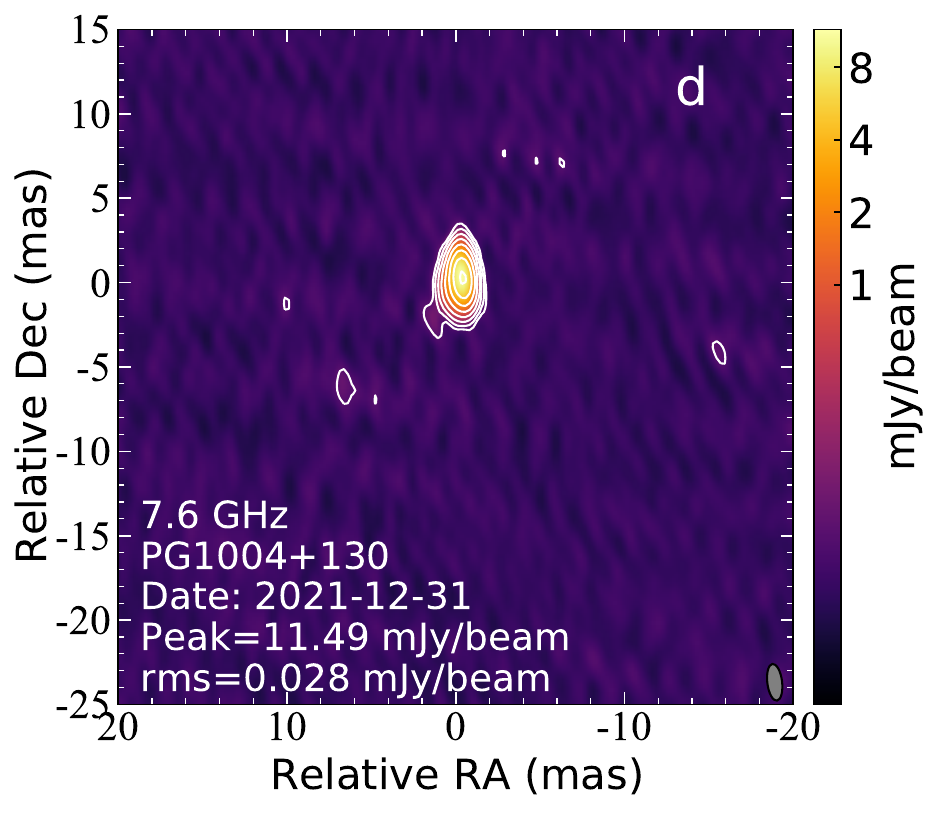}
  \\
  \includegraphics[height=3.8cm]{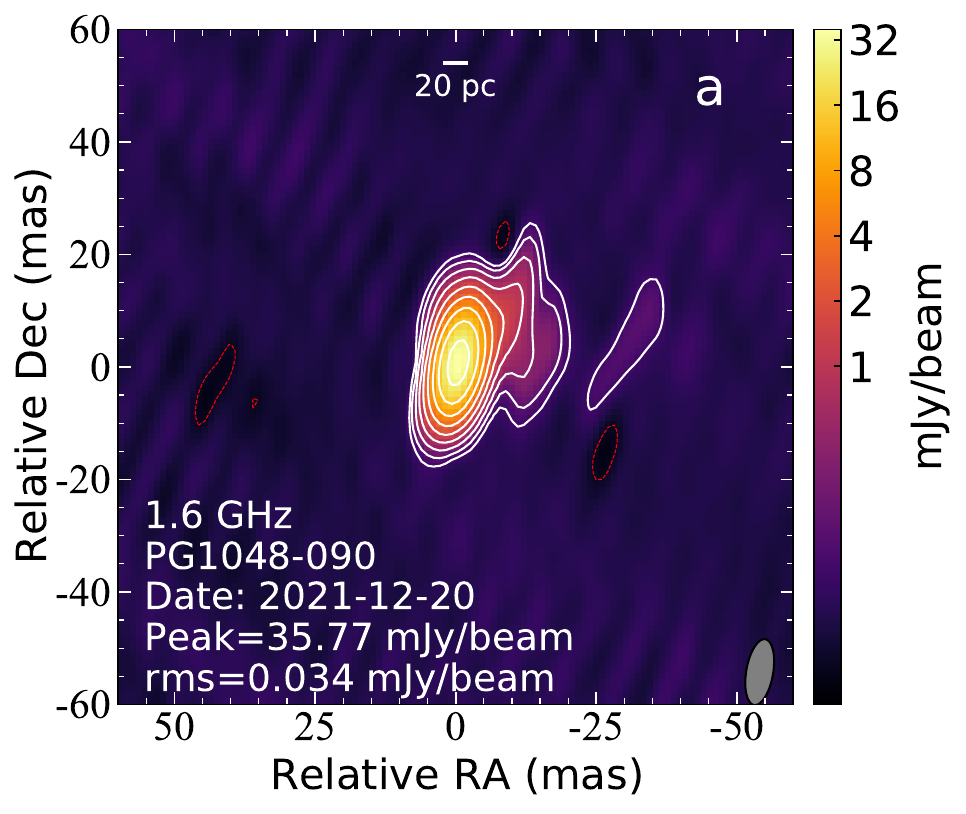}
  \includegraphics[height=3.8cm]{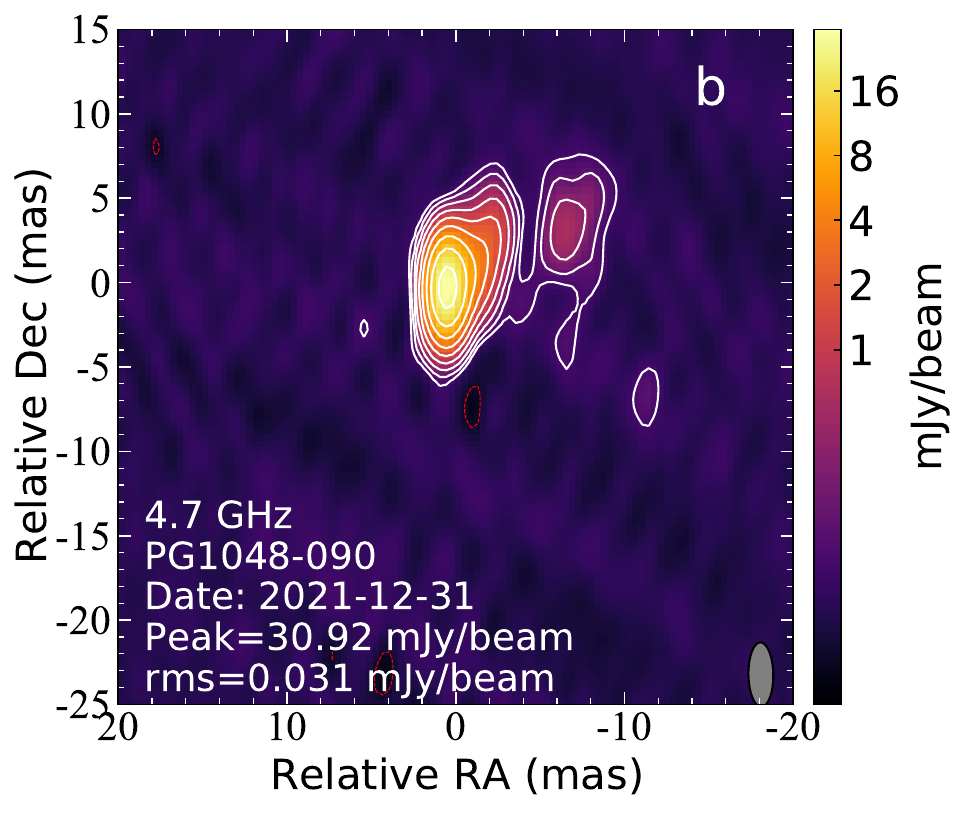}
  \includegraphics[height=3.8cm]{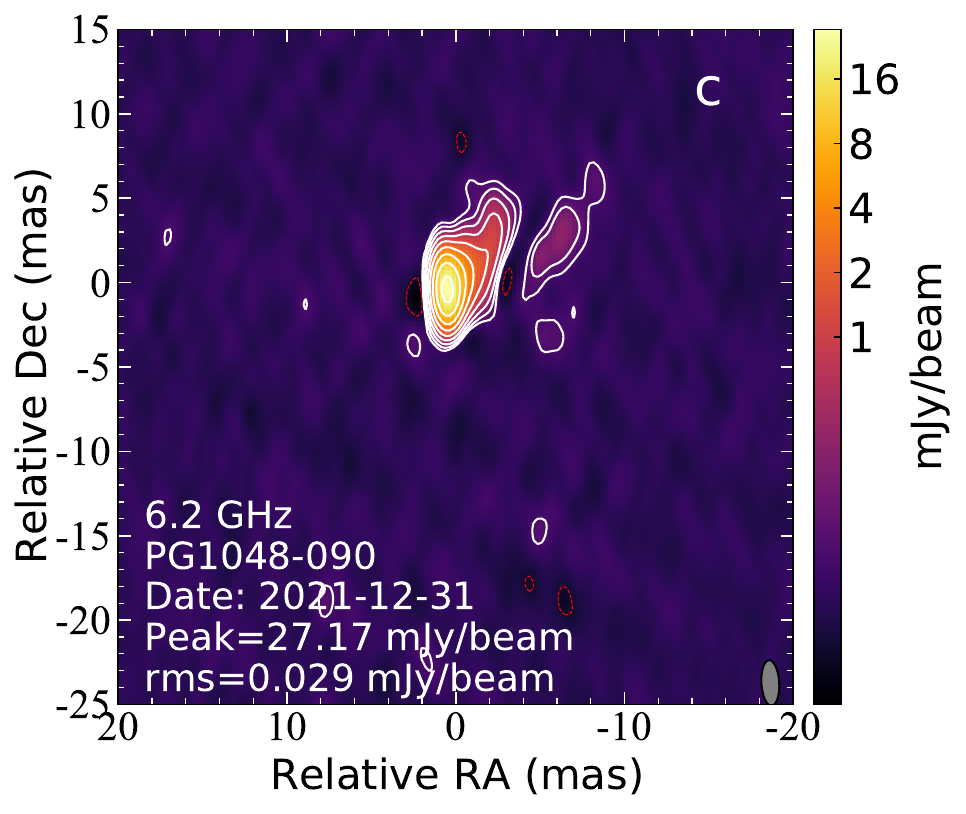}
  \includegraphics[height=3.8cm]{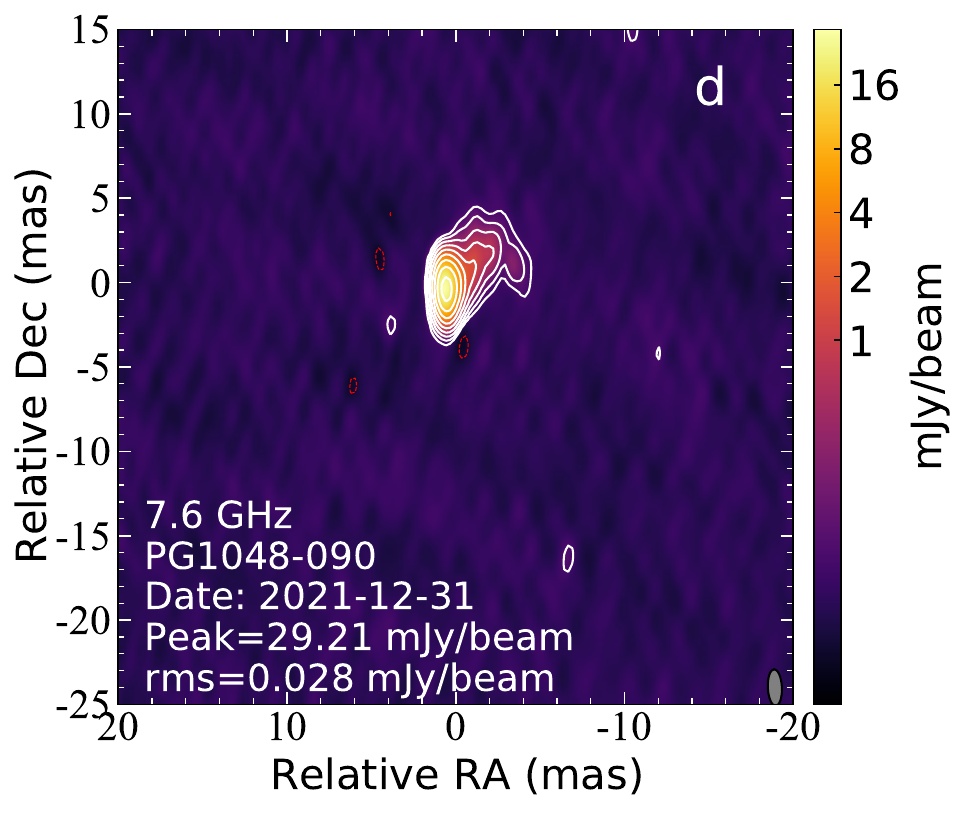} 
\\
  \includegraphics[height=3.8cm]{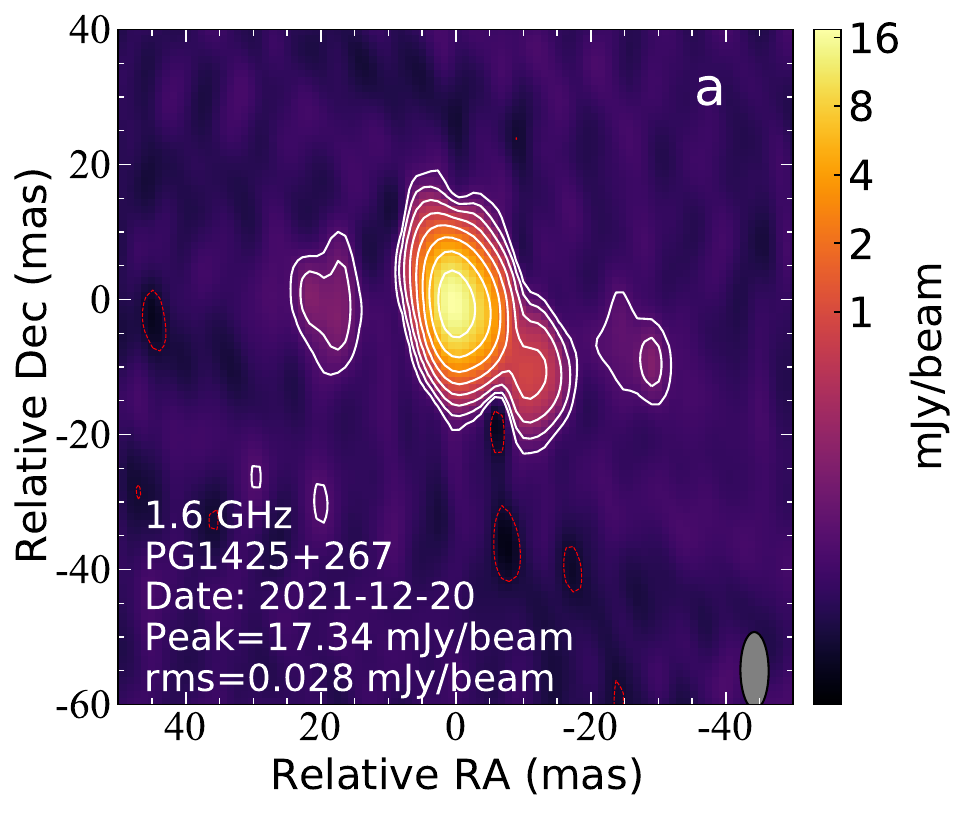}
  \includegraphics[height=3.8cm]{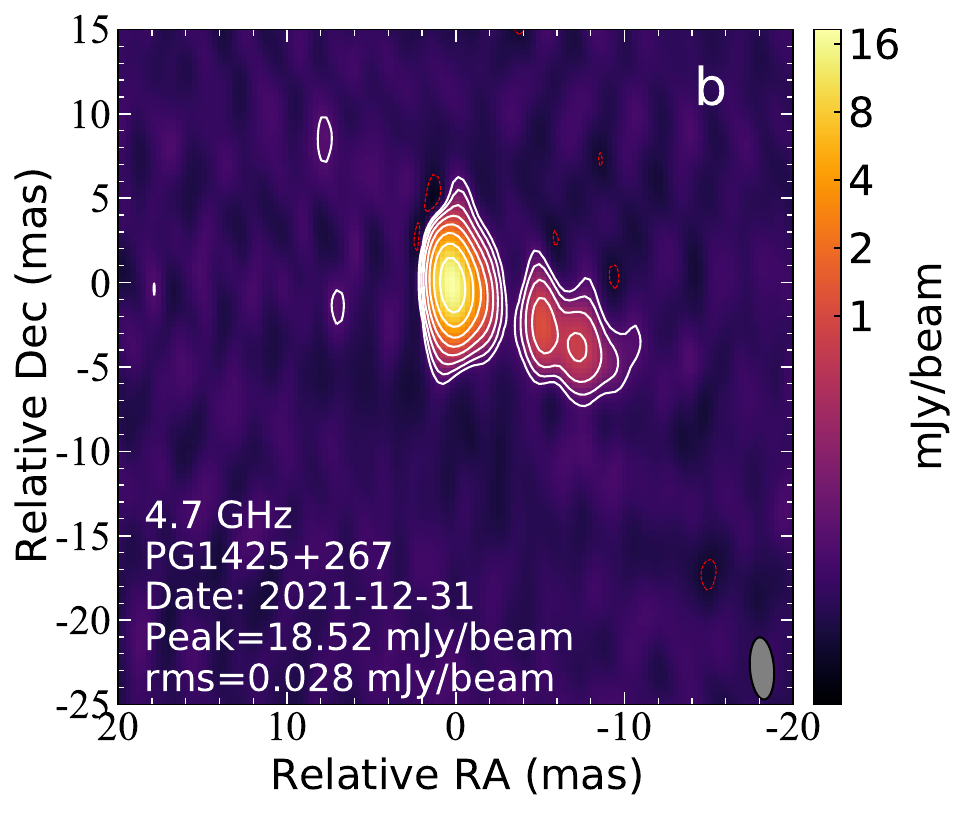}
  \includegraphics[height=3.8cm]{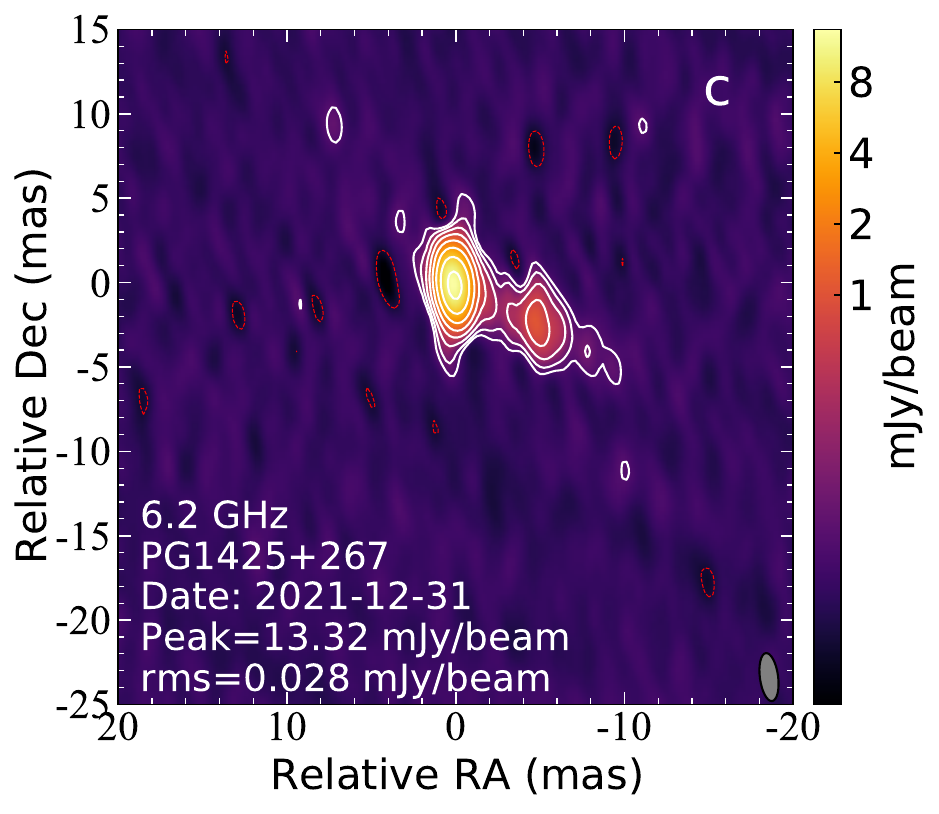}
  \includegraphics[height=3.8cm]{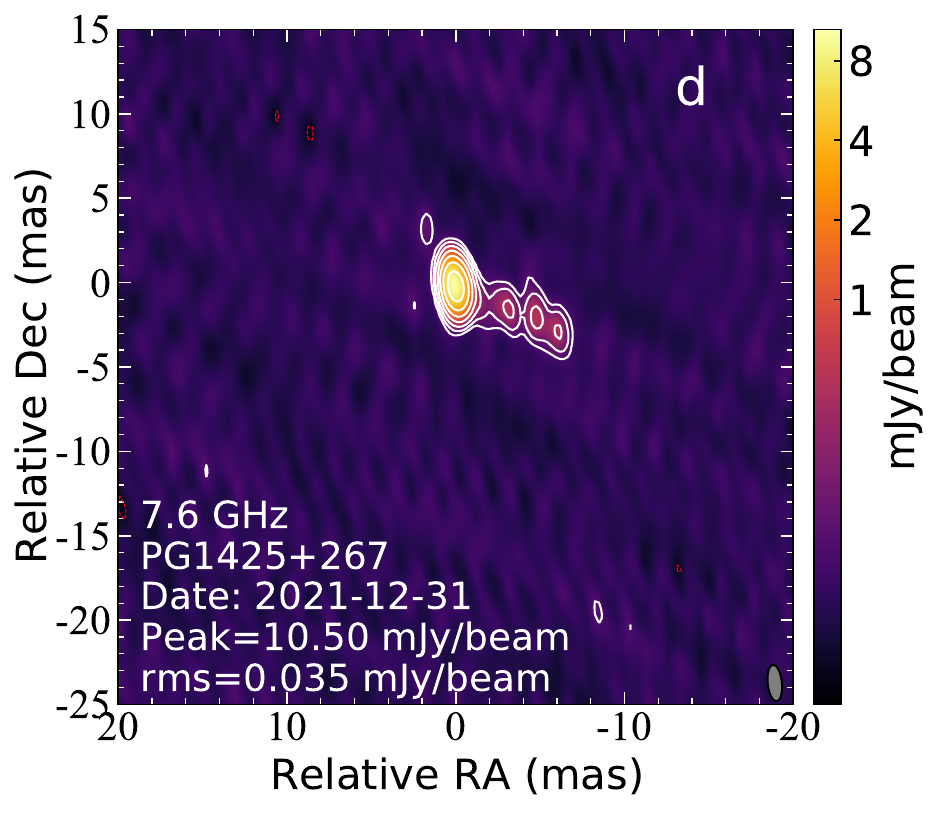}
\\
  \includegraphics[height=3.8cm]{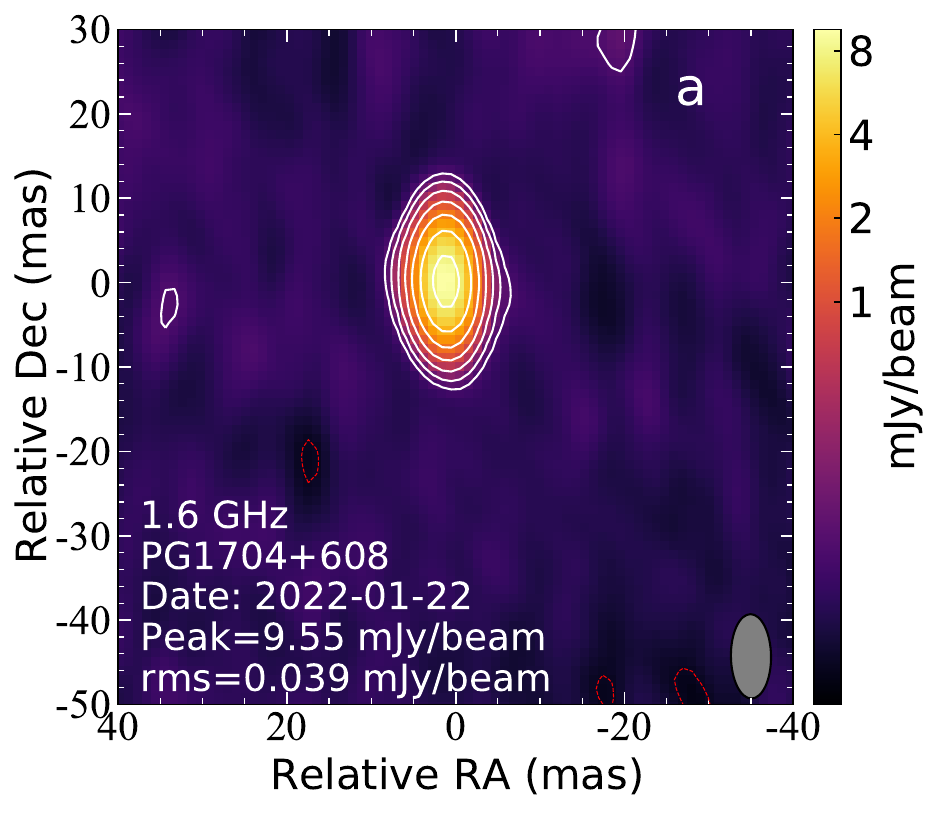}
  \includegraphics[height=3.8cm]{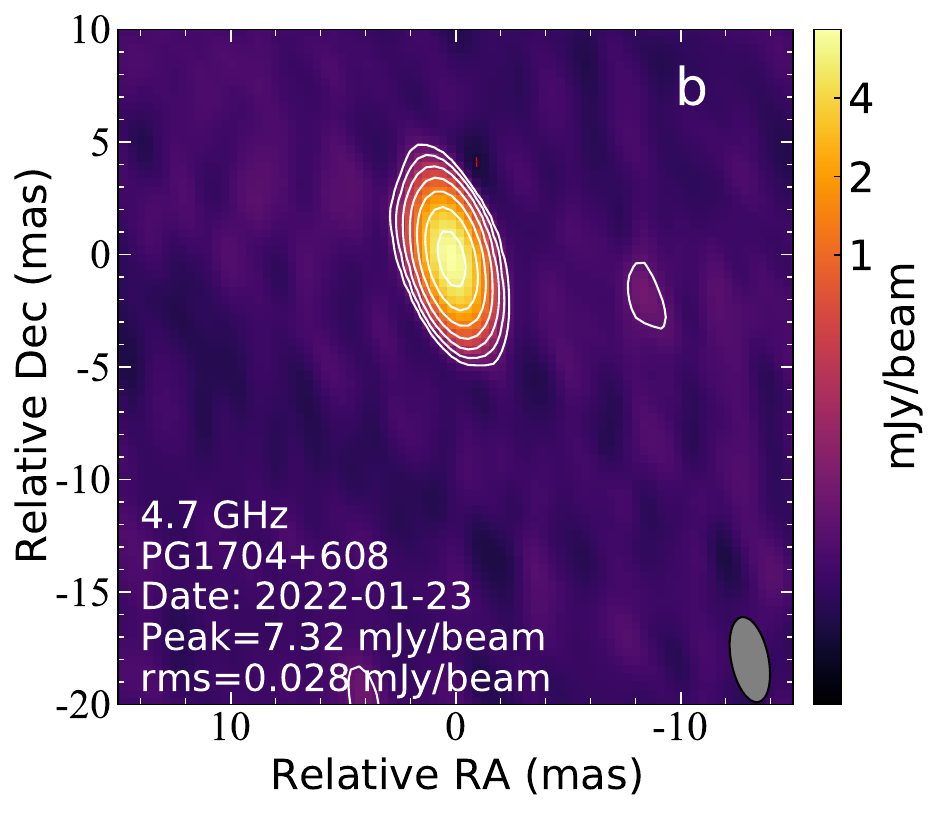}
  \includegraphics[height=3.8cm]{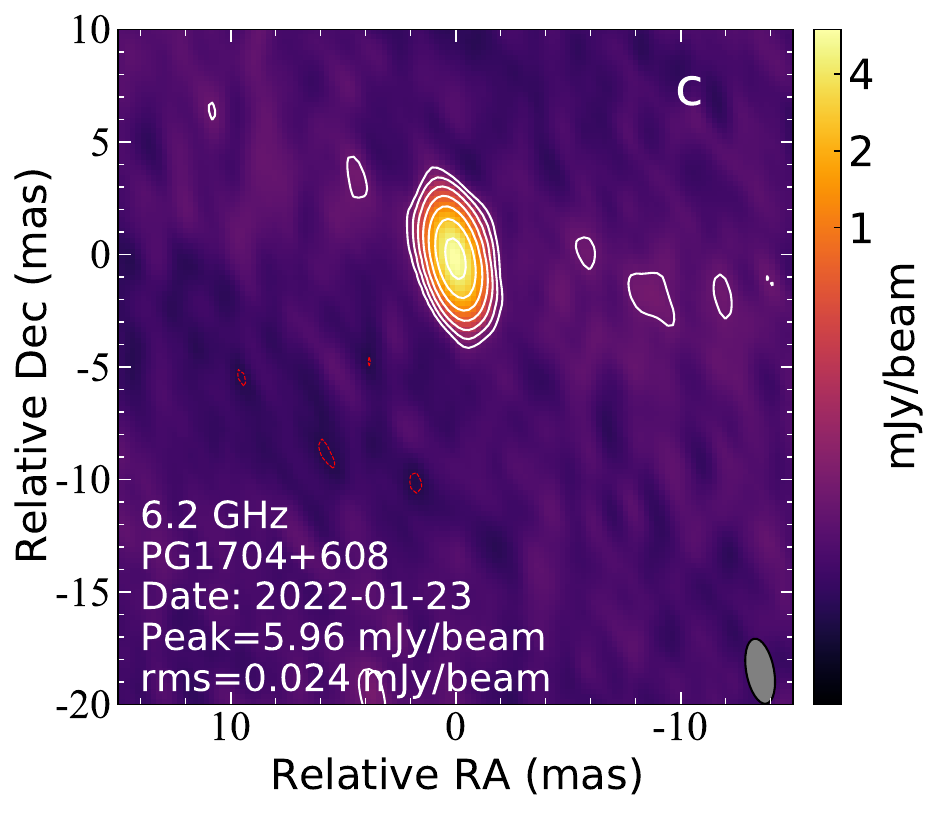}
  \includegraphics[height=3.8cm]{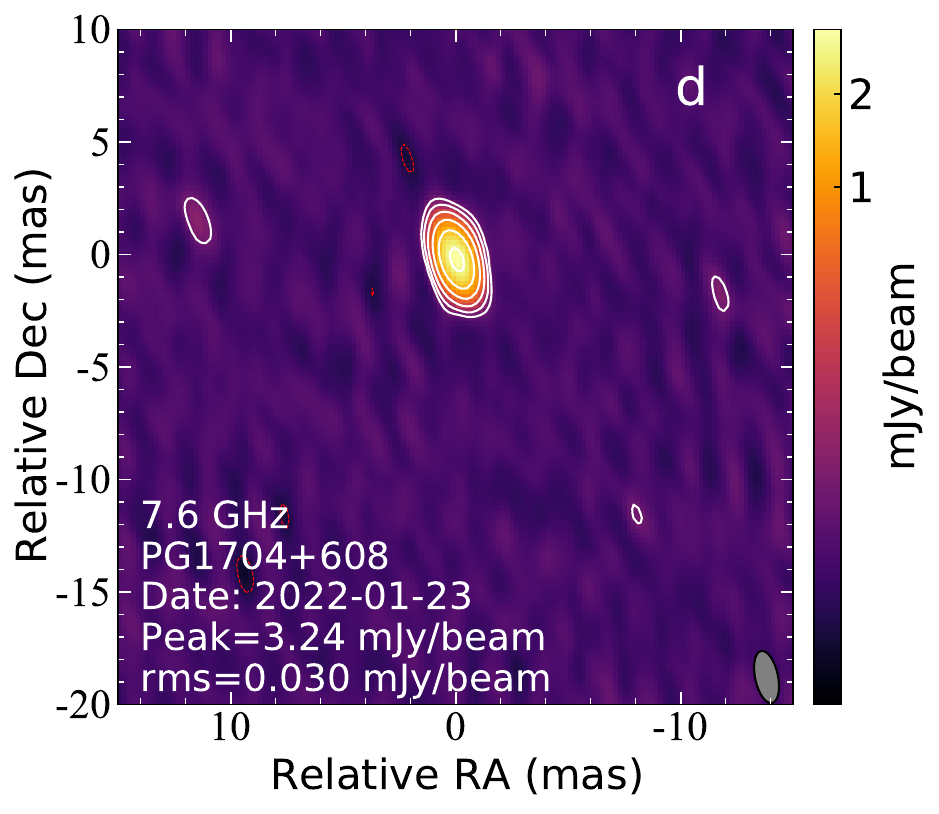}
\end{tabular}
  \caption{VLBA images of radio-loud quasars PG 1004+130, PG~1048-090, PG~1425+267, and PG~1704+608.  }
  \label{fig:RLQ}
\end{figure*}

\begin{figure*}
\centering
  \begin{tabular}{cccc}
  \includegraphics[height=4.5cm]{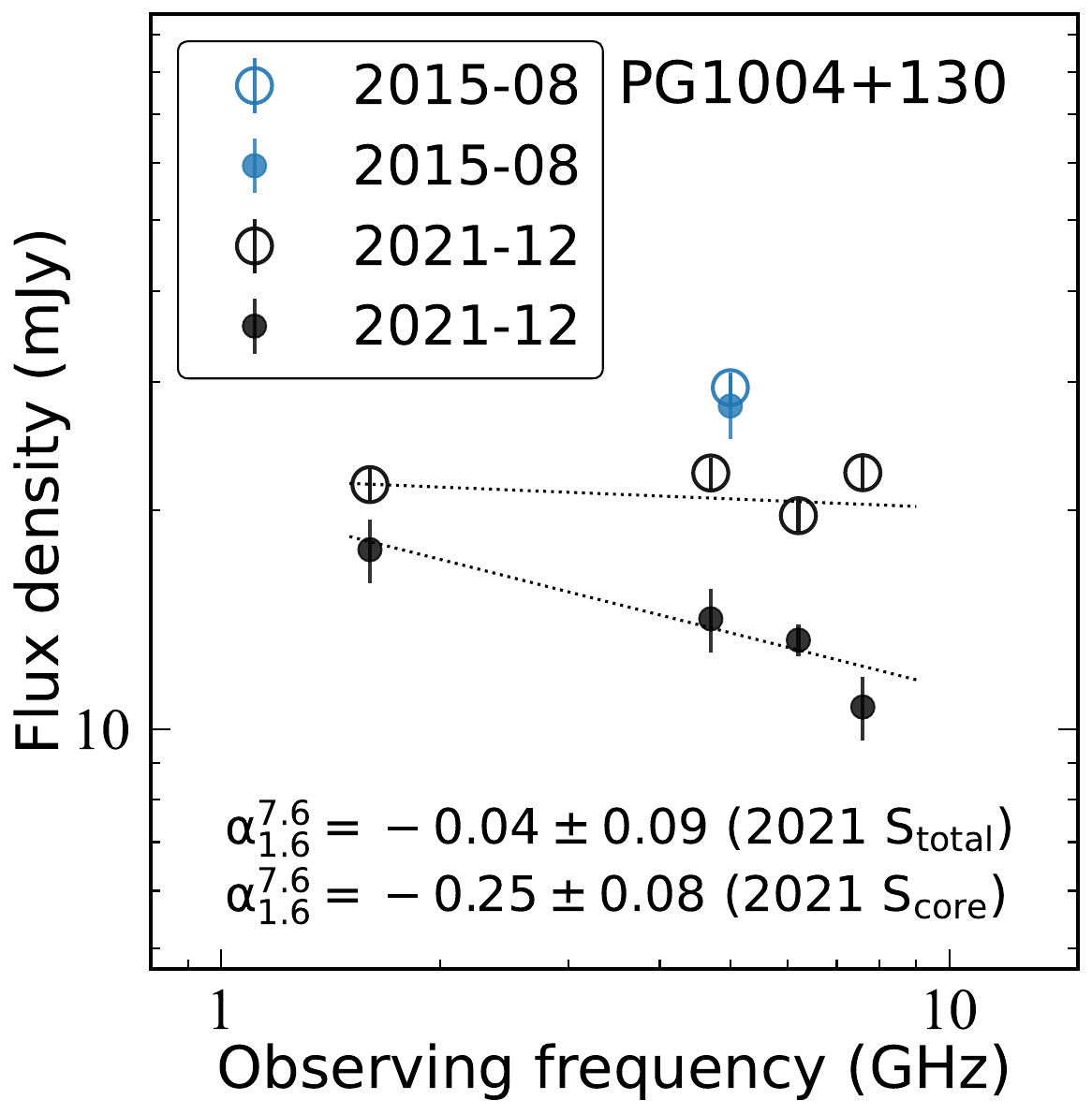}
  \includegraphics[height=4.5cm]{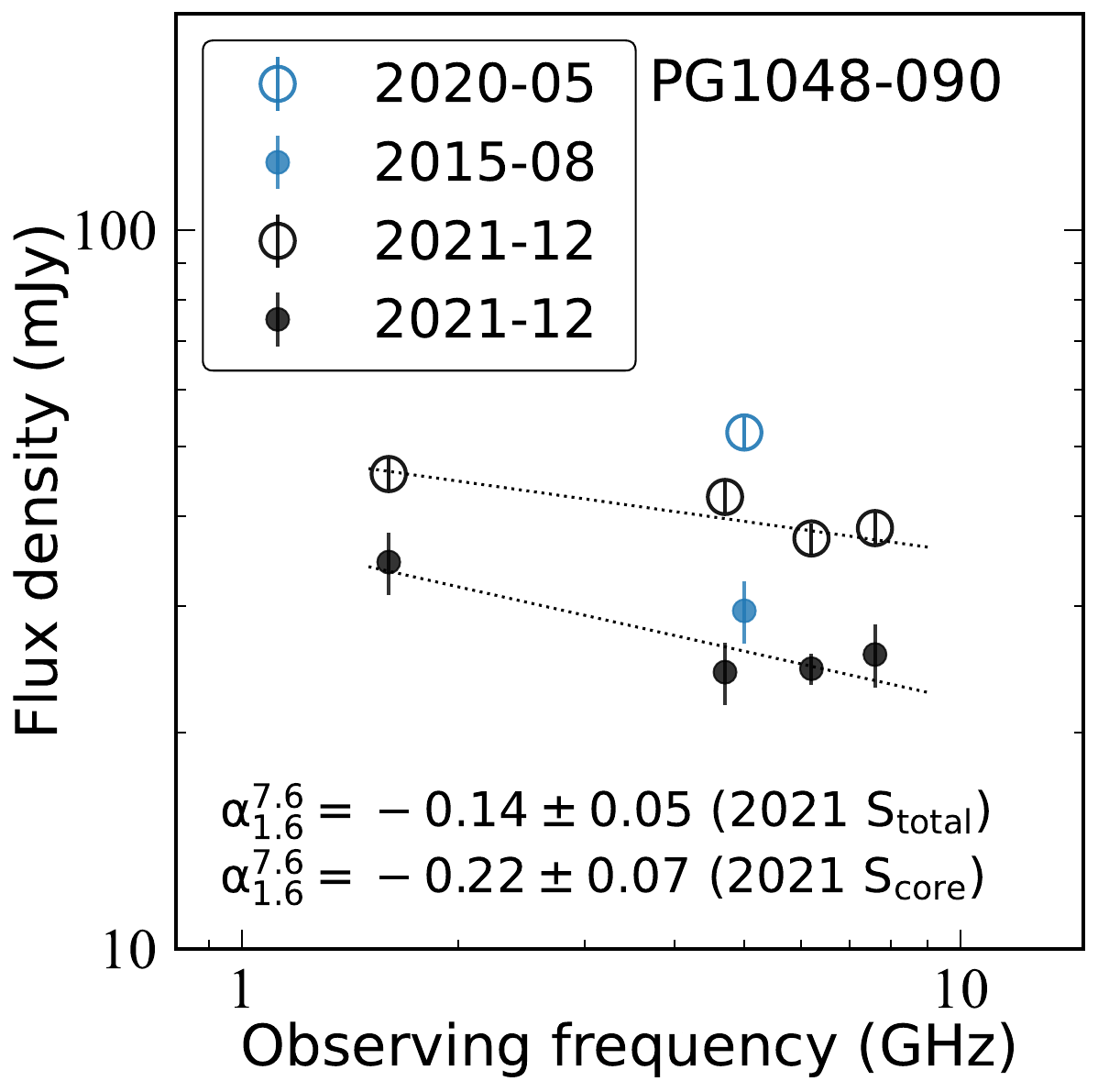}
  \includegraphics[height=4.5cm]{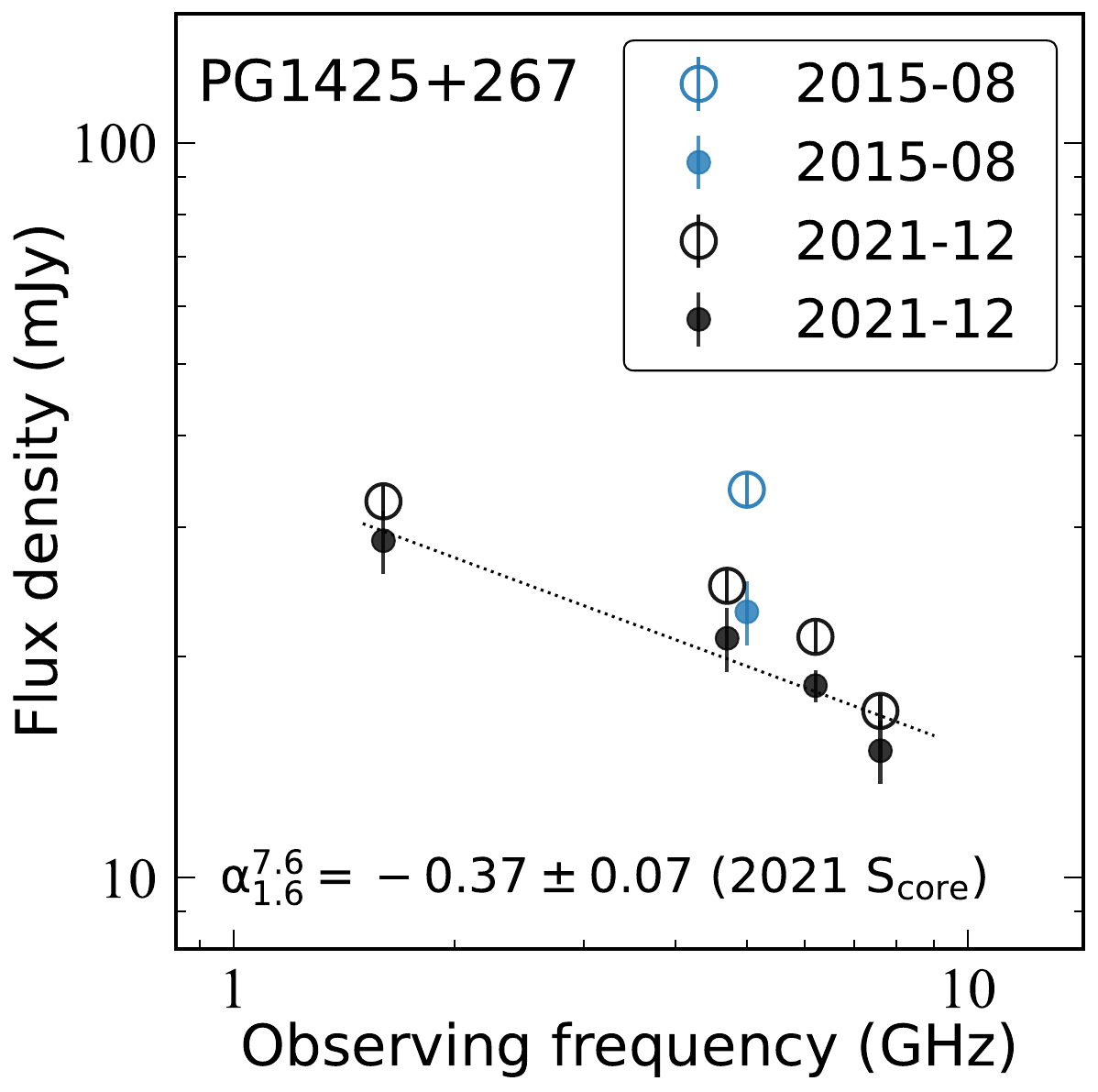}
  \includegraphics[height=4.5cm]{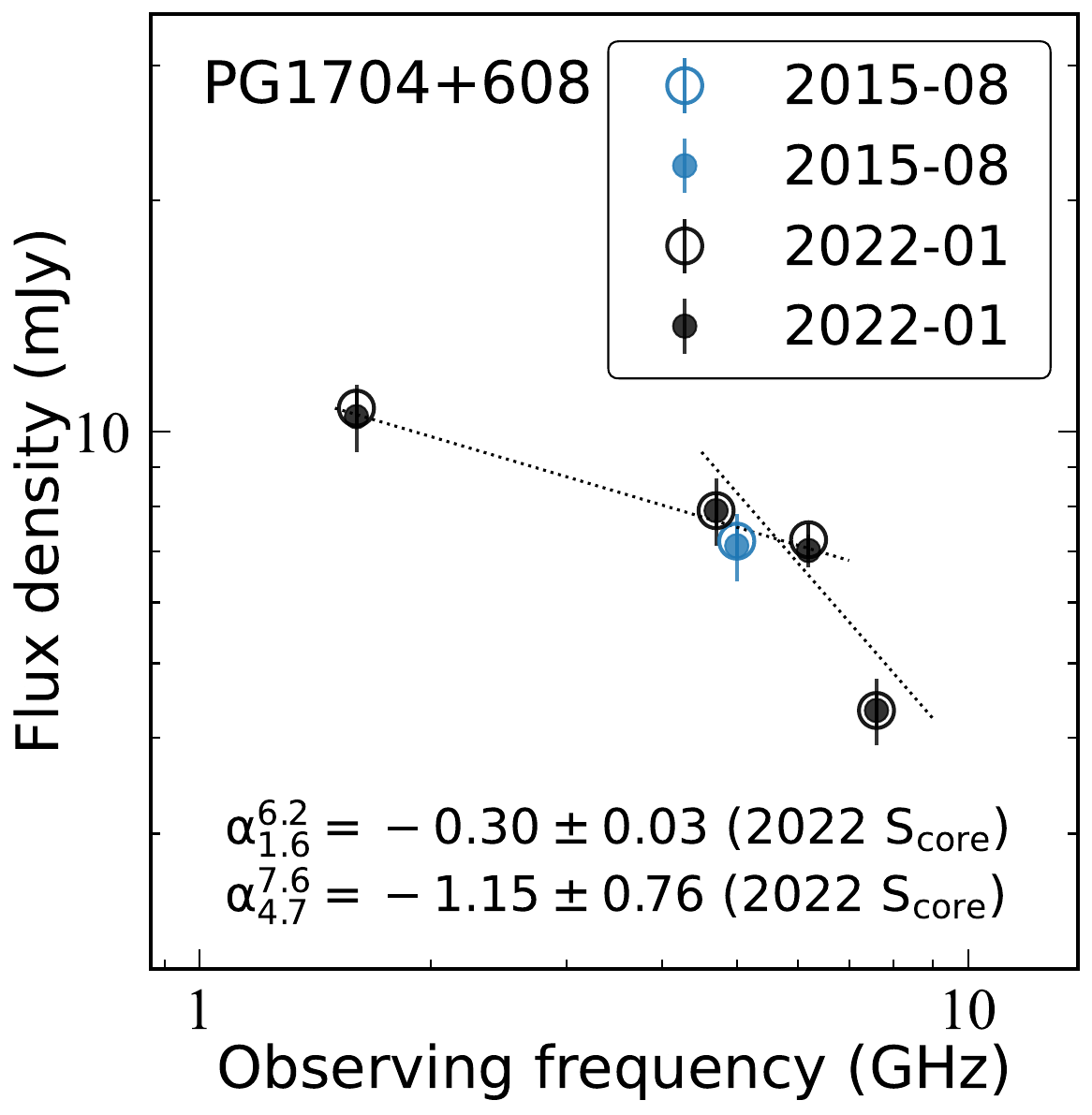} 
  \end{tabular}
  \caption{A compilation of the overall 1.6--7.6 GHz SED of 4 RLQs based on VLBI data including epoch 2015 \citep{2023MNRAS.518...39W} and epoch 2021 (Table \ref{tab:imagepar}). }
  \label{fig:SED-RLQ}
\end{figure*}

We have also studied the radio-emitting properties of the four RLQs. On kiloparsec scales, all four RLQs display Fanaroff-Riley type \textsc{II} radio structures \citep{1993MNRAS.263..425M,1994AJ....108.1163K}. On parsec scales, three RLQs (PG~1004+130, PG1048-090, PG~1425+267) exhibit resolved structures, featuring a bright and compact component located at one end of an elongated structure (Figure \ref{fig:RLQ}). This brightest component exhibits a high brightness temperature ($>10^9$ K), a flat spectrum ($-0.40<\alpha^{7.6}_{1.6}<-0.25$) (Table \ref{table:Spectral_Index}), and shows significant variability (Figure \ref{fig:SED-RLQ}). 
All these properties are naturally consistent with a core-jet structure.
These parsec-scale jets extend up to a projected distance of $\sim$35 mas (176 parsec), consistent with the orientation of their kpc-scale lobes. PG~1704+608 has only a compact core, with insignificant variability and a flat or inverted spectrum with the turnover likely being between 1.6 and 4.7 GHz.

On parsec scales, the radio emission of these RLQs is dominated by the relativistic boosted self-absorbed cores, whereas the radio cores of RQQs are either not dominant or mixed with a significant fraction of the optically-thin jet emission (as discussed above). This difference is also manifested in the brightness temperature: the average $T_{\rm B}$ of the VLBI cores of the RLQs, $\bar{T}_{\rm B,RLQ} = 2.14\times 10^{9}$ K, is 30 times higher than the average $T_{\rm B}$ of the RQQ cores in this sample, $\bar{T}_{\rm B,RQQ} = 6.92\times 10^{7} $ K.

\section{Discussion} \label{sec:discussion}

\subsection{Radio radiation mechanism of RQQ}
The radiation mechanism of the radio emission from radio-quiet quasars remains a subject of active debate. The radio emissions from RQQs originate from a variety of sources, each with its own unique set of observational characteristics that vary at different frequencies, spatial and timescales. These processes include star formation, AGN-driven winds, free-free emission from photo-ionised gas, low-power jets, coronal activity in the innermost accretion disc, and their combinations \citep{2019NatAs...3..387P}. This diversity of these origins and characteristics complicates the process of isolating and studying individual emission mechanisms.
In addition, the presence of multiple origins can lead to signal mixing, adding another layer of complexity to the study of RQQ radio emissions. This signal mixing can obscure the individual characteristics of each source, making it difficult to discern the contribution of each source to the overall radio emission. 
Given these complexities, an approach that focuses solely on a single observational feature within a single observational band is likely to introduce bias. This bias may distort our understanding of RQQs and hinder our ability to provide a comprehensive and accurate interpretation of their radio emission.

To address these issues, our study proposes a multifaceted approach. We propose the use of multiple observational features for cross-validation, which will allow us to more accurately define the dominant emission mechanism of the RQQ at a given scale. By comparing and contrasting data from different observational features, we can gain a more comprehensive understanding of RQQ radio emissions. 
This approach will not only help us identify the dominant emission mechanism at specific scales but also improve our overall understanding of these fascinating astronomical objects.

\begin{figure*}
    \centering
    \includegraphics[width=0.9\textwidth]{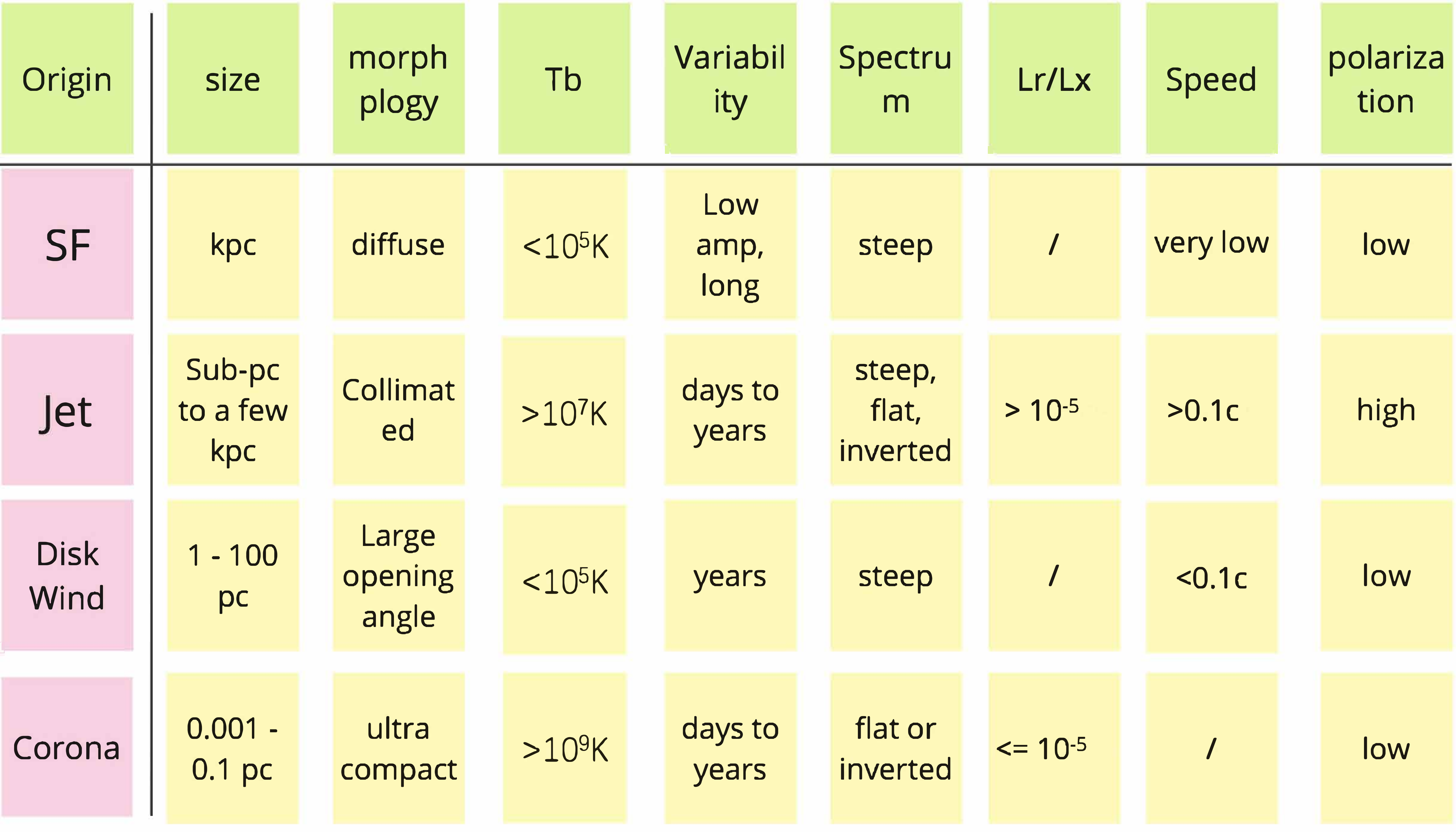}
    \caption{Observational characteristics of different physical mechanisms of radio emission from RQQs. }
    \label{fig:ObsChar}
\end{figure*}

The unique observable features of various physical mechanisms responsible for radio emissions, such as size, morphology, brightness temperature (indicating compactness), component's proper motion velocity, variability, spectrum, the ratio of radio luminosity to X-ray luminosity, and polarisation, serve as initial indicators of the emission source. These characteristics are depicted in Figure \ref{fig:ObsChar}.
The radio emission from star formation has a diffuse morphology that extends throughout the galaxy on kiloparsec scales. It also exhibits low brightness temperatures, typically less than $10^5$ K \citep{1992ARA&A..30..575C}. Only a small fraction of these emissions can be attributed to compact supernovae and supernova remnants clustered in the galactic centre \citep{2006ApJ...647..185L,2015ApJ...799...10B}. The radio spectrum associated with star formation is steep, which can make it difficult to distinguish from extended jet emission using low-resolution observations. However, the morphology and brightness temperature of the radio emission from star formation are very different from that produced by AGN. This difference allows a clear distinction to be made when using high-resolution radio observations. In addition, the very long timescale and barely detectable variability of the radio emission from star formation serves as another distinguishing feature from AGN origins. The study in this paper is based on mas-resolution VLBI observations, so the contribution of star-forming activity to the parsec-scale radio emission is negligible.

In the nuclear region of a galaxy, radio emission produced by AGN comes from a variety of sources, including jets, accretion disk winds, and corona. A key distinguishing feature among these sources is their brightness temperature: the brightness temperature of the wind typically falls below $10^5$ K, while the brightness temperature of the jet base and the corona can exceed $10^7$ K, even reaching the limit of the inverse Compton catastrophe. 
Both the wind and the jet are forms of AGN outflows, but they differ in their degree of collimation. Jets are highly collimated and maintain a narrow shape over a large distance, while winds are less collimated and spread out more widely. This difference is evident in high-resolution VLBI images. Although both jets and winds can display clumpy structures, the clumps in the wind are irregularly shaped and distributed over a wide range of position angles. In addition, the expansion rate of the wind is significantly lower than that of the jet. These differences in morphology, brightness temperature, and expansion rate help to distinguish between these two types of AGN outflows on parsec scales.

The corona and the jet base share many observational characteristics, such as size, brightness, temperature, variability, and radio spectrum. This overlap often makes them difficult to distinguish in the literature. Theoretically, the proper motion velocity of the radio component can distinguish between the jet and the corona. In practice, however, measuring the proper motion of the radio component in RQQs is often challenging.
Morphology is a direct observational feature that can help to distinguish between a corona and a jet: the corona typically has a very compact structure, while the jet is an elongated structure. In this study (see Table \ref{tab:radioprop}), objects such as PG 0003+199, PG 0050+124, PG 0157+001, have rich elongated radio structures and can therefore be classified as jets.
However, there are often some unresolved "naked" cores in RQQs that cannot be distinguished morphologically as either a jet base (core) or a corona.
The radio spectrum of the corona is typically flat or inverted, while the jet can have a variety of manifestations. An observed steep spectrum can be used to rule out the corona as the origin of an RQQ, as in the case of PG 1351+640.

Several studies have identified a correlation between the radio and X-ray luminosities of RQQs \citep{2008MNRAS.390..847L,2023arXiv230713599C}. This correlation could imply that the processes generating synchrotron emission in the corona of RQQs are similar to those observed during stellar coronal mass ejections. The ratio $L_{R}$/$L_{X}$  can provide further insight into the origin of these emissions. When the $L_{R}$/$L_{X}$ ratio is less than or equal to $10^{-5}$, it indicates that the radio emission is primarily from the corona. Conversely, an $L_{R}$/$L_{X}$ ratio greater than $10^{-5}$ indicates that the radio emission is more likely from a jet.

While the proper motion velocity and polarisation degree of radio structures can serve as indicators to distinguish the jet from other radiation origins, practical observations of these features pose significant challenges, limiting their practical application.
Measuring proper motion requires long-term monitoring of bright, compact, and persistent clumps \citep{2003ApJ...591L.103B,2023MNRAS.523L..30W}. This is a time-consuming process and requires that the clumps remain stable over long periods of time, which is not always the case in RQQs.
Similarly, polarisation measurements require extremely sensitive interferometric observations, which are only feasible with advanced facilities such as the Square Kilometre Array (SKA) and the next-generation Very Large Array (ngVLA). This is because the polarisation signals are often very weak compared to the total intensity of the radio emission, and can be easily swamped by noise or systematic errors in the data.

\subsection{Origin of parsec-scale radio emission in these RQQs}
We determined the origin of the parsec-scale radio emission of the 10 RQQs by using a combined indicator of several observational features (see Table \ref{tab:radioprop}). 

PG 0003+199, PG 0050+124, and PG 0157+001 exhibit elongated, narrow, clumpy structures spanning tens of parsecs, with the optical nuclei located in the middle of these radio structures. The overall radio spectrum observed with VLBI is steep, and the brightness temperatures of the clumps exceed $10^{6}$ K but lower than $2.5 \times 10^7$ K. These features suggest that they are weak jets rather than winds or coronae.

PG 0921+525, PG 1149-110, PG 1216+069, and PG 2304+042 are only detected a compact radio component with very high brightness temperatures (near or above $10^8$ K). The first three also exhibit significant variability. These observations align with the characteristics of a corona. However, the steep spectra of PG 0921+525 and PG 1149-110 are more consistent with a jet origin. The high radio luminosity and high radio/X-ray luminosity ratio of PG 1216+069 favour the presence of a relativistic jet.

PG 1351+640 stands out in this sample with the only proper motion measurement with a $v_{j}=0.37\,c$, indicating a mildly relativistic jet \citep{2023MNRAS.523L..30W}. Observations of PG 1700+518 reveal that it displays jet structure at both parsec and kiloparsec scales (the present paper and \citealt{2012MNRAS.419L..74Y}). Additionally, its pc-scale compact structure exhibits an inverted spectrum and its $L_{R}$/$/L_{X}$ ratio is much higher than $10^{-5}$. These results indicate that the radio emission in PG 1700+518 is mainly from the jet. 

As a control sample, all four radio-loud quasars in this paper consistently exhibit the characteristics of relativistic jets. These characteristics include elongated radio structures (Figure \ref{fig:RLQ}) , high brightness temperatures, flat or inverted spectral patterns, and ultra-high $L_{R}$/$L_{X}$  ratios (Figure \ref{fig:LrLx}).

Our study of RQQs is based on a sample with a flux density limit, specifically, $S_{\rm VLA,5GHz}>1$ mJy. When interpreting these observations, it is important to be aware of potential selection effects. 
Our parent sample, as described in Paper I, contains 16 RQQs. In this paper, however, we focus on a subset of 10 sources that were successfully detected by VLBA. These sources were selected for their high brightness temperatures, which potentially allow us to exclude the contribution from star formation.
It is important to note that the six RQQs in \cite{2023MNRAS.518...39W} that are not detected by VLBI could still contain weak jets that remain undetected due to the lack of sufficiently bright and compact emission components. Our study reveals examples of such weak and extended jets in PG 0050+124 and PG 1612+261. While there may be a correlation between the probability of VLBI detections of radio cores and jets and the total flux density, it is not currently possible to make a precise quantitative correlation based on a small sample.

\begin{table*}
\caption{Properties of parsec-scale radio emission of the present PG quasar sample. Columns (1): source name; Columns (2): the size in pc-scale; Columns (3): Morphology; Columns (4): brightness temperature of the core component; Columns (5): speed; Columns (6): variability; Columns (7): Spectrum of the whole VLBI structure (the Spectral Index listed in Table \ref{table:Spectral_Index}); Columns (8): the ratio of radio luminosity in VLBI scale and X-ray luminosity;  Columns (9):  radio luminosity in VLBI scale; Columns (10): Origin of radio radiation in pc-scale.}
\begin{tabular}{c|cccccccc|c} \hline\hline
Source      & Size      & Morphology            & $\log T_{b}$(K) & $v_j$    & Var       & Spectrum     & $\log\frac{L_R}{L_X}$ & $L_{\rm 5GHz}$   & Origin \\ \hline 
            & (mas)     &                       & log (K)         & (c)      &           &              &                       & erg s$^{-1}$     &        \\ \hline 
(1)         & (2)       & (3)                   & (4)             & (5)      & (6)       & (7)          & (8)                   & (9)              & (10)   \\ \hline  
\multicolumn{10}{c}{radio-quiet quasars} \\
PG 0003+199 & 80        & core + two-sided jets & 7.4             & ...      & weak      & steep        & $-$5.76               & 38.04            & jet \\
PG 0050+124 & 20        & core + two-sided jet  & 6.0             & ...      & no        & steep        & $-$5.29               & 38.89            & jet \\
PG 0157+001 & 60        & two-sided jet         & ...             & ...      & ...       & steep        & $-$4.51               & 39.71            & jet \\
PG 0921+525 & 5         & core + one-sided jet  & 8.1             & ...      & strong    & steep        & $-$5.30               & 38.22            & jet \\
PG 1149-110 & 5         & core (+ one-sided jet)& 8.4             & ...      & strong    & steep        & $-$5.48               & 37.35            & jet \\
PG 1216+069 &$<1$       & core                  & 10.2            & ...      & strong    &inverted      & $-$3.99               & 41.11            & jet \\
PG 1351+640 & 5         & core + one-sided jet  & 9.4             &0.37      & strong    & steep        & $-$3.79               & 39.74            & jet \\
PG 1612+261 & 9         & core + one-sided jet  & 6.9             & ...      & strong    & ...          & $-$5.93               & 38.60            & jet \\
PG 1700+518 & 10        & core + one-sided jet  & 8.9             & ...      & weak      & inverted     & $-$2.71               & 40.19            & jet \\ 
PG 2304+042 &$<1$       & core                  & 7.7             & ...      & weak      & inverted     & $-$5.74               & 38.15            & jet/corona  \\ \hline
\multicolumn{10}{c}{radio-loud quasars} \\
PG 1004+130 & 35        & core + one-sided jet  & 9.6             & ...      & strong    & flat         & $-$3.24               & 41.24            & jet \\
PG 1048-090 & 25        & core + one-sided jet  & 9.9             & ...      & weak      & flat         & $-$3.30               & 41.86            & jet \\
PG 1425+267 & 60        & core + one-sided jet  & 8.9             & ...      & weak      & flat         & $-$2.53               & 41.68            & jet \\
PG 1704+608 & 5         & core                  & 9.4             & ...      & no        & inverted     & $-$3.44               & 41.20            & jet \\ \hline

\end{tabular} \label{tab:radioprop}
\end{table*}

\section{Summary}  \label{sec:summary}

In this study, we have performed multi-frequency Very Long Baseline Interferometry (VLBI) observations of a sample consisting of 10 radio-quiet quasars (RQQs) and four radio-loud quasars (RLQs) to investigate their parsec-scale radio emission. Our results reveal a wide range of radio morphologies and spectral properties, indicating that the detected radio emission in these RQQs is primarily associated with compact jets or coronae. The high resolution of our VLBI observations allows us to identify and characterise these compact features, while also resolving other emission structures such as star formation or large-scale extended jets. For nine of the sources, the observed radio morphology, spectrum, and variability strongly support a jet origin for their parsec-scale emission. However, for the remaining source, the compact VLBI components do not show a clear distinction between a jet base and a corona. On the other hand, the radio emission of RLQs on parsec scales is predominantly dominated by relativistic, self-absorbed cores, in contrast to the radio cores of RQQs, which is either not dominant or mixed with a significant fraction of optically thin jet emission. Our study highlights the importance of a multifaceted approach that incorporates multiple observational features for cross-validation, allowing a more precise determination of the dominant emission mechanism in RQQs at specific spatial and temporal scales. We also discuss the potential selection effects inherent in our sample and highlight the need for further research in this area.

\section*{Acknowledgements}
This work is supported by the National SKA Program of China (grant number 2022SKA0120102).  
AW and TA thank Jun Yang for his help in observation preparation and data Processing.
TA is supported by the Tianchi Talents Program of Xinjiang Uygur Autonomous Region. 
X-P. Cheng is supported by the Brain Pool Programme through the National Research Foundation of Korea (NRF) funded by the Ministry of Science and ICT (2019H1D3A1A01102564).
LCH was supported by the National Science Foundation of China (11721303, 11991052, 12011540375) and the China Manned Space Project (CMS-CSST-2021-A04, CMS-CSST-2021-A06).
WAB acknowledges the support from the National Natural Science Foundation of China under grant No.11433008 and the Chinese Academy of Sciences President's International Fellowship Initiative under grants No. 2021VMA0008 and 2022VMA0019. 
The VLBA observations were sponsored by Shanghai Astronomical Observatory through an MoU with the NRAO. 
The National Radio Astronomy Observatory is a facility of the National Science Foundation operated under cooperative agreement by Associated Universities, Inc. Scientific results from data presented in this publication are derived from the following VLBA project codes: BA114, BW138.
The VLBI data processing made use of the compute resource of the China SKA Regional Centre prototype, funded by the Ministry of Science and Technology of China and the Chinese Academy of Sciences. 

\section*{DATA AVAILABILITY}
The correlated data used in this article are available in the NRAO VLBA data archive (\url{https://archive.nrao.edu/archive/archiveproject.jsp}). The calibrated visibility data underlying this article can be requested from the corresponding author.




\bibliographystyle{mnras}
\bibliography{ref}  


\appendix
\onecolumn

\section{Individual radio quiet quasars}

\subsection{PG~0003+199 (Mrk~335)}


PG 0003+199 (Mrk~335; $z=0.025785$: \citealt{1999ApJS..121..287H}) is a low-luminosity Narrow Line Seyfert 1 (NLS1) galaxy with a central black hole (mass of $\sim10^7 M_\odot$, \citealt{2004ApJ...613..682P,2014ApJ...782...45D}) in a super-Eddington accretion state \citep{2013PhRvL.110h1301W}. 
It is one of the brightest AGN in X-rays and has been extensively observed at optical and X-ray wavelengths  \citep[e.g.][]{2008ApJ...681..982G,2013ApJ...766..104L,2015MNRAS.446..633G,2018MNRAS.478.2557G}, exhibiting remarkable X-ray variability \citep{2008ApJ...681..982G,2012ApJS..199...28G,2018MNRAS.478.2557G,2020MNRAS.499.1266T,2020A&A...643L...7K}.  
Observations at radio wavelengths, especially high-resolution imaging before and after large X-ray flares, can help to reveal the jet structure in the heart of the galaxy and the connection between the jet and the X-ray variability.

Recently, \citet{2022MNRAS.510.1043B} obtained a 45-GHz VLA image of Mrk~335, revealing a compact component with a size of less than 50 parsec and a total flux density of $\sim$0.6 mJy. 
The spectral indices of $\alpha_{\rm 5GHz}^{\rm 8.5GHz} = -0.86$ \citep{2019MNRAS.482.5513L} and $\alpha_{\rm 5GHz}^{\rm 45GHz} = -0.83$ \citep{2022MNRAS.510.1043B} are consistent with a typical non-thermal synchrotron radiation spectrum, suggesting that the total radio emission is dominated by optically-thin jets or winds in the nuclear region of this galaxy \footnote{The spectral index $\alpha$ is defined as $S_\nu \propto \nu^\alpha$}.  
Panels \textit{b} and \textit{c} in Figure~\ref{fig:0003} show the 4.7-GHz image obtained from the new VLBA observation, which is in good agreement with that shown in Fig.~2 in \citet{2023MNRAS.518...39W}.  All these images display an elongated structure along the northeast-southwest direction with a maximum extension of $\sim$30 mas ($\sim$16 parsec). This radio structure is resolved into several bright knots. 

In the 7.6 GHz image (not shown), no radio feature above $5\sigma$ is detected.  We have therefore combined the 4.7 and 7.6 GHz visibility data to obtain an image at the intermediate frequency of 6.2 GHz (Figure \ref{fig:0003}, panel $d$). It has a higher resolution and higher sensitivity than the 4.7-GHz image. The central compact component corresponding to the optical \textit{Gaia} nucleus is the brightest in the 6.2-GHz image, while the other components become weaker, and their brightness decrease at 6.2 GHz is due to either being more resolved by the higher resolution or their steeper spectra.

The central compact component has a size of less than 1 mas ($\sim 0.5$ parsec) and a brightness temperature of $> 10^{7}$ K. Its spectral index between 4.7 and 6.2 GHz is $\alpha=-0.37\pm0.03$ (Fig.~\ref{fig:0003}-$e$), supporting it as a flat-spectrum radio core. The other components have steep spectra and show a collimated morphology and symmetry with respect to the core; we therefore identify them as jet knots. In the 1.6 GHz image (Fig.~\ref{fig:0003}-$a$), the core is surrounded by brighter jet emission, so its fitted size and flux density cannot be accurately derived. Therefore only 4.7 and 6.2 GHz data points have been fitted in Fig.~\ref{fig:0003}-$e$. The extended jet in the 1.6 GHz image is much larger. About 42 mas (22 parsec) north of the nucleus, there is a more diffuse component, probably a hot spot in the jet. The southern jet is brighter and corresponds to the advancing jet; it is oriented southwest up to $\sim$20 mas, but beyond 20 mas the jet turns south. The entire ridge of the jet appears S-shaped.

The total size of the radio structure shown in the 1.6-GHz VLBI image is about 80 mas ($\lesssim41$ parsec), which is consistent with the radio structure obtained by \citet{2022MNRAS.510.1043B} in their 45-GHz VLA image. Comparing the flux densities of $S_{\rm VLA}^{\rm 45GHz}$ and $S_{\rm VLBA}^{\rm 1.6GHz}$, we obtain a spectral index of $\alpha^{\rm 45GHz}_{\rm 1.6GHz} = -0.74 \pm 0.03$, which is in general agreement with the spectral index measured from the VLA observations \citep{2019MNRAS.482.5513L,2022MNRAS.510.1043B} and with the spectral index of the entire VLBI structure of $\alpha^{6.2}_{1.6} = -1.50\pm0.09$. This overall spectral index is much steeper than that of the radio core, reinforcing the dominance of the optically thin jet emission in the radio structure on both kpc and parsec scales.

A comparison of the results of our two 5-GHz VLBA observations shows variability in the radio core of  PG~0003+199 (Fig.~\ref{fig:0003}-$e$). The flux density changed from $0.23\pm0.02$ mJy in 2015 to $0.31\pm0.02$ mJy in 2021, increasing by 35 per cent. The total flux density obtained in the VLBA images increased from $1.06\pm0.10$ mJy in 2015 to $1.49\pm0.07$ mJy in 2021, an increase of approximately 40 per cent. The VLA monitoring at 8.4 GHz from January 1997 to April 1999 did not find significant variability of the total flux density \citep{2005ApJ...618..108B}. The flux density measured using VLBA at 1.5 GHz in 2018 \citep{2021MNRAS.508.1305Y} is $6.46\pm0.06$ mJy. In our 2021 observation at 1.6 GHz we obtained a flux density of $6.99\pm0.35$ mJy, indicating that there is no significant variability at 1.5 GHz. The two 5-GHz VLBA images show differences in detail. For example, the northern jet was weakened in the 2021 image, and the southern jet became brighter. However, due to the complex structure of the jet and the long interval between the two observations, we cannot relate the jet clumps from one to the other or determine the jet's proper motion.

A well-known observational feature of PG~0003+199 is the extremely strong X-ray variability (up to a factor of 50), which is thought to be caused by changes in the column density and covering factor of the X-ray absorber \citep[e.g., the clumpy accretion disc winds, ][]{2020A&A...643L...7K} or by the formation of jet-like corona in the X-ray flare state \citep{2019MNRAS.484.4287G}. The interactions of high-speed winds with the ISM should produce notable flux and structural changes in the radio core. The size of the radio core in our VLBA images is $\lesssim0.5$ parsec, close to the size of a typical broad line region. However, the VLBA images do not reveal any notable change that would be expected from a high-velocity, clumpy accretion disc wind or extended corona. Verifying the presence of winds or diffuse outflows requires extremely high-sensitivity VLBI images, e.g., using the High Sensitivity Array.

In summary, the 1.6 GHz VLBA image of PG~0003+199 shows continuous radio emission on a scale of $\sim40$ parsec, and the 5 GHz VLBA image reveals compact clumps with high brightness temperatures of $2.4\times10^{7}$ K, aligned along a straight line. The bright clumps show variability at 5 GHz over an interval of 6.32 years, but the low-frequency VLBI structure (dominated by optically thin jets) shows no significant variability. These observational results, such as the morphology, brightness temperature, spectral index, and variability, suggest that the parsec-scale radio emission detected in the VLBI images of PG~0003+199 originates from the jet. The discovery of active radio jets in black holes with high Eddington accretion ratios challenges the conventional understanding of the radiation model of AGN. It will be worthwhile to study PG~0003+199 in more depth to explore the coupling between the jet and the accretion disc.

\subsection{PG~0050+124 (Mrk~1502, I~Zw~1)}

The host spiral galaxy of PG~0050+124 (I~Zwicky~001, Mrk~1502, UGC~00545, $z=0.061$) is a  signature of galactic mergers, e.g., two tidal tails \citep{1999A&A...349..735Z}. \citet{2019ApJ...876..102H} formed a black hole mass of about $10^7 M_\odot$ for PG~0050+124 using the optical reverberation mapping technique and inferred that it is in  a super-Eddington accretion state (similar to PG~0003+199 in our sample). Another study using X-ray reverberation also measured a similar black hole mass \citep{2021Natur.595..657W}. One of the most intriguing observational features of PG~0050+124 is the multi-component outflows, including ultra-high velocity outflows of up to $0.26\,c$ \citep{2022MNRAS.516.5171R}. These outflows play a crucial role in AGN feedback.

In the early VLA images \citep[e.g. ][]{1994AJ....108.1163K,1995MNRAS.276.1262K}, PG~0050+124 had an unresolved morphology. The more recent VLA A-array 45 GHz image \citep{2022MNRAS.510.1043B} displays a marginal extension of about 200 parsec in length from the radio core to the north. The 5-GHz VLBA image in our Paper 1 shows a fragmented structure: a $4\sigma$ component near the position of the optical nucleus; two clumps at about 11 mas southeast of the optical nucleus, with the brightness of 5 and 4 times the noise level, respectively. In order to classify the radio structure of PG~0050+124, we carried out longer-integration imaging observations of this source with the sensitive EVN+eMERLIN. The high-sensitivity images of PG~0050+124 are shown in Fig.~\ref{fig:0050}.

EVN has large sensitive antennas; our EVN+e-MERLIN observation time of 5.9 h is seven times longer than the VLBA exposure time in Paper 1; EVN+e-MERLIN has better (u,v) coverage in short spacing to recover diffuse emission. These factors allow the EVN image (Fig.~\ref{fig:0050}) to have very high sensitivity and good image quality. The \textit{rms} noise is estimated to be  about 9~$\mu$Jy beam$^{-1}$, which is about 2.5 times lower than the noise of the VLBA image at a similar frequency in our Paper 1 \citep{2023MNRAS.518...39W}. Thanks to its high sensitivity, the EVN+eMERLIN image reveals a richer jet structure than the VLBA image.

We set the observation pointing to the location of the optical nucleus obtained from the \textit{Gaia} observations \citep{2023A&A...674A...1G}, which corresponds to the image centre marked by the cross in the figure. In the previous 5 GHz VLBA image \citep{2023MNRAS.518...39W}, we detected only two knotty components, one close to the optical nucleus and the other located about 11.4 mas to the east of the nucleus. 
In contrast, in our high-sensitivity EVN images, we detected a continuous core-jet structure extending eastward from the nucleus by $\sim20$ mas (projected size about 23 parsec). The brightness temperature of the core is $\sim10^{6}$ K. The flat spectrum of the core is $\alpha^{4.8}_{1.4}=0.00\pm0.29$ \citep[see Fig. \ref{fig:0050}--c panel]{2022ApJ...936...73A}. In addition to the core, which is compact and bright, the other bright component (brightness temperature $6.03\times10^{5}$ K) is a jet knot with steep spectral of $\alpha^{4.8}_{1.4}=-1.21\pm0.19$ \citep{2022ApJ...936...73A} ($\sim$11 mas east of the core). 
\cite{2022ApJ...936...73A} presented the 1.4 and 4.8 GHz image of PG 0050+124, which shows a similar east-west extended jet of comparable length with the 5-GHz jet revealed by our EVN image. The high resolution of the EVN image shows the details of the jet structure. The jet appears to exhibit a sinusoidal-shape morphology, possibly due to hydrodynamic instabilities in the surface layer of the weak jet \citep{2000ApJ...533..176H} caused by the interactions of the jet with the ISM of the host galaxy (see a similar oscillatory jet structure in 3C~48, \citealt{2010MNRAS.402...87A}). To the west of the core at about 13.6 mas, there is a faint component of about 0.05 \mJyb, which is located symmetrically with the bright eastern jet knot with respect to the core and appears to be a bright knot in the counter jet.

\subsection{PG~0157+001 (Mrk~1014)}

PG~0157+001 (Mrk~1014, $z = 0.163$) is in the late stages of a galaxy merger, characterised by a prominent tidal tail to the northeast of its host galaxy \citep[][]{1998ApJ...492..116S,1999MNRAS.308..377M,2000AJ....119..991S} and an excess of infrared emission \citep[e.g. ][]{1988ApJ...335L...1S}. On the other hand, it has been identified as a radio-quiet quasar \citep{1983ApJ...269..352S,1989AJ.....98.1195K}, suggesting that it is in an evolutionary transition from a ULIRG to a quasar \citep{1988ApJ...328L..35S}.

The radio emission of PG~0157+001 shows a triple structure along the east-west direction in the VLA map at the sub-arcsec resolution, with a total size of about 2\farcs6 \citep{1993MNRAS.263..425M,2006A&A...455..161L}. The western lobe is much brighter than the eastern one. The radio structure seems to correlate with the optical emission line structure \citep{2002ApJ...574L.105B}. \citet{2006A&A...455..161L} interpret the western lobe as a shock created by the collision of the jet with the ISM. Other VLA images with lower sensitivity only show the central VLA core and the western lobe \citep{1994AJ....108.1163K,1998MNRAS.297..366K}.
In the recent VLA images at 45 GHz \citep{2022MNRAS.510.1043B}, only a faint component is detected at the location of the optical nucleus.

In our 5-GHz VLBA image of Paper I, no VLBI component was detected near the position of the \textit{Gaia} optical nucleus, but several clumps were detected about 25 mas ($\sim$70 parsec) east of the optical nucleus. The new 5-GHz VLBA image presented in this paper (Fig.~\ref{fig:0157}-$b$ and Fig.~\ref{fig:0157}-$c$) confirms the presence of the clumps at a distance of 20--30 mas east of the optical nucleus. However, these clumps are not detected at 7.6 GHz. We have combined the 4.7 and 7.6 GHz data to produce a 6.2-GHz image, see Figure~\ref{fig:0157}-$d$, showing similar morphology to the 4.7-GHz image. 
The eastern clumps show an extremely steep spectral index $\alpha^{6.2}_{4.7} = -2.44 \pm 0.06$, consistent with optically-thin hotspots in the jet. 
In addition, a faint clump is detected on the intermediate scale (at about 18 mas from the optical nucleus) in Figure~\ref{fig:0157}-$c$.
In Figure~\ref{fig:0157}-$a$, we present the 1.6-GHz VLBA image, which clearly shows the two-sided jet structure, with the eastern jet being brighter. We estimate the spectral index of the eastern jet clump to be $\alpha^{4.7}_{1.6} = -1.70\pm0.09$, in general agreement with the in-band spectral index $\alpha^{6.2}_{4.7}=-2.44 \pm 0.06$ from the simultaneous observations (Fig.~\ref{fig:0157}-$e$).
The brightness temperature of the eastern clump is in the range of (0.8--32.0)$\times10^{6}$~K. PG 0157+001 was also detected by \cite{2023arXiv230713599C}, and their observed result is close to our observation.
The direction of the extended radio structure in the VLBI images aligns with the kpc-scale structure shown in the VLA images. The parsec-scale radio emission may be associated with the eastern lobe revealed in the VLA images.  Because of the low brightness temperature of the VLBA component and the large viewing angle of the jet, the brightness asymmetry of the two-sided parsec-scale jets is therefore unlikely to be caused by the Doppler boosting effect,  but may instead be related to the difference in jet-ISM interactions due to the homogeneity of the ISM distribution.

Comparing the 5 GHz VLBA images observed in 2015, 2021 and 2022, we find that the total flux density increased from $0.99\pm0.05$ mJy in 2015 to $1.56\pm0.08$ mJy in 2021, and further to $1.79\pm0.30$ mJy in 2022 (Fig.~\ref{fig:0157}-$e$). Due to the complex radio structure of this source and the different (u,v) coverage and sensitivity of the three VLBA observations, we cannot ascertain whether the variability is real or not due to observational effects. If there is an intrinsic variability, it is likely to have occurred in the eastern hotspot.

\subsection{PG 0921+525 (Mrk~110)}

The early 5-GHz VLA B-array image of PG~0921+525 (Mrk~110, $z = 0.03529$) shows a highly curved radio structure that initially extends to the northwest and then turns to the north \citep{1994AJ....108.1163K}. In the subsequent 1.4-GHz VLA A-array image, the northern extended structure reaches about 2\farcs5 \citep{1998MNRAS.297..366K}. The latest 5-GHz tapered VLA A-array image reveals extended radio emission up to 4\farcs5 to the south of the core, in addition to the extended radio emission to the north \citep{2022A&A...658A..12J}. 
The spectral index of the VLA core is $-0.33\pm0.01$.   The origin of these arcsecond-scale radio emissions may be starbursts or AGN jets or a combination of both. In the VLBI images (Paper I and Fig. \ref{fig:0921} of this paper), we detect an unresolved component near the optical nucleus position with a 5-GHz flux density of about 1 mJy and a brightness temperature of $\sim10^8$~K, identifying it as the radio core. Using our broad C-band VLBA data, we can measure the spectral index of this VLBI component as $\alpha^{7.6}_{4.7} = -0.51\pm0.09$ (Fig. \ref{fig:0921}-$e$).  
The steeper spectral index of the radio core tends to support it being the base of a weak jet rather than a corona. Due to the mixing of the radio core with the optically thin jet emission, we need to detect the genuine optically thick radio core at higher frequencies.
In the 1.6-GHz VLBA image (Fig.~\ref{fig:0921}-$a$), no extended jet is detected, while the spectral index between 1.6 GHz and 7.6 GHz is slightly steeper, $\alpha^{7.6}_{1.6} = -0.66\pm0.04$. In addition, in our new VLBA images (Fig.~\ref{fig:0921}-b and Fig.~\ref{fig:0921}-c), a compact jet-like feature is found to the southwest of the core, again supporting the core-jet structure. The core-jet structure in PG 0921+525 was also detected by \cite{2023arXiv230713599C} at L and C band.
This compact jet appears to be aligned with the southwest kpc-scale jet \citep{2022A&A...658A..12J}.

PG~0921+525 exhibits variability at both the optical \citep{1997AJ....114..565J} and the X-ray wavelength \citep{2006ApJ...651L..13D}. The variability, on the order of minutes to hours, limits the size of the X-ray emission region to a few Schwarzschild radii \citep{2006ApJ...651L..13D}. Its black hole has an accretion rate close to 10 per cent of the Eddington limit.
Comparing our two VLBA 5-GHz observations separated by 6.5 years, we find that the flux density of the radio core has increased from $1.04\pm0.05$ mJy to $1.12\pm0.06$ mJy, with no significant change (Fig. \ref{fig:0921}-$e$).
However, a moderate-level ($\sim10$ per cent) variability for PG~0921+525 was found by \citet{2022MNRAS.510..718P}  on the days-to-weeks timescale at 5 GHz.

\subsection{PG~1149$-$110}

High-resolution ($0.44\arcsec \times 0.33\arcsec$) 5 GHz VLA images show a triple structure extending 1\farcs5 along the east-west direction \citep{2006A&A...455..161L}. In other higher-frequency VLA images, only the central component is detected \citep{1998MNRAS.297..366K,2022MNRAS.510.1043B}, and the spectral index of the central VLA core is $\alpha = -0.65$. In previously published VLBI images \citep{2022ApJ...936...73A,2023MNRAS.518...39W}, PG~1149-110 shows an unresolved component with a flat spectrum of $\alpha^{4.8}_{1.4} = -0.30\pm0.11$ close to the optical nucleus and can be identified as the radio core (jet base). Our new 4.7-GHz VLBA image (Figure~\ref{fig:1149}-$b$) clearly shows a core and a jet-like extension to the northwest. The jet-like feature is not detected in the 6.2 and 7.6 GHz images, indicating that it has a steep spectrum $\alpha^{7.6}_{4.7} < -1.83$. The compact unresolved component has a steep spectral index of $\alpha^{7.6}_{1.6} = -0.74\pm0.03$ (Fig.~\ref{fig:1149}-$e$) and a brightness temperature of up to $10^8$~K. The 1.6 GHz image in Fig.~\ref{fig:1149}-$a$  shows an unresolved component, indicating that the radio emission is concentrated within 4.4 parsec (the resolution of the 1.6-GHz VLBA). There is significant variability in the VLBA core, with the 5 GHz flux density increasing by a factor of about 1.7 from $0.33\pm0.02$ mJy in 2015 to $0.57\pm0.03$ mJy in 2021 (Fig.~\ref{fig:1149}-$e$). Comparing the VLBA observations made by \citet{2022ApJ...936...73A} and by us shows that the flux density at 1.6 GHz rises from $1.06 \pm 0.11$ mJy in 2020 to $1.28 \pm 0.13$ mJy in 2021. The radio flux density variation is mainly associated with the VLBA core.

\subsection{PG~1216+069}
\label{context:PG1216}

PG~1216+069 ($z=0.3313$) shows compact structure in the VLA images \citep{1993MNRAS.263..425M,1994AJ....108.1163K}. It remains unresolved in the VLBA images, but shows significant variability (Fig. \ref{fig:1216}): on 21 January 2000, $S_{\rm 5GHz}$ = $6.5 \pm 0.3$~mJy \citep{2005ApJ...621..123U}; on 7 August 2015, $S_{\rm 4.9GHz}$ = $1.20\pm0.12$~mJy with a core brightness temperature $>10^8$ K \citep{2023MNRAS.518...39W}; in our new 4.7 GHz VLBA observations, $S_{\rm 4.7GHz} = 8.17 \pm 0.41$ mJy. The VLBI core has a flux density increase by a factor of 6.8 in the 6.5 years from August 2015 to December 2021, which is the largest variability in our RQQ sample. The radio properties, including compact radio structure, high brightness temperature, and extreme variability, rule out the possibility of the radio core being from accretion disk wind. The 8.4 GHz VLBA image of \citet{1998MNRAS.299..165B} shows a possible extension from the core to the south, but only a compact core is detected in our 7.6 GHz image, and our observation was in a flux-enhanced state. Therefore, their elongated radio structure on the parsec scale needs to be treated with caution. 

From our 2021 observations, PG 1216+069 shows a peaked spectrum (black data points in Fig. \ref{fig:1216}-$e$), possibly due to synchrotron self-absorption (SSA) or free-free absorption (FFA). We fitted the radio spectrum of the core with an SSA model \citep{1971PhT....24i..57P,1999A&A...349...45T} and the best-fitted parameters are: the maximum flux density at the turnover frequency $F_{m}=8.99$ mJy, the turnover frequency $\nu_{m}=6.95$ GHz, the spectral indices of the optically thin region $\alpha_{thin}=-3.59$ and the spectral indices of the optically thick region $\alpha_{thick}=1.41$. \cite{2005ApJ...621..123U} also reported an inverted shape spectrum on parsec-scale with a turnover frequency of about 3 GHz (red data points in Fig. \ref{fig:1216}-$e$), which is different from the shape of our VLBA result. 
Similarly, the radio spectrum of its total flux density measured by the VLA also showed an inverted shape between 4.89--14.9 GHz, with the turnover frequency around 8.4 GHz \citep{1996AJ....111.1431B}. PG~1216+069 has significant variability and a changing spectrum, which warrants further study of the relationship between the variability and the evolution of the radio spectrum.

\subsection{PG~1351+640}


PG 1351+640 is not resolved in either the 5 GHz VLA D-array image \citep{1994AJ....108.1163K} or the 1.4 GHz VLA B-array image (the VLA FIRST survey, \citealt{1995ApJ...450..559B}). The most attractive aspect of this source is that in our 5 GHz VLBA image in Paper 1 \citep{2023MNRAS.518...39W}, a structure analogous to Compact Symmetric Objects (CSOs, \citealt{1982A&A...106...21P,1994ApJ...432L..87W}) is presented with two compact components separated by about 8 parsec. This dual-component structure is confirmed in our new VLBA images: in Fig.~\ref{fig:1351}, the 4.7, 6.2 and 7.6 GHz images all show two compact components; in the 1.6 GHz image, although the source is not resolved due to the relatively low resolution, the image shows an elongation in the southeast-northwest direction. In the VLBA image of 9 June 1996 at 8.4 GHz, PG~1351+640 shows a ``core-jet''-like morphology with a weaker jet-like feature extending to the northwest of the core \citep{1998MNRAS.299..165B}, very similar to our results. However, this extension is only about 2.5 mas, which is  about half of the two-component separation in our images. In the earlier (6 February 2000) VLBA images at 1.42, 2.27, and 4.99 GHz, there is only one unresolved core in the PG~1351+640 image \citep{2005ApJ...621..123U}. The position of the optical nucleus is closer to the southeastern component. 

We further measured the spectral indices of two components (see Fig. \ref{fig:1351}-$e$), $\alpha^{7.6}_{1.6}=-0.04\pm0.12$ (southeast component) and $\alpha^{7.6}_{1.6}=-1.45\pm0.12$ (northwest component). Our 1.6 GHz observation, separated by 21 days from the 4.7 GHz observation, yields a spectral index of $\alpha^{4.7}_{1.6} = -0.97\pm0.03$ for the total parsec-scale emission. From the position of the optical nucleus and the flat or inverted radio spectrum, the southeast component appears to be the radio core and the northwest component is the jet. In all VLBI observations, the brightness temperature of the radio core  exceeds $10^7$ K and reaches up to $2.5 \times 10^9$ K, which is at the same level as the brightness temperature of the RLQs in our sample (Table~\ref{tab:imagepar}),  supporting that its radio emission is associated with the optically thick part of the jet base. No extended jet component is found to the southeast of the radio core (Fig. \ref{fig:1351}-a). The one-sided jet morphology may be due to the Doppler boosting effect, although the Doppler factor obtained directly from the VLBI data is not very high.

PG~1351+640 shows prominent variability in several bands. In X-rays, it is strongly variable on timescales of several years \citep{2008A&A...491..425G}. It also varies in the ultraviolet in spectral lines and in the continuum on timescales of about one month, but no significant variation has been found in the optical band \citep{1985MNRAS.216..529T}. Its total radio flux density is highly variable by a factor of more than 4 within a period of 3--4 years at 5 and 15 GHz \citep{1989ApJS...70..257B}. The VLBI core is also highly variable, increasing from 1.98$\pm$0.20 mJy in August 2015 to 2.43$\pm$0.12 mJy in December 2021.
The variability nature of PG~1351+640 strengthens its identification of it as a core-jet source rather than a disk wind.

\subsection{PG~1612+261}


PG~1612+261 (Ton~256, $z=0.131$) shows a triple structure in the VLA images at 5 and 8.4 GHz aligned along the northeast-southwest direction \citep{1993MNRAS.263..425M,1998MNRAS.297..366K,2000ApJ...530..704K,2006A&A...455..161L}, with the eastern jet being brighter but shorter than the western jet. The total flux density of PG~1612+261 has a steep spectrum between 5 and 8.5 GHz with $\alpha^{8.5}_{5.0} = -1.57$, indicating that the VLA core could be dominated by optically thin emission \citep{2019MNRAS.482.5513L}. In the 5-GHz VLBA image of \citet{2023MNRAS.518...39W}, PG~1612+261 shows a faint compact source with a peak flux density of 0.17~mJy. In \citet{2022ApJ...936...73A}, the source was not detected at 4.8 GHz (below $5\sigma$), but the 1.4 GHz VLBA image clearly showed an elongated structure in the east-west direction, which appears to be a core and two-sided jet structure. The brighter component in the centre corresponds to the location of the optical counterpart. Due to the weakness of PG~1612+261 at 5 GHz, in the new VLBA observations, we only detected it at 1.6 GHz. Figure \ref{fig:1612}-$a$ displays our 1.6 GHz VLBA image, showing two components. From the comparison of the 1.6 and 5 GHz images, we can infer that the west component corresponds to the radio core, with an estimated spectral index of $\alpha^{4.9}_{1.6}=-0.28\pm0.03$ (Fig. \ref{fig:1612}-$b$). Recent VLBI observations (\citealt{2022ApJ...936...73A,2023MNRAS.518...39W} and the present paper) show the presence of variability in the parsec-scale emission structure within a few years, supporting that the radio emission comes from the jet rather than a wind.

\subsection{PG~1700+518} \label{sec:1700}


In the VLA image at 8.4 GHz, the radio structure shows two components (VLA-East and VLA-West) separated by about 4 kpc \citep{1998MNRAS.297..366K}. In the VLBI image, VLA-East is fully resolved and is interpreted as a star-forming region or a shock resulting from jet-ISM interactions \citep{2012MNRAS.419L..74Y}. The total flux density of VLA-West shows a steep spectral index $\alpha = -1.02\pm0.01$ between 1.5 and 14.9 GHz \citep{1989ApJS...70..257B}. A part of the optically thin component  is resolved in the VLBI images. VLA-West is resolved into a core and double lobes separated by $0\farcs5$ in Fig.~1 of \citet{2012MNRAS.419L..74Y}. In our VLBA images (Fig.~\ref{fig:1700} in this paper and Fig. 2 in Paper I), only a compact component is detected, corresponding to the component C in \citet{2012MNRAS.419L..74Y}. This component has a high brightness temperature above $3.16\times10^{7}$ K and is identified as the radio core of the AGN. Interestingly, it has a steep spectrum between 4.7 and 7.6 GHz with $\alpha^{7.6}_{4.7} = -1.64\pm0.76$, but a flat spectrum between 1.6 and 4.7 GHz with $\alpha^{4.7}_{1.6} = -0.21\pm0.03$(Fig.~\ref{fig:1700}-$e$). In contrast, the spectral index of the kpc-scale radio core observed with the VLA is $\alpha^{14.9}_{1.49} = -1.02\pm0.01$ \citep{1989ApJS...70..257B}. The discrepancy between the positions of the radio peak and the optical nucleus, shown in Figure~\ref{fig:1700}, suggests that the detected VLBA component may be a bright knot in the jet or corona rather than a genuine radio core (jet base).

We conducted a comparison between two 5-GHz VLBA images acquired in 2015 (1.35 mJy, Paper I) and 2021 (1.33 mJy, this paper) and found no significant variability (Fig. \ref{fig:1700}-$e$). However, a decrease in flux density was observed in the 2022 epoch \citep[0.75 mJy, ][]{2023arXiv230713599C}. This clear decline in flux density is also apparent at 1.6 GHz, reducing from 1.9 mJy \citep[November 2010, ][]{2012MNRAS.419L..74Y} to 1.44 mJy (December 2021, Paper I), and further down to 1.07 mJy \citep[April 2022, ][]{2023arXiv230713599C}.
\citet{2012MNRAS.419L..74Y} tapered the 1.6-GHz EVN data and found two faint radio lobes, separated by about 0\farcs5, with a near-symmetric distribution with respect to the radio core. We have tapered the 1.6 GHz VLBA data in the same way but did not find these two faint features. The radio structure of PG~1700+518 at 1.6 GHz from \cite{2023arXiv230713599C} is also dominated by a compact core.

In Paper I, we found a weak extension to the southeast of the compact radio core, which is confirmed in the new 5 GHz VLBA images in this paper, located 2 mas east and 1.5 mas south of the optical nucleus position. This southeastward extension is also seen in the 1.6-GHz VLBA image and extends farther. 

PG~1700+518 ($z=0.292$) is a well-known broad absorption line quasar with high-speed broad Balmer lines associated with outflows from rotating accretion disk winds \citep{2007Natur.450...74Y}. In AGN, winds and jets can co-exist \citep[e.g. III~Zw~2,][]{2023ApJ...944..187W}. The driving mechanism of the fast outflow, whether wind or jet or both, is still uncertain. Exploring the starting point of the wind requires sub-parsec resolution.  Assuming that the wind and jet are co-axial, high-resolution VLBI images can provide independent constraints on the wind geometry. The outer part of the 1.6-GHz extended jet extends eastward, which means that there is a jet bending from the east to the southeast (the SE component in \citealt{2012MNRAS.419L..74Y}) between 10 mas (43 parsec) and 270 mas (1.2 kpc). 
The bending of the jet could be caused by the jet colliding with clouds in the torus or BLR, or by the pressure gradient of the rotating accretion disk wind. 

\subsection{PG~2304+042}

On the kpc scale, PG~2304+042 was detected only a single compact component \citep{1993MNRAS.263..425M}.  PG~2304+042 shows a compact core component in our VLBA images (Fig.~\ref{fig:2304}), which has a brightness temperature $>3.2\times10^{7}$ K. The spectral index of the radio core is $\alpha^{4.7}_{1.6} = -0.26\pm0.03$ consistent with a flat spectrum of $\alpha^{4.8}_{1.4} = -0.09\pm0.12$ \citep{2022ApJ...936...73A} but is steepened abruptly between 4.7 and 7.6 GHz with a spectral index of $\alpha^{7.6}_{4.7}=-1.36\pm0.40$ (Fig. \ref{fig:2304}-$e$). A similar spectrum shape can be seen in PG~1700+518 (Section~\ref{sec:1700}). There is a weak extension at $\sim2.5$ mas from the core to the northwest in the 4.7 and 7.6 GHz images, but the 1.6 GHz image shows some weak extension to the northeast $\sim14$ mas from the core. \citet{2005ApJ...618..108B} showed that PG~2304+042 has an extreme  variability at 8.6 GHz. But the VLBI core shows no significant variability at C band between epochs 2015 and 2021.

\section{Radio loud quasars}

Figure~\ref{fig:RLQ} shows the images of the four RLQs (PG~1004+130, PG~1048-090, PG~1425+267 and PG~1704+608) at four frequencies, respectively. All three sources show core-jet structures except PG~1704+608, which remains unresolved at all frequencies. The 1.6 GHz image of PG~1425+267 shows a faint component of $\sim20$ mas to the east of the radio core. It may be associated with the counter jet if it is a real feature. Other than that, all other images of PG~1425+267 show a core and a one-sided jet. The brightness temperatures of the radio cores are between $2.0\times10^{8}$ and $7.9\times10^{9}$ K, which is higher than the average brightness temperature of the present RQQ sample by a factor of 3--115. The Doppler boosting factors estimated from core brightness temperatures for these four RLQs do not favour an extreme relativistic beaming effect, suggesting that their jets are only mildly or moderately relativistic. These RLQs exhibit significant variability on VLBI scales (Fig. \ref{fig:SED-RLQ}): the core variability in the 5 GHz images observed in 2015 and 2021/2022 is in the range of 8--49 per cent; the variability of the total flux density ranges from 10 per cent to 26 per cent. The radio variability of RLQs is related to the active relativistic jet base.

\bsp	
\label{lastpage}

\end{document}